\documentclass[a4paper,11pt]{article}
\usepackage{jheppub}
\usepackage{amsthm}
\usepackage{amssymb}
\usepackage{amsthm}
\usepackage{amstext}
\usepackage{amsmath}
\usepackage{breqn}
\usepackage{epstopdf}
\setkeys{breqn}{breakdepth={0}}
\usepackage{amsfonts}
\usepackage{bbm}
\usepackage[german,english]{babel}

\usepackage{verbatim}

\usepackage{caption,subcaption}
\usepackage{graphicx} 
\usepackage{epstopdf}
\usepackage{float}

\usepackage{enumerate}
\usepackage{algorithm}
\usepackage{algorithmic}

\usepackage{tikz}
\usetikzlibrary{positioning}
\usetikzlibrary{arrows.meta}

\usepackage{listings}
\usepackage{xcolor}
\definecolor{codegreen}{rgb}{0,0.6,0}
\definecolor{codegray}{rgb}{0.5,0.5,0.5}
\definecolor{codepurple}{rgb}{0.58,0,0.82}
\definecolor{backcolour}{rgb}{0.93,0.93,0.93}

\theoremstyle{plain}

\theoremstyle{definition}

\allowdisplaybreaks[4]

\title{A study of Feynman integrals with uniform transcendental weights and the symbology from dual conformal symmetry}
\preprint{USTC-ICTS/PCFT-22-17}

\author[a,b,d]{Song He}
\author[a,e]{Zhenjie Li}
\author[c,d]{Rourou Ma}
\author[c,d]{Zihao Wu}
\author[a,e]{Qinglin Yang}
\author[c,d]{Yang Zhang}

\affiliation[a]{CAS Key Laboratory of Theoretical Physics, Institute of Theoretical Physics, Chinese Academy of Sciences, Beijing 100190, China}
\affiliation[b]{School of Fundamental Physics and Mathematical Sciences, Hangzhou Institute for Advanced Study \&  ICTP-AP, UCAS, Hangzhou 310024, China}
\affiliation[c]{Interdisciplinary Center for Theoretical Study, University of Science and Technology of China,
Hefei, Anhui 230026, China}
\affiliation[d]{Peng Huanwu Center for Fundamental Theory, Hefei, Anhui 230026, China}
\affiliation[e]{School of Physical Sciences, University of Chinese Academy of Sciences, No.19A Yuquan Road, Beijing 100049, China}

\abstract{
Multi-loop Feynman integrals are key objects for the high-order correction computations in high energy phenomenology. These integrals with multiple scales, may have complicated symbol structures. We show that the dual conformal symmetry sheds light on the alphabet and symbol structures of multi-loop Feynman integrals. In this paper,
first, as a cutting-edge example, we derive the two-loop four-external-mass Feynman integrals with uniform transcendental (UT) weights, based on the latest developments on UT integrals. Then we show that all the symbol letters can be nicely obtained from those of closely-related dual conformal integrals, by sending a dual point to infinity. Certain properties of the symbol such as first two entries and extended Steinmann relations are also studied from analogous properties of dual conformal integrals. 
}

\emailAdd{songhe@mail.itp.ac.cn, lizhenjie@itp.ac.cn, marr21@mail.ustc.edu.cn, wuzihao@mail.ustc.edu.cn, yangqinglin@itp.ac.cn,
yzhphy@ustc.edu.cn}

\begin{document}

\maketitle
\section{Introduction}
With the start of the run III stage of the Large Hadron Collider (LHC), demands for the high-order perturbative computations of the precision physics get higher. For the high-order perturbation computations in quantum field theory, multiloop Feynman integrals are the key objects which are difficult to reduce or evaluate. 

Frequently multiloop Feynman integrals with multiple scales (Mandelstam variables or mass parameters) appear in high-order perturbative computations. These integrals, even if they can be expressed in term of polylogarithm functions, their symbol structure \cite{Goncharov:2010jf} would be very complicated: the alphabet (the list of symbol letters) may be very long and contains complicated square roots of the kinematics variables; The adjacency condition of letters in the symbols can be intricate. On the other hand, the information on the alphabet and symbol of multi-loop Feynman integrals are invaluable. It enables the simplification algorithm of polylogarithm functions \cite{Goncharov:2010jf, Duhr:2011zq, Duhr:2012fh}, the bootstrap of supersymmetric Yang-Mills amplitudes \cite{Dixon:2011pw,Dixon:2013eka,Dixon:2014voa,Caron-Huot:2016owq,Caron-Huot:2019vjl}, the bootstrap of Feynman integrals \cite{Chicherin:2017dob},  the finite-field interpolation of the canonical differential equation \cite{Peraro:2016wsq} and a lot of other important applications. Therefore, it is of great interests to study the alphabet and symbol structures of multi-loop Feynman integrals.

In the literature there are many interesting studies of the theoretical properties of alphabet and symbol structures, for example, the so-called first entry-condition \cite{Gaiotto:2011dt}, the relation between symbols and Landau Singularity \cite{Dennen:2015bet,Dennen:2016mdk}, the relation between the second entry condition and the Steinmann relation \cite{Caron-Huot:2016owq, Dixon:2016nkn}, and the final entry condition \cite{Caron-Huot:2011dec}. The cluster algebra structure was applied for the computation of the symbol structure in $\mathcal N=4$ SYM theory \cite{Golden:2013xva,Harrington:2015bdt}. Recently, the ref.~\cite{Chicherin:2020umh,He:2021eec, He:2021esx} studied multi-loop Feynman integrals' alphabet from the cluster algebra structure. Besides these exciting developments, the multi-scale multi-loop Feynman integrals' alphabet and symbol structures still have a lot of mysteries. In this paper, we try to illustrate the alphabet and symbol structures for Feynman integrals with uniform transcendental (UT) weight \cite{Henn:2013pwa,Henn:2014qga}, from the viewpoint of dual conformal symmetry \cite{Drummond:2006rz,Drummond:2007aua,Drummond:2008vq,Henn:2011xk}.

The method \cite{Henn:2013pwa,Henn:2014qga} of applying UT Feynman integrals and the corresponding differential equation is currently a standard way of evaluating Feynman integrals analytically. In this paper, we derive a cutting-edge example of the UT Feynman integral basis for the two-loop double box diagrams with four different massive external legs, and then show that its alphabet and symbol structure can be well understood from the dual conformal properties. This family is particularly interesting because it is a sub-family of the two-loop eight-point massless diagrams, and also a sub-family of the two-loop five-point diagrams with three massive external legs. So, this family is important both for  formal theories like $\mathcal N=4$ SYM and also the phenomenology. It contains two Mandelstam variables as well as four mass parameters, thus the kinematics is very complicated. It is not an easy task to find the corresponding UT basis, so we turn to the latest UT basis determination methods \cite{Dlapa:2021qsl, Henn:2021cyv} with the leading singularity analysis in the Baikov representation. The complete UT basis for this family is found, and the canonical differential equation is calculated analytically via the finite-field method \cite{Peraro:2016wsq}. From the canonical differential equation, we obtained the full alphabet. Furthermore, the symbols of these UT integrals are calculated. As expected for other UT integrals with multiple scales, the alphabet is long and contains many square roots in the kinematic variables. The symbol structure for this family is also complicated. 

Therefore, in this paper, we propose to analyze the symbology of these Feynman integrals, namely the alphabet and properties of the symbols, from certain  {\it dual conformal invariant} (DCI) Feynman integrals \cite{Arkani-Hamed:2010pyv} by sending a dual point to infinity. Similar considerations have appeared in~\cite{Chicherin:2020umh} for one-mass five-point process, whose symbology is closely related to that of eight-point DCI integrals. The symbology of DCI integrals is much better understood by explicit computations \cite{Caron-Huot:2018dsv,Bourjaily:2018aeq,Henn:2018cdp,He:2020uxy,He:2020lcu,Kristensson:2021ani} as well as various other ``predictions", such as Landau analysis~\cite{Landau:1959fi,Dennen:2015bet,Dennen:2016mdk,Mizera:2021icv}, cluster algebras~\cite{Chicherin:2020umh,He:2021eec,He:2021esx,He:2021non} and very recently analysis based on twistor geometries known as Schubert problems~\cite{Yang:2022gko}. Moreover, properties of the symbols such as conditions on first two entries and (extended) Steinmann relations have also been studied more extensively for DCI integrals (see \cite{He:2021mme} and references therein). By sending a generic dual point, which are not null separated from adjacent points, to infinity, dual conformal symmetry is broken and we obtain non-DCI integrals if the latter are finite. There is strong evidence~\cite{Chicherin:2020umh} that such limits are still relevant for IR divergent integrals in dimensional regularization, thus we find it very rewarding to study the implications of DCI results for general, non-DCI integrals.

For our integrals with four external masses, it is necessary to consider fully massive DCI pentagon kinematics, which depends on five cross-ratios and reduces to four-mass (non-DCI) kinematics by sending any dual point to infinity. In such limits, the five cross-ratios reduce to ratios of kinematic invariants, $s, t, m_1^2, \cdots, m_4^2$. Remarkably we find that all the $68$ letters can be explained from symbol letters of related DCI integrals. Out of all $12$ non-trivial rational letters, $11$ of them turn out to be (the square of) leading singularities associated with one- and two-loop DCI integrals, including four-mass boxes and double-boxes (the last one is a Gram determinant which can be obtained from following odd letters). The remaining $50$ letters are parity odd with respect to the $11$ square roots, which fall into two classes. There are $16$ of them that has only one square root, which can all be identified with algebraic letters of four-mass boxes and two-loop generalizations; for the $34$ ``mixed" odd letters that depends on two square roots, we find their origin by studying twistor geometries of corresponding one- and two-loop DCI integrals. The appearance of these letters in the matrix of canonical DE also exhibit nice patterns. Moreover, we also find that, to all orders in $\epsilon$, the first two entries of the symbols of such integrals can always be written as linear combinations of five four-mass box functions (and $\log \log$ terms) which nicely explain the first-two entry conditions. Relatedly, we have also checked that extended Steinmann relations are satisfied by all the integrals, which follows from the canonical DE.

This paper is organized as: in section \ref{sec:DE-UT-Symbol} we review the UT integrals, canonical differential equation and symbol. In section \ref{sec:int4m}, we introduce the main example of two-loop double box diagram with four different external mass, derive the UT basis, canonical differential equation and the symbols. In section \ref{sec:DCI}, we will first review basics of DCI integrals, focusing on one- and two-loop integrals which depend on fully-massive pentagon kinematics; we then explain that by taking any dual point to infinity, letters of DCI integrals nicely become those of double-box {\it etc.}, including all square roots and odd letters; finally, we will comment on how certain properties of the symbols can be explained from DCI considerations.  In section \ref{sec:summary}, we summarize this paper and provide an outlook for the future application of the DCI properties on the multi-loop UT integrals. More technical details are given in the appendices of this paper.

\section{A review of differential equations, UT integrals and symbol}
\label{sec:DE-UT-Symbol}
\subsection{The canonical differential equation}
Differential equation method is one of the most popular method of evaluating Feynman integrals \cite{Kotikov:1990kg,Fleischer:1998nb,Kotikov:2000pm}. The differential equation of a Feynman integral basis $I$, via the integration-by-parts reduction, reads  
\begin{equation}\label{DE_normal}
    \frac{\partial }{\partial x_i}I=A_i(\epsilon,x) I,
\end{equation}
where $x_i$'s are kinematic variables and $A_i$'s are matrices for the differential equations. 

Usually, solving the differential equation in \eqref{DE_normal} is difficult. A better choice of the integral basis will make it dramatically easier. One of the best choices of integral basis consists of integrals with uniformal transcendental (UT) weights \cite{Henn:2013pwa}. These UT integrals have the following form in terms of $\epsilon$ expansion as, 
\begin{equation}\label{eq:UTdef}
 I= \epsilon^k \sum_{i=0}^\infty I^{(n)} \epsilon^n,
\end{equation}
where $I^{(n)}$ is a pure weight-$n$ transcendental function, i.e. $\mathcal T(I^{(n)})=n$ and $\mathcal{T}(\frac{\partial}{\partial x}I^{(n)})=n-1$, where $\mathcal T$ stands for the transcendental weight of a function. The transcendental weight of some common functions are shown in Table \ref{table-tw}, where $\text{Li}_n(x)$, $H(\{a_1,\ldots a_n\},x) $ and $G(\{a_1,\ldots
a_n\},x) $ are, respectively, the weight $n$ classical, harmonic and Goncharov polylogarithm functions .
\begin{table}
\centering

\begin{tabular}{|c|c||c|c|}
\hline
$f(x)$&$\mathcal T(f(x))$&$f(x)$&$\mathcal T(f(x))$\\
\hline
rational number&0&rational function&0\\
\hline
$\pi$&1&Log$(x)$&1\\
\hline
$\zeta_n$&$n$&$\text{Li}_n(x)$&$n$\\
\hline
$H(a_1,\cdots,a_n;x)$&$n$&$G(a_1,\cdots,a_n;x)$&$n$\\
\hline

\end{tabular}
\caption{transcendental weight of some functions}\label{table-tw}
\end{table}

The differential equations for a basis formed by UT integrals \eqref{eq:UTdef} are called canonical differential equations. For the matrices of canonical differential equations, the $\epsilon$ parameter factorizes out as \cite{Henn:2013pwa},
\begin{equation}\label{DEUT}
    \frac{\partial }{\partial x_i}I=\epsilon A_i(x) I.
\end{equation}
This is called the canonical differential equation. Consequently, we can solve the differential equations order by order in $\epsilon$ expansion as,
\begin{equation}
\begin{aligned}
&\frac{\partial}{\partial x_i}I^{(k)}=A_i I^{(k-1)}\ .
\end{aligned}
\end{equation}
From the integrability condition, the canonical differential equation matrices for UT basis satisfies,
\begin{equation}
    [A_i,A_j]=0,\quad\frac{\partial }{\partial x_i}A_j-\frac{\partial }{\partial x_j}A_i=0.
\end{equation}
From the second property, we know that,
\begin{equation}
    A_i=\frac{\partial }{\partial x_i}\tilde{A}\ ,
\end{equation}
where the matrix $\tilde{A}$ takes the form as
\begin{equation}\label{Atilde}
\tilde{A}=\sum_{i=1}^N a_i\log(W_i)\ ,
\end{equation}
where $a_i$'s are matrices of rational numbers, and $W_i$'s are algebraic functions of $x_i$'s, called \textit{symbol letters}. With the form of \eqref{Atilde}, the canonical equation \eqref{DEUT} can be solved analytically with the information of the boundary condition. On the other hand, one can easily derive the \textit{symbol} \cite{Goncharov:2010jf} of the solutions of the differential equation. 

The symbol $\mathcal{S}$ of a function is defined as,
\begin{equation}
    \mathcal{S}(\text{log} R)\equiv S[R],
\end{equation}
and
\begin{equation}
    \mathcal S( F ) \equiv  \sum_i c_i\mathcal S(F_i) \otimes S[R_i],
\end{equation}
if
\begin{equation}
  dF =\sum_ic_i F_i d\log R_i,
\end{equation}
where $R$ and $R_i$ are rational functions and $S[R_1,\cdots,R_n]\otimes S[R]\overset{\text{def}}{=}S[R_1,\cdots,R_n,R]$. With these, one can immediately derive the symbols of the solutions from the canonical differential equations as
\begin{equation}\label{eq:symbol_from_Atilde}
    \mathcal S(  I^{(n)})=\sum_{i_1,\cdots,i_n=1}^N   a_{i_n}\cdots a_{i_1}  I^{(0)} S[W_{i_1},\ldots,W_{i_n}]  \ ,
\end{equation}

\subsection{UT integral determination}
For applying the method of canonical differential equation, finding a corresponding UT basis is not a trivial task. There are many methods and algorithms designed for determining a UT basis, including the leading singularity analysis \cite{Henn:2013pwa,Henn:2014qga,Dlapa:2021qsl}, the Magnus and Dyson Series \cite{Argeri:2014qva}, the dlog integrand construction \cite{Wasser:2022kwg,Wasser:2018qvj,Chicherin:2018old,Henn:2020lye}, the intersection theory \cite{Frellesvig:2020qot, Chen:2020uyk, Chen:2022lzr,Chen:2022fyw}, the initial algorithm \cite{Dlapa:2020cwj}, the Lee's algorithm \cite{Lee:2014ioa,Lee:2017oca,Lee:2020zfb} and so on. Based on these methods or algorithms, some public available packages were designed including {\sc Canonica} \cite{Meyer:2016slj,Meyer:2017joq}, {\sc Fuchsia} \cite{Gituliar:2017vzm}, {\sc epsilon} \cite{Prausa:2017ltv}, {\sc initial} \cite{Dlapa:2020cwj} and {\sc libra} \cite{Lee:2020zfb}.
In this paper, we are using the leading singularity analysis  method in Baikov representation \cite{Schabinger:2018dyi, Chicherin:2018old, Dlapa:2021qsl, Henn:2021cyv}  as well as other methods to determine a UT basis. 

Here we briefly introduce the leading singularity method in Baikov representation \cite{Baikov:1996cd,Baikov:1996rk,Baikov:2005nv}. For a Feynman integral defined as
\begin{equation}
    G_{\alpha_1,\cdots,\alpha_n}\equiv\int \prod_{i=1}^{L} \frac{{\rm d}^D l_i}{i \pi^{D/2}}\frac{1}{D_1^{\alpha_1}\cdots D_n^{\alpha_n}},
\end{equation}
in the Baikov representation, it is
\begin{equation}\label{eq:BaikovRep}
     G_{\alpha_1,\cdots,\alpha_n}=C_E^LU^{\frac{E-D+1}{2}}\int_\Omega {\rm d}z_1\cdots{\rm d}z_nP(z)^{\frac{D-L-E-1}{2}}\frac{1}{{z_1}^{\alpha_1}\cdots{z_n}^{\alpha_n}}.
\end{equation}
In this representation, $E$ and $L$ are the number of independent external momenta and loop momenta respectively, $C_E^L$ is a constant irrelevant to kinematic variables, and $U$ and $P$ are Gram diterminants as
\begin{equation}
U={\rm det}G
\begin{pmatrix}
&p_1,&\dots&p_E\\
&p_1,&\dots&p_E
\end{pmatrix},
\end{equation}
\begin{equation}
P={\rm det}G
\begin{pmatrix}
l_1,&\dots&l_L,&p_1,&\dots&p_E\\
l_1,&\dots&l_L,&p_1,&\dots&p_E
\end{pmatrix}.
\end{equation}
The idea of using Baikov leading singularity method to determine UT integrals is based on an conjecture \cite{Chen:2020uyk} that an integral with constant leading singularity should be a UT integral. In the Baikov representation, the leading singularity of an integral can be derived using the cut \cite{Primo:2016ebd,Primo:2017ipr,Frellesvig:2017aai}, which is to replace the integration by taking residues at $z_i\to 0$, schematically as
\begin{equation}
    {\rm LS} =\oint_{z_i\to 0} {\rm d}z_1\cdots{\rm d}z_nP(z)^{\frac{D-L-E-1}{2}}\frac{1}{{z_1}^{\alpha_1}\cdots{z_n}^{\alpha_n}}.
\end{equation}

For multi-loop integrals, a more practical way to derive the leading singularities is the loop-by-loop Baikov representations \cite{Frellesvig:2017aai,Harley:2017qut}. Take the double box with $4$ external masses diagram (to be introduced in section \ref{sec:example_2L} in detail) as an example. The propagator and kinematic information are given in   \eqref{eq:4m_kinematics} and \eqref{eq:dbox-propagators}. Consider the scalar integral in the top sector in \eqref{2L_scalar_top_sector_integral}, which is
\begin{equation}
    J_{1}= \int\frac{{\rm d}^D l_1}{i \pi^{D/2}}\frac{{\rm d}^D l_2}{i \pi^{D/2}}\frac{1}{D_1\cdots D_7},
\end{equation}
one can compute the leading singularity from full Baikov representation introduced in \eqref{eq:BaikovRep}. In comparison, a more computationally economical way is to derive the Baikov representation loop by loop: First we treat $l_1$ as loop momentum and $l_2$, $p_1$, $p_2$, and $p_4$ as external momenta, so that
\begin{equation}
\begin{aligned}
    J_{1}=& \int\frac{{\rm d}^D l_2}{i \pi^{D/2}}\frac{1}{D_4D_5 D_6}\int\frac{{\rm d}^D l_1}{i \pi^{D/2}}\frac{1}{D_1D_2D_3 D_7}\\
    =&C \int\frac{{\rm d}^D l_2}{i \pi^{D/2}}\frac{1}{D_4D_5 D_6}U_1^{\frac{E-D+1}{2}}\int_\Omega {\rm d}z_1{\rm d}z_2{\rm d}z_3{\rm d}z_7P_1(z)^{\frac{D-L-E-1}{2}}\frac{1}{z_1z_2z_3z_7}.
\end{aligned}
\end{equation}
Here, $E=3$, $L=1$, $D=4$, $C$ is some constant irrelevant to kinematics and 
\begin{equation}
    P_1=\det G
    \begin{pmatrix}
        l_1,&p_1,&p_2,&l_2\\
        l_1,&p_1,&p_2,&l_2
    \end{pmatrix}.
\end{equation}
Then the leading singularity for the ``left'' loop integral is
\begin{equation}
    \oint {\rm d}z_1{\rm d}z_2{\rm d}z_3{\rm d}z_7P_1(z)^{\frac{D-L-E-1}{2}}\frac{1}{z_1z_2z_3z_7}
    =\frac{4}{\Delta},
\end{equation}
where 
\begin{equation}
\Delta^2=\frac{1}{16} \left(m_1^4 D_4^2+\left(m_2^2 D_6-s D_9\right){}^2-2 m_1^2 D_4 \left(m_2^2 D_6+s D_9\right)\right).
\end{equation}
With the definition $D_9=(l_2-p_1)^2$, we can similarly derive the leading singularity for the ``right'' loop as
\begin{equation}
\begin{aligned}
{\rm LS}(J_{1})=&{\rm LS}\Big(\int\frac{{\rm d}^D l_2}{i \pi^{D/2}}\frac{4}{D_4D_5 D_6 \Delta}\Big)\\
=&\oint_0 {\rm d}z_4{\rm d}z_5{\rm d}z_6{\rm d}z_9P_2(z)^{\frac{D-L-E-1}{2}}\frac{1}{z_4z_5z_6\Delta(z)}=\frac{16}{s r_1},
\label{eq:dbox_scalar_LS}
\end{aligned}
\end{equation}
where
\begin{equation}
    r_1^2=s^2 t^2-2 s t m_1^2 m_3^2+m_1^4 m_3^4-2 s t m_2^2 m_4^2-2 m_1^2 m_2^2 m_3^2 m_4^2+m_2^4 m_4^4.
\end{equation}
Requiring the leading singularity of this integral to be constant, we guess that $s r_1 J_{1}$ is a UT integral candidate. Later, from the differential equations, we verify that it indeed has the uniform transcendental weights.

\section{Feynman integrals with four different external masses}
\label{sec:int4m}
In this paper, we study the one and two-loop Feynman integrals with four different external masses. These integrals are interesting because they appear in the eight-point massless scattering processes which is the focus of the amplitude study of formal theories, and also in the five-point scattering processes with three external masses. These integrals have rich structures related to dual conformal invariance.

The kinematic condition for these integrals is
\begin{equation}\label{eq:4m_kinematics}
\begin{aligned}
&p_1^2=m_1^2,\quad p_2^2=m_2^2,\quad p_3^2=m_3^2,\quad p_4^2=m_4^2,
&(p_1+p_2)^2=s,\quad(p_2+p_3)^2=t,
\end{aligned}
\end{equation}
with $p_1+p_2+p_3+p_4=0$.

\subsection{One loop UT integral basis}
The propagators of one-loop Feynman integral family with four different external masses are:
\begin{equation}
\begin{aligned}
&D_1=l_1^2,\quad
D_2=(l_1-p_1)^2,\quad
D_3=(l_1-p_1-p_2)^2,
&D_4=(l_1+p_4)^2.\quad
\end{aligned}
\end{equation}
\begin{figure}[H]
\centering
\includegraphics[width=0.45\textwidth]{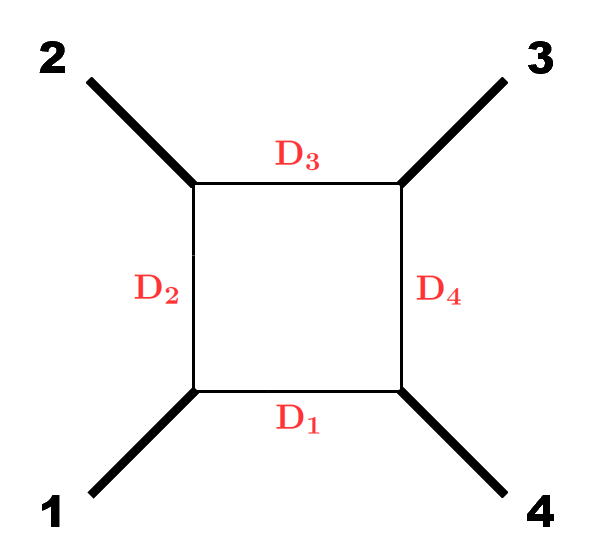}
\caption{1-loop diagram with 4 external massive legs}
\label{Figure_dbox}
\end{figure}

There are $11$ master integrals after the IBP reduction.
\begin{align}
& G_{1,1,1,1},\quad
G_{1,1,1,0},\quad
G_{1,1,0,1},\quad
G_{1,0,1,1},\quad
G_{0,1,1,1},\nonumber \\
& G_{1,2,0,0},\quad
G_{1,0,2,0},\quad
G_{1,0,0,2},\quad
G_{0,1,2,0},\quad
G_{0,1,0,2},\quad
G_{0,0,1,2}.
\end{align}
 These master integrals are almost UT integrals up to some rational function factors in kinematic variables. It is not difficult to determine these factor by looking at the differential equation, then the UT integral basis of this four masses box diagram family is: 
\begin{align}
I_1=& r_1 G_{1,1,1,1},\quad
I_2= r_2 G_{1,1,1,0},\quad
I_3= r_3 G_{1,1,0,1},\\
I_4=& r_4 G_{1,0,1,1},\quad
I_5= r_5 G_{0,1,1,1},\quad
I_6= \frac{m_1^2 G_{1,2,0,0}}{\epsilon },\\
I_7=& \frac{s G_{1,0,2,0}}{\epsilon },\quad
I_8= \frac{m_4^2 G_{1,0,0,2}}{\epsilon },\quad
I_9= \frac{m_2^2 G_{0,1,2,0}}{\epsilon },\\
I_{10}=& \frac{t G_{0,1,0,2}}{\epsilon },\quad
I_{11}= \frac{m_3^2 G_{0,0,1,2}}{\epsilon }.
\end{align}
Where we used the following roots, $r_1\sim r_5$,
\begin{equation}\label{defi:r1-5}
\begin{aligned}
r_1^2=& s^2 t^2-2 s t m_1^2 m_3^2+m_1^4 m_3^4-2 s t m_2^2 m_4^2-2 m_1^2 m_2^2 m_3^2 m_4^2+m_2^4 m_4^4,\\
r_2^2=& s^2-2 s m_1^2+m_1^4-2 s m_2^2-2 m_1^2 m_2^2+m_2^4,\\
r_3^2=& t^2-2 t m_1^2+m_1^4-2 t m_4^2-2 m_1^2 m_4^2+m_4^4,\\
r_4^2=& s^2-2 s m_3^2+m_3^4-2 s m_4^2-2 m_3^2 m_4^2+m_4^4,\\
r_5^2=& t^2-2 t m_2^2+m_2^4-2 t m_3^2-2 m_2^2 m_3^2+m_3^4.
\end{aligned}
\end{equation}

\subsection{Two loop UT integral basis}\label{sec:example_2L}
The propagators of two-loop Feynman integral family with four different external masses are:
\begin{gather}\label{eq:2L_prop}
D_1=l_1^2,\quad
D_2=(l_1-p_1)^2,\quad
D_3=(l_1-p_1-p_2)^2,\nonumber\\
D_4=(l_2+p_1+p_2)^2,\quad
D_5=(l_2+p_1+p_2+p_3)^2,\quad
D_6=l_2^2,\nonumber\\
D_7=(l_1+l_2)^2,\quad
D_8=(l_1-p_1-p_2-p_3)^2,\quad
D_9=(l_2+p_1)^2.
\label{eq:dbox-propagators}
\end{gather}

The top sector, $(1,1,1,1,1,1,1,0,0)$ for this family, is a double box diagram shown in Fig.\ref{Figure_dbox}. Here $D_8$ and $D_9$ are irreducible scalar products (ISPs).
\begin{figure}[H]
\centering
\includegraphics[width=0.7\textwidth]{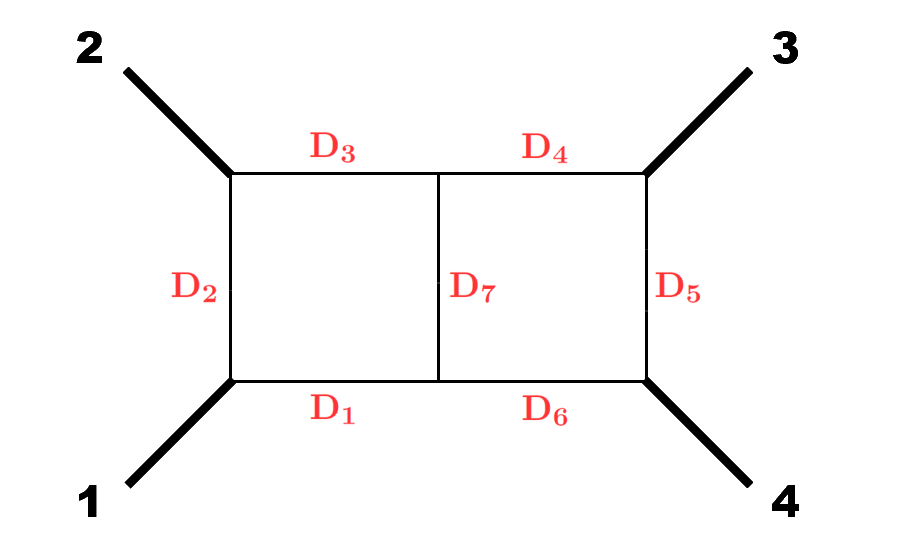}
\caption{2-loop diagram with 4 external massive legs}
\label{Figure_dbox}
\end{figure}
From the stand IBP reduction procedure, we see that the number of master integrals for this family is $74$. 
The diagrams for all master integrals are given in the appendix \ref{appendix:2l4pMIs}. Some of the master integrals in the family were calculated in \cite{Usyukina:1993ch}.

Using the Baikov leading singularity analysis techniques described in the previous section, as well as other methods that will be explained later, we have derived the UT basis for this diagram, as follows:
\begin{align}
I_1=& s r_1 G_{1,1,1,1,1,1,1,0,0},\label{2L_scalar_top_sector_integral}\\
I_2=& r_4 \left(-m_1^2 G_{1,1,1,0,1,1,1,0,0}-m_2^2 G_{1,1,1,1,1,0,1,0,0}+s G_{1,1,1,1,1,1,1,0,-1}\right),\\
I_3=& r_2 \left(-m_3^2 G_{0,1,1,1,1,1,1,0,0}-m_4^2 G_{1,1,0,1,1,1,1,0,0}+s G_{1,1,1,1,1,1,1,-1,0}\right),\\
I_4=&s G_{1,1,1,1,1,1,1,-1,-1}+\frac{1}{2} s \left(s-m_1^2-m_2^2\right) G_{1,1,1,1,1,1,1,-1,0}+ \nonumber\\
&\frac{1}{2} s \left(s-m_3^2-m_4^2\right) G_{1,1,1,1,1,1,1,0,-1}+\ldots\\
I_5=& r_9 G_{0,1,1,1,1,1,1,0,0},\,
I_6= r_8 G_{1,1,0,1,1,1,1,0,0},\,
I_7= r_{11} G_{1,1,1,0,1,1,1,0,0},\\
I_8=& r_{10} G_{1,1,1,1,1,0,1,0,0},\,
I_9= r_2 r_4 G_{1,1,1,1,1,1,0,0,0},\,
I_{10}= \frac{r_1 G_{0,1,0,1,1,1,2,0,0}}{\epsilon },\\
I_{11}=& \frac{r_4 G_{0,2,0,1,1,1,1,0,-1}}{\epsilon },\,
I_{12}= \frac{m_2^2 r_4 G_{0,1,2,0,1,1,1,0,0}}{\epsilon },\,
I_{13}= \frac{m_2^2 r_3 G_{0,2,1,0,1,1,1,0,0}}{\epsilon },\\
I_{14}=& \frac{m_4^2 r_2 G_{0,1,1,0,1,2,1,0,0}}{\epsilon },\,
I_{15}= \frac{m_4^2 r_5 G_{0,1,1,0,2,1,1,0,0}}{\epsilon },\,
I_{16}= \frac{r_1 G_{0,1,1,0,1,1,2,0,0}}{\epsilon },\\
I_{17}=& r_6 G_{0,1,1,0,1,1,1,0,0},\\
I_{18}=& -\frac{\left(s t-m_1^2 m_3^2+m_2^2 m_4^2\right) G_{0,1,1,0,1,1,2,0,0}}{2 \epsilon }-\frac{\left(s-m_1^2+m_2^2\right) m_4^2 G_{0,1,1,0,1,2,1,0,0}}{\epsilon }\nonumber\\&-\frac{m_2^2 \left(s-m_3^2+m_4^2\right) G_{0,1,2,0,1,1,1,0,0}}{\epsilon }+\frac{s m_2^2 m_4^2 G_{0,1,2,0,1,2,1,0,0}}{\epsilon ^2},\\
I_{19}=& r_2 G_{0,1,1,1,0,1,1,0,0},\,
I_{20}= r_5 G_{0,1,1,1,1,0,1,0,0},\,
I_{21}= \frac{m_2^2 r_4 G_{0,1,2,1,1,1,0,0,0}}{\epsilon },\\
I_{22}=& r_4 G_{1,0,1,0,1,1,1,0,0},\,
I_{23}= r_4 G_{1,0,1,1,1,0,1,0,0},\,
I_{24}= s r_4 G_{1,0,1,1,1,1,1,0,0},\\
I_{25}=& r_3 G_{1,1,0,0,1,1,1,0,0},\,
I_{26}= r_2 G_{1,1,0,1,0,1,1,0,0},\,
I_{27}= \frac{m_1^2 r_4 G_{2,1,0,1,1,0,1,0,0}}{\epsilon },\\
I_{28}=& \frac{m_1^2 r_5 G_{1,2,0,1,1,0,1,0,0}}{\epsilon },\,
I_{29}= \frac{m_3^2 r_2 G_{1,1,0,2,1,0,1,0,0}}{\epsilon },\,
I_{30}= \frac{m_3^2 r_3 G_{1,1,0,1,2,0,1,0,0}}{\epsilon },\\
I_{31}=& \frac{r_1 G_{1,1,0,1,1,0,2,0,0}}{\epsilon },\,
I_{32}= r_7 G_{1,1,0,1,1,0,1,0,0},\\
I_{33}=& -\frac{\left(s t+m_1^2 m_3^2-m_2^2 m_4^2\right) G_{1,1,0,1,1,0,2,0,0}}{2 \epsilon }-\frac{\left(s+m_1^2-m_2^2\right) m_3^2 G_{1,1,0,2,1,0,1,0,0}}{\epsilon }\nonumber\\&-\frac{m_1^2 \left(s+m_3^2-m_4^2\right) G_{2,1,0,1,1,0,1,0,0}}{\epsilon }+\frac{s m_1^2 m_3^2 G_{2,1,0,2,1,0,1,0,0}}{\epsilon ^2},\\
I_{34}=& \frac{m_1^2 r_4 G_{1,2,0,1,1,1,0,0,0}}{\epsilon },\,
I_{35}= \frac{r_1 G_{1,1,1,0,1,0,2,0,0}}{\epsilon },\,
I_{36}= \frac{r_2 G_{1,1,1,0,2,0,1,-1,0}}{\epsilon },\\
I_{37}=& \frac{m_4^2 r_2 G_{1,1,1,0,2,1,0,0,0}}{\epsilon },\,
I_{38}= s r_2 G_{1,1,1,1,0,1,1,0,0},\,
I_{39}= \frac{m_3^2 r_2 G_{1,1,1,2,1,0,0,0,0}}{\epsilon },\\
I_{40}=& \frac{r_4 G_{0,0,1,0,1,1,2,0,0}}{\epsilon },\,
I_{41}=\frac{s m_3^2 G_{0,0,2,0,1,1,2,0,0}}{\epsilon ^2} -\frac{3 \left(s+m_3^2-m_4^2\right) G_{0,0,1,0,1,1,2,0,0}}{\epsilon },\\
I_{42}=& \frac{r_3 G_{0,1,0,0,1,1,2,0,0}}{\epsilon },\,
I_{43}= \frac{t m_1^2 G_{0,2,0,0,1,1,2,0,0}}{\epsilon ^2}-\frac{3 \left(t+m_1^2-m_4^2\right) G_{0,1,0,0,1,1,2,0,0}}{\epsilon },\\
I_{44}=& \frac{r_2 G_{0,1,0,1,0,1,2,0,0}}{\epsilon },\,
I_{45}= \frac{m_1^2 m_2^2 G_{0,2,0,1,0,1,2,0,0}}{\epsilon ^2}-\frac{3 \left(-s+m_1^2+m_2^2\right) G_{0,1,0,1,0,1,2,0,0}}{\epsilon },\\
I_{46}=& \frac{r_5 G_{0,1,0,1,1,0,2,0,0}}{\epsilon },\,
I_{47}= \frac{t m_2^2 G_{0,2,0,1,1,0,2,0,0}}{\epsilon ^2}-\frac{3 \left(t+m_2^2-m_3^2\right) G_{0,1,0,1,1,0,2,0,0}}{\epsilon },\\
I_{48}=& \frac{r_2 G_{0,1,1,0,0,1,2,0,0}}{\epsilon },\,
I_{49}= \frac{s m_1^2 G_{0,1,1,0,0,2,2,0,0}}{\epsilon ^2}-\frac{3 \left(s+m_1^2-m_2^2\right) G_{0,1,1,0,0,1,2,0,0}}{\epsilon },\\
I_{50}=& \frac{r_5 G_{0,1,1,0,1,0,2,0,0}}{\epsilon },\,
I_{51}= \frac{t m_3^2 G_{0,1,1,0,2,0,2,0,0}}{\epsilon ^2}-\frac{3 \left(t-m_2^2+m_3^2\right) G_{0,1,1,0,1,0,2,0,0}}{\epsilon },\\
I_{52}=& \frac{m_2^2 m_4^2 G_{0,2,1,0,2,1,0,0,0}}{\epsilon ^2},\,
I_{53}= \frac{s m_2^2 G_{0,1,2,1,0,2,0,0,0}}{\epsilon ^2},\,
I_{54}= \frac{m_2^2 m_3^2 G_{0,2,1,1,2,0,0,0,0}}{\epsilon ^2},\\
I_{55}=& \frac{r_4 G_{1,0,0,1,1,0,2,0,0}}{\epsilon },\,
I_{56}= \frac{s m_4^2 G_{2,0,0,1,1,0,2,0,0}}{\epsilon ^2}-\frac{3 \left(s-m_3^2+m_4^2\right) G_{1,0,,1,1,0,2,0,0}}{\epsilon },\\
I_{57}=& \frac{r_4 G_{1,0,1,0,1,0,2,0,0}}{\epsilon },\,
I_{58}= \frac{m_3^2 m_4^2 G_{1,0,1,0,2,0,2,0,0}}{\epsilon ^2}-\frac{3 \left(-s+m_3^2+m_4^2\right) G_{1,0,1,0,1,0,2,0,0}}{\epsilon },\\
I_{59}=& \frac{s m_4^2 G_{2,0,1,0,2,1,0,0,0}}{\epsilon ^2},\,
I_{60}= \frac{s^2 G_{1,0,2,1,0,2,0,0,0}}{\epsilon ^2},\,
I_{61}= \frac{s m_3^2 G_{2,0,1,1,2,0,0,0,0}}{\epsilon ^2},\\
I_{62}=& \frac{r_3 G_{1,1,0,0,1,0,2,0,0}}{\epsilon },\,
I_{63}=\frac{t m_4^2 G_{1,1,0,0,2,0,2,0,0}}{\epsilon ^2} -\frac{3 \left(t-m_1^2+m_4^2\right) G_{1,1,0,0,1,0,2,0,0}}{\epsilon },\\
I_{64}=& \frac{m_1^2 m_4^2 G_{1,2,0,0,2,1,0,0,0}}{\epsilon ^2},\,
I_{65}= \frac{r_2 G_{1,1,0,1,0,0,2,0,0}}{\epsilon },\\
I_{66}=& \frac{s m_2^2 G_{1,1,0,2,0,0,2,0,0}}{\epsilon ^2}-\frac{3 \left(s-m_1^2+m_2^2\right) G_{1,1,0,1,0,0,2,0,0}}{\epsilon },\\
I_{67}=& \frac{s m_1^2 G_{2,1,0,2,0,1,0,0,0}}{\epsilon ^2},\,
I_{68}= \frac{m_1^2 m_3^2 G_{1,2,0,1,2,0,0,0,0}}{\epsilon ^2},\,
I_{69}= \frac{m_3^2 G_{0,0,2,0,2,0,1,0,0}}{\epsilon ^2},\\
I_{70}=& \frac{t G_{0,2,0,0,2,0,1,0,0}}{\epsilon ^2},\,
I_{71}= \frac{m_2^2 G_{0,2,0,2,0,0,1,0,0}}{\epsilon ^2},\,
I_{72}= \frac{m_4^2 G_{2,0,0,0,2,0,1,0,0}}{\epsilon ^2},\\
I_{73}=& \frac{s G_{0,0,2,0,0,2,1,0,0}}{\epsilon ^2},\,
I_{74}= \frac{m_1^2 G_{0,2,0,0,0,2,1,0,0}}{\epsilon ^2}.\label{UT_I74}
\end{align}

In order to define the UT basis, besides the roots appearing in the one-loop integrals, we need other $6$ roots, $r_6\sim r_{11}$,
\begin{align}
r_6^2=& s^2+2 s t+t^2-2 s m_1^2-2 t m_1^2+m_1^4-2 s m_3^2-2 t m_3^2+2 m_1^2 m_3^2+m_3^4-4 m_2^2 m_4^2,\\
r_7^2=& s^2+2 s t+t^2-2 s m_2^2-2 t m_2^2+m_2^4-4 m_1^2 m_3^2-2 s m_4^2-2 t m_4^2+2 m_2^2 m_4^2+m_4^4,\\
r_8^2=& s^2 t^2-2 s^2 t m_1^2+s^2 m_1^4+2 s t m_1^2 m_4^2-2 s m_1^4 m_4^2-2 s t m_2^2 m_4^2+2 s m_1^2 m_2^2 m_4^2\nonumber\\&-4 s m_1^2 m_3^2 m_4^2+m_1^4 m_4^4-2 m_1^2 m_2^2 m_4^4+m_2^4 m_4^4,\\
r_9^2=& s^2 t^2-2 s^2 t m_2^2+s^2 m_2^4-2 s t m_1^2 m_3^2+2 s t m_2^2 m_3^2+2 s m_1^2 m_2^2 m_3^2-2 s m_2^4 m_3^2\nonumber\\&+m_1^4 m_3^4-2 m_1^2 m_2^2 m_3^4+m_2^4 m_3^4-4 s m_2^2 m_3^2 m_4^2,\\
r_{10}^2=& s^2 t^2-2 s^2 t m_3^2+2 s t m_2^2 m_3^2-4 s m_1^2 m_2^2 m_3^2+s^2 m_3^4-2 s m_2^2 m_3^4+m_2^4 m_3^4\nonumber\\&-2 s t m_2^2 m_4^2+2 s m_2^2 m_3^2 m_4^2-2 m_2^4 m_3^2 m_4^2+m_2^4 m_4^4,\\
r_{11}^2=& s^2 t^2-2 s t m_1^2 m_3^2+m_1^4 m_3^4-2 s^2 t m_4^2+2 s t m_1^2 m_4^2-4 s m_1^2 m_2^2 m_4^2+2 s m_1^2 m_3^2 m_4^2\nonumber\\&-2 m_1^4 m_3^2 m_4^2+s^2 m_4^4-2 s m_1^2 m_4^4+m_1^4 m_4^4.
\end{align}

Here we remark on how we find this UT integral bases:
\begin{enumerate}
    \item  Use the standard IBP programs \cite{Klappert:2020nbg, Smirnov:2021rhf}, we derived the analytic differential equation for the master integrals. For instance, with FIRE6 \cite{Smirnov:2021rhf}, the computation took about two days with a node of $50$ cores. The resulting differential equation is saved for the latter computation to determine some sub-sector the UT integral candidates.
    \item Apply the Baikov leading singularity method to determine the UT integrals in the sectors with $7$, $6$ and $5$ propagators. 
    \begin{itemize}

    \item (Top sector) The top sector is $(1,1,1,1,1,1,1,0,0)$ and there are $4$ master integrals in this sector. From the experience between UT integrals and the UV finiteness condition, we tend to use 
    \begin{gather}
           G_{1,1,1,1,1,1,1,0,0},\quad G_{1,1,1,1,1,1,1,-1,0},\quad G_{1,1,1,1,1,1,1,0,-1},\quad G_{1,1,1,1,1,1,1,-1,-1}
    \end{gather}
    as UT integral candidates. If necessary, some subsector integrals would also be used. 
    
    From the Baikov leading singularity example \eqref{eq:dbox_scalar_LS} in the previous section, we estimate that $s r_1 G_{1,1,1,1,1,1,1,0,0}$ is a UT integral. To make a UT integral candidate from $G_{1,1,1,1,1,1,1,-1,0}$, based on the experience in \cite{Dlapa:2021qsl}, we can consider the Baikov integral with the sub-maximal cut. Given $G_{1,1,1,1,1,1,1,-1,0}$, the ansatz for a UT integral can be set as, 
    \begin{align*}
         & f_1 G_{1,1,1,1,1,1,1,-1,0} +f_2 G_{0,1,1,1,1,1,1,0,0} +f_3 G_{1,0,1,1,1,1,1,0,0} +f_4 G_{1,1,0,1,1,1,1,0,0} \\
        & +f_5 G_{1,1,1,0,1,1,1,0,0} +f_6 G_{1,1,1,1,0,1,1,0,0} +f_7 G_{1,1,1,1,1,0,1,0,0} +f_8 G_{1,1,1,1,1,1,0,0,0}\,.
    \end{align*}
    
    We  use right-to-left Baikov representation on this ansatz to simplify the calculation of leading singularities. Require the leading singularities to be rational numbers, and then the coefficients $f_1\sim f_8$ are fixed up to some constants:
    \begin{align*}
        & f_1=c_1 r_2 s,\quad f_2=-c_1 r_2 m_3^2,\quad f_3=c_2 r_4 s,\quad f_4=-c_1 r_2 m_4^2\\
        & f_5=c_3 r_{11},\quad f_6=c_4 r_2 s,\quad f_7=c_5 r_{10},\quad f_8=c_6 r_2 r_4\,.
        \end{align*}
    $c_1\sim c_6$ are rational numbers, and $c_1 \neq 0$ since this ansatz is from $G_{1,1,1,1,1,1,1,-1,0}$. We can set $c_2\sim c_6$ to $0$, and $c_1=1$ for the simplicity. Finally, we get a UT integral candidate: 
    \begin{equation}
    I_3= r_2 \left(-m_3^2 G_{0,1,1,1,1,1,1,0,0}-m_4^2 G_{1,1,0,1,1,1,1,0,0}+s G_{1,1,1,1,1,1,1,-1,0}\right).\
    \end{equation}

    With the leading singularity analysis in the Baikov representation, it is also easy to see that
    \begin{gather}
       \tilde I_4 \equiv s G_{1,1,1,1,1,1,1,-1,-1}+\frac{1}{2} s \left(s-m_1^2-m_2^2\right) G_{1,1,1,1,1,1,1,-1,0}+ \nonumber\\
\frac{1}{2} s \left(s-m_3^2-m_4^2\right) G_{1,1,1,1,1,1,1,0,-1}
    \end{gather}
    has constant leading singularities, from the {\it maximal cut} of the Baikov representation. However, later on, from the differential equation computation, we see $\tilde I_4$ itself is not a UT integral. To upgrade $\tilde I_4$ to a UT integral, subsector integrals should be added. It is a nontrivial computation to determine those subsector integrals, so we keep $\tilde I_4$ at this stage and later describe how to get subsector integrals.
    
    We remark that the $D=4$ leading singularity analysis of integrals in this sector was calculated in the ref.~ \cite{Johansson:2013sda}.
    
    \item (6-propagator sector) It is straightforward to find the UT candidates for 6-propagator sectors by the Baikov leading singularity analysis. For example, the sector $(0,1,1,1,1,1,1,0,0)$ contains one master integral $G_{0,1,1,1,1,1,1,0,0}$. By a maximal cut of the loop by loop Baikov representation of $G_{0,1,1,1,1,1,1,0,0}$, it's easy to get a UT integral candidate,
    \begin{equation}
       I_5= r_9 G_{0,1,1,1,1,1,1,0,0}.
   \end{equation}
    
    \item (5-propagator sector) We take the sector  $(1,1,0,1,1,0,1,0,0)$ as an example. This sector contains $7$ master integrals, which is the sector with the largest number of master integrals in this family. To find all the corresponding $7$ UT integrals would be challenging.  From the loop-by-loop Baikov leading singularity analysis, it is easy to see that 
    \begin{equation}
        I_{32}=\frac{r_7 G_{1,1,0,1,1,0,1,0,0}}{\epsilon}
    \end{equation}
    is a UT integral candidate. Furthermore, based on the algorithm in \cite{Dlapa:2021qsl}, we can look for reducible integrals in the super sectors of the sector $(1,1,0,1,1,0,1,0,0)$. In this way, we obtained $6$ UT integrals candidates, namely $I_{j}$, $j=27,\ldots, 32$. To get the last UT candidate for this sector, we start with $\tilde I_{33}=s m_1^2 m_3^2 G_{2,1,0,2,1,0,1,0,0}/\epsilon^2$, an integral with two double propagators. Later on, from the differential equation, we upgrade $\tilde I_{33}$ to a UT integral.
    
      \end{itemize}
    
    \item For sectors with $3$ or $4$ propagators, it is relatively easy to find the UT integral candidates. We apply the strategy from \cite{Henn:2013pwa,Henn:2014qga} to add double propagators and an extra factor $\epsilon^{-1}$ for each bubble sub-diagram, in order to get UT candidates.
    
    \item Then we transfer the original DE to a DE for the $74$ UT integral candidates. The computation for this transformation is difficult, due to the complicated kinematics. Here our strategy in the transformation, is to set $s$, $t$, $m_{1}$, $m_{2}$, $m_{3}$ and $m_{4}$ as some random integer values while keep $\epsilon$ analytically. Then we can identify the entries of the differential equation matrix which are {\it not} proportional to $\epsilon$. Explicitly, such entries are located in the row and columns of the new differential equation corresponding to $\tilde I_{4} $ and $\tilde I_{33}$. This indicates that $\tilde I_{4} $ and $\tilde I_{33}$ should be upgraded to UT real integrals. 
   
    We observe that entries in the DE matrix's $33$-rd row, which correspond to $\tilde I_{33}$, are either proportional to $\epsilon$ or has the form $a+b \epsilon$. For the latter form, a simple transformation like that in \cite{Meyer:2016slj}, 
    \begin{equation}
    I_{33}\equiv\tilde I_{33}+\sum_{j=27,j\not=33}^{74}    I_j \int \bigg(A_{33,j}^{(n)}\big|_{\epsilon\to 0}\bigg) dx_n
    \label{eq:transformation}
    \end{equation}
    can remove all the $\epsilon^0$ terms in the $33$rd row. Here $x_1,\ldots, x_6$ stand for the variables $s$, $t$, $m_1$, $m_2$, $m_3$ and $m_4$. $A_{i,j}^{(n)}$ is the differential equation matrix element for the derivative in $x_n$. To apply this transformation, we need to first evaluate $A_{i,j}^{(n)}$ analytically for $i=33$. The resulting integral,
    \begin{align}
 I_{33}=& -\frac{\left(s t+m_1^2 m_3^2-m_2^2 m_4^2\right) G_{1,1,0,1,1,0,2,0,0}}{2 \epsilon }-\frac{\left(s+m_1^2-m_2^2\right) m_3^2 G_{1,1,0,2,1,0,1,0,0}}{\epsilon }\nonumber\\&-\frac{m_1^2 \left(s+m_3^2-m_4^2\right) G_{2,1,0,1,1,0,1,0,0}}{\epsilon }+\frac{s m_1^2 m_3^2 G_{2,1,0,2,1,0,1,0,0}}{\epsilon ^2}
     \end{align}
    is indeed a UT integral.
    
    The upgrading of $\tilde I_4$ is more involved. The DE matrix's $4$-th row contains several entries which are not linear in $\epsilon$, so a transformation like \eqref{eq:transformation} is not enough to make the DE canonical. Here our strategy is an induction over the number of mass parameters. We take the limit when some or all of the mass parameter to zero, and reduce $\tilde I_4$ to the known UT integral basis \cite{Henn:2013pwa,Henn:2014lfa,Dlapa:2021qsl} in this limit. The reduction coefficients are not all rational constants, so we can by-hand add some integrals to $\tilde I_4$ to make the coefficients to be constants, in the partially massless or completely massless limit. After this, the resulting integral should be "closer" to a UT integral and indeed the corresponding DE matrix row is completely linear in $\epsilon$. Then we use a similarity transformation like $\eqref{eq:transformation}$ again, to make the whole differential equation matrix proportional to $\epsilon$.
 \end{enumerate}
The computer readable UT integrals definitions, as well as the corresponding definition of the square roots, can be found in the ``output'' folder of the auxiliary files, named ``dbox4m\_UT.txt'' and ``rootdef.txt'', respectively.

\subsection{Canonical differential equations and the alphabet}\label{sec:alphabet}
With $\tilde I_4$ and $\tilde I_{33}$ upgraded, numerically, we see that the differential equation is canonical. Then the next computation is to get the analytic canonical differential equation. Here we apply the finite-field package {\sc FiniteFlow} \cite{Peraro:2016wsq} for this computation, since it is well known that the canonical DE usually has much simpler coefficients than the ordinary DE and the finite field reconstruction is extremely efficient. For this family, the canonical differential equation reconstruction with {\sc FiniteFlow} only takes about $14$ seconds on a workstation with $50$ cores. The analytic expressions of the differential equations can be found in the ``output'' folder of the auxiliary files, named `` UTDE.txt'', which is a list of 6 differential equations, with respect to the kinematic variables $s$, $t$, and $m_1^2\sim m_4^2$, respectively.

Schematically, the canonical differential equations for the UT basis can be written as
\begin{equation}
\partial_{x_i} I =\epsilon (\partial_{x_i} \tilde{A}) I,
\end{equation}
where
\begin{equation}\label{eq:Atilde_def}
\tilde{A}=\sum_i a_i\log(W_i)\,.
\end{equation}
Here $a_i$'s are matrices of rational numbers irrelevant to $x_i$'s. $W_i$ are the symbol letters. After integrating the expressions of the differential equations, we derived $\tilde{A}$. We found that it contains 68 symbol letters. Among the letters, $W_1\sim W_{18}$ are even letters, which are polynomials of the $6$ kinematic variables. The rest $50$ letters are odd under the simultaneous sign change of all square roots $r_1,\ldots,r_{11}$. Letters $W_{19}\sim W_{34}$ contains one square root each, and $W_{35}\sim W_{68}$ contains 2 square roots each.
Specifically, the even letters are
\begin{equation}
\begin{aligned}
W_1=& m_1^2,\,
W_2= m_2^2,\,
W_3= m_3^2,\,
W_4= m_4^2,\,
W_5= s,\,
W_6= t,\\
W_7=& r_5^2,\,
W_8= r_3^2,\,
W_9= r_2^2,\,
W_{10}= r_4^2,\,
W_{11}= r_7^2,\,
W_{12}= r_6^2,\\
W_{13}=& r_1^2,\,
W_{14}= r_8^2,\,
W_{15}= r_9^2,\,
W_{16}= r_{10}^2,\,
W_{17}= r_{11}^2,\\
W_{18}=& s^2 t+s t^2-s t m_1^2-s t m_2^2+s m_1^2 m_2^2-s t m_3^2-s m_1^2 m_3^2-t m_1^2 m_3^2+m_1^4 m_3^2\\&+t m_2^2 m_3^2-m_1^2 m_2^2 m_3^2+m_1^2 m_3^4-s t m_4^2+t m_1^2 m_4^2-s m_2^2 m_4^2-t m_2^2 m_4^2\\&-m_1^2 m_2^2 m_4^2+m_2^4 m_4^2+s m_3^2 m_4^2-m_1^2 m_3^2 m_4^2-m_2^2 m_3^2 m_4^2+m_2^2 m_4^4,
\end{aligned}
\end{equation}
The odd letters that contain one square root are,
\begin{equation}
\begin{aligned}
W_{19}=& \frac{f_{19}+r_6}{f_{19}-r_6},\,
W_{20}= \frac{f_{20}+r_7}{f_{20}-r_7},\,
W_{21}= \frac{f_{21}+r_2}{f_{21}-r_2},\,
W_{22}= \frac{f_{22}+r_4}{f_{22}-r_4},\\
W_{23}=& \frac{f_{23}+r_5}{f_{23}-r_5},\,
W_{24}= \frac{f_{24}+r_3}{f_{24}-r_3},\,
W_{25}= \frac{f_{25}+r_1}{f_{25}-r_1},\\
W_{26}=& \frac{f_{26}+\left(m_1^2-m_2^2\right) r_2}{f_{26}-\left(m_1^2-m_2^2\right) r_2},\,
W_{27}= \frac{f_{27}+\left(m_3^2-m_4^2\right) r_4}{f_{27}-\left(m_3^2-m_4^2\right) r_4},\\
W_{28}=& \frac{f_{28}+\left(m_2^2-m_3^2\right) r_5}{f_{28}-\left(m_2^2-m_3^2\right) r_5},\,
W_{29}= \frac{f_{29}+\left(m_1^2-m_4^2\right) r_3}{f_{29}-\left(m_1^2-m_4^2\right) r_3},\\
W_{30}=& \frac{f_{30}+\left(m_1^2 m_3^2-m_2^2 m_4^2\right) r_1}{f_{30}-\left(m_1^2 m_3^2-m_2^2 m_4^2\right) r_1},\\
W_{31}=& \frac{f_{31}+\left(s t-s m_3^2+m_2^2 m_3^2-m_2^2 m_4^2\right) r_{10}}{f_{31}-\left(s t-s m_3^2+m_2^2 m_3^2-m_2^2 m_4^2\right) r_{10}},\\
W_{32}=& \frac{f_{32}+\left(-s t+m_1^2 m_3^2+s m_4^2-m_1^2 m_4^2\right) r_{11}}{f_{32}-\left(-s t+m_1^2 m_3^2+s m_4^2-m_1^2 m_4^2\right) r_{11}},\\
W_{33}=& \frac{f_{33}+\left(s t-s m_1^2+m_1^2 m_4^2-m_2^2 m_4^2\right) r_8}{f_{33}-\left(s t-s m_1^2+m_1^2 m_4^2-m_2^2 m_4^2\right) r_8},\\
W_{34}=& \frac{f_{34}+\left(-s t+s m_2^2+m_1^2 m_3^2-m_2^2 m_3^2\right) r_9}{f_{34}-\left(-s t+s m_2^2+m_1^2 m_3^2-m_2^2 m_3^2\right) r_9},
\end{aligned}
\end{equation}
where
\begin{equation}
\begin{aligned}
f_{19}=& -s-t+m_1^2+m_3^2,\,
f_{20}= -s-t+m_2^2+m_4^2,\,
f_{21}= s-m_1^2-m_2^2,\\
f_{22}=& s-m_3^2-m_4^2,\,
f_{23}= t-m_2^2-m_3^2,\,
f_{24}= t-m_1^2-m_4^2,\\
f_{25}=& s t-m_1^2 m_3^2-m_2^2 m_4^2,\\
f_{26}=& -s m_1^2+m_1^4-s m_2^2-2 m_1^2 m_2^2+m_2^4,\\
f_{27}=& -s m_3^2+m_3^4-s m_4^2-2 m_3^2 m_4^2+m_4^4,\\
f_{28}=& -t m_2^2+m_2^4-t m_3^2-2 m_2^2 m_3^2+m_3^4,\\
f_{29}=& -t m_1^2+m_1^4-t m_4^2-2 m_1^2 m_4^2+m_4^4,\\
f_{30}=& -s t m_1^2 m_3^2+m_1^4 m_3^4-s t m_2^2 m_4^2-2 m_1^2 m_2^2 m_3^2 m_4^2+m_2^4 m_4^4,\\
f_{31}=& s^2 t^2-2 s^2 t m_3^2+2 s t m_2^2 m_3^2-2 s m_1^2 m_2^2 m_3^2+s^2 m_3^4-2 s m_2^2 m_3^4+m_2^4 m_3^4\\&-2 s t m_2^2 m_4^2+2 s m_2^2 m_3^2 m_4^2-2 m_2^4 m_3^2 m_4^2+m_2^4 m_4^4,\\
f_{32}=& s^2 t^2-2 s t m_1^2 m_3^2+m_1^4 m_3^4-2 s^2 t m_4^2+2 s t m_1^2 m_4^2-2 s m_1^2 m_2^2 m_4^2\\&+2 s m_1^2 m_3^2 m_4^2-2 m_1^4 m_3^2 m_4^2+s^2 m_4^4-2 s m_1^2 m_4^4+m_1^4 m_4^4,\\
f_{33}=& s^2 t^2-2 s^2 t m_1^2+s^2 m_1^4+2 s t m_1^2 m_4^2-2 s m_1^4 m_4^2-2 s t m_2^2 m_4^2+2 s m_1^2 m_2^2 m_4^2\\&-2 s m_1^2 m_3^2 m_4^2+m_1^4 m_4^4-2 m_1^2 m_2^2 m_4^4+m_2^4 m_4^4,\\
f_{34}=& s^2 t^2-2 s^2 t m_2^2+s^2 m_2^4-2 s t m_1^2 m_3^2+2 s t m_2^2 m_3^2+2 s m_1^2 m_2^2 m_3^2-2 s m_2^4 m_3^2\\&+m_1^4 m_3^4-2 m_1^2 m_2^2 m_3^4+m_2^4 m_3^4-2 s m_2^2 m_3^2 m_4^2.
\end{aligned}
\end{equation}

The odd letters that contain 2 square roots are
\begin{align}
W_{35}&= \frac{f_{35}+r_2 r_4}{f_{35}-r_2 r_4},\,\label{letterW35}
W_{36}= \frac{f_{36}+r_2 r_5}{f_{36}-r_2 r_5},\,
W_{37}= \frac{f_{37}+r_2 r_6}{f_{37}-r_2 r_6},\,
W_{38}= \frac{f_{38}+r_2 r_3}{f_{38}-r_2 r_3},\\
W_{39}&= \frac{f_{39}+r_2 r_7}{f_{39}-r_2 r_7},\,
W_{40}= \frac{f_{40}+r_3 r_4}{f_{40}-r_3 r_4},\,
W_{41}= \frac{f_{41}+r_4 r_6}{f_{41}-r_4 r_6},\,
W_{42}= \frac{f_{42}+r_4 r_5}{f_{42}-r_4 r_5},\\
W_{43}&= \frac{f_{43}+r_4 r_7}{f_{43}-r_4 r_7},\,
W_{44}= \frac{f_{44}+r_2 r_9}{f_{44}-r_2 r_9},\,
W_{45}= \frac{f_{45}+r_2 r_8}{f_{45}-r_2 r_8},\,
W_{46}= \frac{f_{46}+r_1 r_2}{f_{46}-r_1 r_2},\\
W_{47}&= \frac{f_{47}+r_2 r_{10}}{f_{47}-r_2 r_{10}},\,
W_{48}= \frac{f_{48}+r_2 r_{11}}{f_{48}-r_2 r_{11}},\,
W_{49}= \frac{f_{49}+r_4 r_{11}}{f_{49}-r_4 r_{11}},\,
W_{50}= \frac{f_{50}+r_4 r_{10}}{f_{50}-r_4 r_{10}},\\
W_{51}&= \frac{f_{51}+r_1 r_4}{f_{51}-r_1 r_4},\,
W_{52}= \frac{f_{52}+r_4 r_8}{f_{52}-r_4 r_8},\,
W_{53}= \frac{f_{53}+r_4 r_9}{f_{53}-r_4 r_9},\,
W_{54}= \frac{f_{54}+r_3 r_5}{f_{54}-r_3 r_5},\\
W_{55}&= \frac{f_{55}+r_5 r_6}{f_{55}-r_5 r_6},\,
W_{56}= \frac{f_{56}+r_3 r_6}{f_{56}-r_3 r_6},\,
W_{57}= \frac{f_{57}+r_5 r_7}{f_{57}-r_5 r_7},\,
W_{58}= \frac{f_{58}+r_3 r_7}{f_{58}-r_3 r_7},\\
W_{59}&= \frac{f_{59}+r_1 r_5}{f_{59}-r_1 r_5},\,
W_{60}= \frac{f_{60}+r_5 r_9}{f_{60}-r_5 r_9},\,
W_{61}= \frac{f_{61}+r_5 r_{10}}{f_{61}-r_5 r_{10}},\,
W_{62}= \frac{f_{62}+r_1 r_3}{f_{62}-r_1 r_3},\\
W_{63}&= \frac{f_{63}+r_3 r_8}{f_{63}-r_3 r_8},\,
W_{64}= \frac{f_{64}+r_3 r_{11}}{f_{64}-r_3 r_{11}},\,
W_{65}= \frac{f_{65}+r_1 r_8}{f_{65}-r_1 r_8},\,
W_{66}= \frac{f_{66}+r_1 r_9}{f_{66}-r_1 r_9},\\
W_{67}&= \frac{f_{67}+r_1 r_{10}}{f_{67}-r_1 r_{10}},\,
W_{68}= \frac{f_{68}+r_1 r_{11}}{f_{68}-r_1 r_{11}}.\label{letterW68}
\end{align}
The definition of the polynomials $f_{35}\sim f_{68}$ is in the appendix \ref{appendix:f_polynomial}. 

The computer readable results of the symbol letters and $\tilde A$ can be found in the ``output'' folder of the auxiliary files, named ``letterdef.txt'' and ``Atilde.txt'', respectively.

\subsection{Symbol structures}\label{subsec:symbol}
The result of the canonical differential equations \eqref{eq:Atilde_def} are derived in the last subsection. The integration to get the analytic expressions for the UT integrals is not easy due to the boundary condition. However, deriving the corresponding symbol letters are relatively easier, which can already help us to acquire interesting properties of the analytic expressions.
The integrals' symbols can be easily derived from \eqref{eq:Atilde_def}, using \eqref{eq:symbol_from_Atilde}, or equivalently, the recursion
\begin{equation}\label{eq:integrate_symbol}
   \mathcal S(I^{(m+1)}_i)=\sum_{k} (a_k)_{ij} \Big(\mathcal S(I^{(m)}_j)\otimes S[W_k]\Big),
\end{equation}
with the lowest order boundary condition
\begin{equation}\label{eq:0-order symbol}
   \mathcal S(I^{(0)}_i)=I^{(0)}_i.
\end{equation}
Having the specific results for the constant matrices $a_i$ defined in \eqref{eq:Atilde_def}, in order to derive the symbols for the integrals, we also need the results for $I^{(0)}_i$. These are the coefficients at the order $\epsilon^{-2 L}$, being rational constants, where $L$ is the loop order.

The coefficients $I^{(0)}_i$ can be derived analytically from the infrared and collinear regions of Feynman integrals.  The resulting coefficients at $\epsilon^{-4}$ order, namely $I_i^{(0)}$ satisfy that $I^{(0)}_4=-\frac{3}{2}$, $I^{(0)}_i=1$ for
\begin{equation}
    i\in \{{18,33,41,43,45,47,49},51\sim54,{56},58\sim61,{63,64},66\sim74\},
\end{equation}
and $I^{(0)}_i=0$ for other choices of $i$. These coefficients are consistent with the numeric results obtained using {\sc{FIESTA}} \cite{Smirnov:2008py,Smirnov:2015mct,Smirnov:2021rhf}.

With results of $I^{(0)}_i$ and $a_i$ defined in \eqref{eq:Atilde_def}, we can use \eqref{eq:integrate_symbol} and \eqref{eq:0-order symbol} to derive the symbols of the UT integrals and higher $\epsilon$ orders. Using this method, we derived the symbol letters of the UT integrals up to the weight-4 order. The result can be found in ``output'' folder of the auxiliary files, named ``UTSymbols.txt'', where ``UTSymbol[$k$]'' stands for $I_i^{(k)}$. From the results, we have observed the following phenomena:

Firstly, the complexity of the symbol expressions goes rapidly up while the order increases. At the weight-1 order , the symbols are quite simple. There are only $28$ integrals having non-zero symbols at this order with totally $64$ terms. For example, we have $\mathcal S (I_4^{(1)})=3S[W_5]$ with 1 term. Among the symbols at this order, the ones with most terms are
\begin{equation}
   \mathcal S (I_{18}^{(1)})=S[W_1] - S[W_2] + S[W_3] - S[W_4] - 2 S[W_5]
\end{equation}
and 
\begin{equation}
   \mathcal S (I_{33}^{(1)})=-S[W_1] + S[W_2] - S[W_3] + S[W_4] - 2 S[W_5],
\end{equation}
with 5 terms each. At the weight-2  order, there are $56$ integrals of non-zero symbols with totally $360$ terms. The symbols having the most number of terms are $\mathcal S (I_{18}^{(2)})$ and $\mathcal S (I_{33}^{(2)})$, with $31$ terms each. At the weight-3  order, there are $56$ integrals of non-zero symbols with totally $2898$ terms. The symbols having the most number of terms are also $\mathcal S (I_{18}^{(3)})$ and $\mathcal S (I_{33}^{(3)})$, with $213$ terms each. At the weight-2 order, there are $74$ integrals of non-zero symbols with totally $35673$ terms. The symbols having the most number of terms are again $\mathcal S (I_{18}^{(4)})$ and $\mathcal S (I_{33}^{(4)})$, with $2139$ terms each.

Secondly, many integrals have vanishing symbols at certain $\epsilon$ orders. At the weight-1 order, we have shown that only $28$ integrals are with non-zero symbols. Integrals with the zero symbol at the weight-0 order also have the zero symbol at the order the weight-1 order. The integrals with the zero symbol at weight $1$ and $2$, are $I_i$ for 
\begin{equation}\label{eq:46 halfly vanishing integrals}
    i \in \{1\sim3,5\sim17,19\sim32,34\sim40,{42,44,46,48,50,55,57,62,65}\}.
\end{equation}
At the weight-2 order only $18$ integrals have the zero symbol. They are $I_i$ for
\begin{equation}\label{eq:18 vanishing integrals}
    i \in \{1\sim3,5\sim9,{17,19,20},22\sim26,{32,38}\}.
\end{equation}
These integral also have the zero symbol at the weight-3 order. At the the weight-4 order, all integrals have nonzero symbols. Note that, \eqref{eq:18 vanishing integrals} is a subset of \eqref{eq:46 halfly vanishing integrals}. 

Thirdly,  letters cannot appear at arbitrary positions of the symbols. To state this, we the following terminology: The symbols of UT integrals consists of terms proportional to some $S[W_{i_1},W_{i_2},\cdots]$, of which we call $W_{i_1}$ the first letter (or letter at the first entry), $W_{i_2}$ the second letter (or letter at second entry) and so on.
As we can see from the results, among the $68$ symbol letters, only $W_1\sim W_6$ can appear at the first entries, which are exactly the 6 kinematic variables $s$, $t$, $m_1^2$, $m_2^2$, $m_3^2$ and $m_4^2$. The letters appearing at the second entries are $W_1\sim W_6$ and $W_{21}\sim W_{30}$.

Moreover, some pairs of letters can never be next to each other in the symbols. We can explain this from \eqref{eq:integrate_symbol}. For constant matrices $a_{i}$ and $a_{j}$, if $a_i\ a_j=a_j\ a_i=0$, then according to \eqref{eq:integrate_symbol}, $W_i$ and $W_j$ cannot appear at adjacent entries, which means, symbols of the UT integrals contain no term   like $S[\cdots,W_i,W_j,\cdots]$ or $S[\cdots,W_j,W_i,\cdots]$, at all orders of $\epsilon$. Then we call that the adjacency between letters $W_i$ and $W_j$ is forbidden. Among all possible adjacencies of 68 letters, which are $68\times69/2=2346$ in total, only $919$ of them are allowed and the rest $1427$ are forbidden. The explicit allowed and forbidden letter pairs can be found in ``output'' folder of the auxiliary files, named ``AdjLetters.txt'' and ``NonAdjLetters.txt'', respectively.

\subsection{Check with known dual conformal invariant integral result}
Some of the symbol  derived from canonical differential equations in section \ref{subsec:symbol} can be verified with results from the dual conformal invariance (DCI) point of view. We take $I_1$ \eqref{2L_scalar_top_sector_integral}, which is the scalar double-box integral (proportional to $G[1,1,1,1,1,1,1,0,0]$) as an example. In the ref.~\cite{He:2021esx}, the symbol of this integral is derived from the DCI analysis, which is given as
\begin{equation}\label{eq:f^(L)}
    f^{(L)}=\sum_{m=L}^{2 L} \frac{m !(-1)^m (\log (-z \bar{z}))^{2 L-m}}{L !(m-L) !(2 L-m) !}\left(\mathrm{Li}_{m}(z)-\mathrm{Li}_{m}(\bar{z})\right).
\end{equation}
(An introduction to DCI integrals is to be given in the next section.)

Taking the loop number $L=2$, we get the scalar double box UT integral $f^{(2)}$ at the right hand side. In \eqref{eq:f^(L)}, the integral is in two variables $z$ and $\bar z$. They are related to the kinematic variables $s$, $t$, and $m_1\sim m_4$ as
\begin{equation}\label{eq:z-def-1}
    \frac{z \bar{z}}{(1-z)(1-\bar{z})}=\frac{x_{13}^{2} x_{57}^{2}}{x_{15}^{2} x_{37}^{2}}, \quad \frac{1}{(1-z)(1-\bar{z})}=\frac{x_{17}^{2} x_{35}^{2}}{x_{15}^{2} x_{37}^{2}} ,
\end{equation}
where
\begin{equation}\label{eq:z-def-2}
x_{13}^2=m_1^2,\,x_{57}^2=m_3^2,\,x_{15}^2=s,\,x_{37}^2=t,\,x_{17}^2=m_4^2,\,x_{35}^2=m_2^2.
\end{equation}
Thus, we are able to derive the symbol expressions for $f^{(2)}$ in terms of $z$ and $\bar z$ and use the relations above to rewrite it in terms of $s$, $t$, and $m_1\sim m_4$. 
Before this, we need to mention that the DCI result \eqref{eq:f^(L)} is derived at dimension $d=4$ and this integral is UV and IR finite. 

From the expression \eqref{eq:f^(L)}, we can derive $\mathcal S(f^{(2)})$. In the right hand side of \eqref{eq:f^(L)}, the presenting transcendental functions are logarithm functions $\log(z)$ and poly logarithm functions $\text {Li}_m(z)$. The corresponding symbols are
\begin{equation}\label{eq:log symbol}
    \mathcal S(\log(z))=S[z],
\end{equation}
and
\begin{equation}
    \mathcal S(\text {Li}_m( z))=-S[1-z,\underbrace{z,\cdots,z}_{(m-1)}].\\
\end{equation}
To derive the symbol $\mathcal S(f^{(2)})$ from the symbols of above functions, we may need the following properties of symbols: For two functions $f_1$ and $f_2$ whose symbols are
\begin{equation}
    \mathcal S(f_i)=\sum_j c_{ij} S_{ij}[\dots],
\end{equation}
where $i=1,2$, we have the following relations
\begin{equation}
    \mathcal S(f_1+f_2)=\mathcal S(f_1)+\mathcal S(f_2),
\end{equation}
\begin{equation}\label{eq:symbol-times}
    \mathcal S(f_1 f_2)=\sum_{i,j}c_{1i}c_{2j}\text{ Shuffle}(S_{1i}[\dots], S_{2j}[\dots]).
\end{equation}
The operation \textit{shuffle} for two symbol monomials $S[a_1,\dots,a_m]$ and $[b_1,\dots,b_n]$ 
results in a summation of symbol monomials with a sequence of $(m+n)$ letters which are some permutations of $\{a_1,\dots,a_m,b_1,\dots,b_n\}$ keeping the relative orders of $a$'s and $b$'s unchanged. 
Using \eqref{eq:log symbol}$\sim$\eqref{eq:symbol-times}, as well as \eqref{eq:f^(L)}, we can calculate the symbol of $f^{(2)}$ as
\begin{equation}
\begin{aligned}
\mathcal S(f^{(2)})=&
S[1-z,z,z,{\bar{z}}]+S[1-z,z,{\bar{z}},z]-S[1-z,z,{\bar{z}},{\bar{z}}]+S[1-z,{\bar{z}},z,z]\\&
-S[1-z,{\bar{z}},z,{\bar{z}}]-S[1-z,{\bar{z}},{\bar{z}},z]+S[z,1-z,z,z]-S[z,1-z,z,{\bar{z}}]\\&
-S[z,1-z,{\bar{z}},z]-S[z,z,1-z,z]+S[z,z,1-{\bar{z}},{\bar{z}}]+S[z,1-{\bar{z}},z,{\bar{z}}]\\&
+S[z,1-{\bar{z}},{\bar{z}},z]-S[z,1-{\bar{z}},{\bar{z}},{\bar{z}}]-S[z,{\bar{z}},1-z,z]+S[z,{\bar{z}},1-{\bar{z}},{\bar{z}}]\\&
+S[1-{\bar{z}},z,z,{\bar{z}}]+S[1-{\bar{z}},z,{\bar{z}},z]-S[1-{\bar{z}},z,{\bar{z}},{\bar{z}}]+S[1-{\bar{z}},{\bar{z}},z,z]\\&
-S[1-{\bar{z}},{\bar{z}},z,{\bar{z}}]-S[1-{\bar{z}},{\bar{z}},{\bar{z}},z]+S[{\bar{z}},1-z,z,z]-S[{\bar{z}},1-z,z,{\bar{z}}]\\&
-S[{\bar{z}},1-z,{\bar{z}},z]-S[{\bar{z}},z,1-z,z]+S[{\bar{z}},z,1-{\bar{z}},{\bar{z}}]+S[{\bar{z}},1-{\bar{z}},z,{\bar{z}}]\\&
+S[{\bar{z}},1-{\bar{z}},{\bar{z}},z]-S[{\bar{z}},1-{\bar{z}},{\bar{z}},{\bar{z}}]-S[{\bar{z}},{\bar{z}},1-z,z]+S[{\bar{z}},{\bar{z}},1-{\bar{z}},{\bar{z}}].
\end{aligned}
\end{equation}
This result can also be found in ``output'' folder of the auxiliary files, named ``f2Symbol.txt''.

In order to compare with the symbol from DCI with that from the canonical differential equations of the section \ref{subsec:symbol}, we consider the relations between the two languages introduced in \eqref{eq:z-def-1} and \eqref{eq:z-def-2}. After some simple calculations, we get the relations of the corresponding symbols,

\begin{equation}\label{eq:symbol-f2-in-z}
\begin{aligned}
    \text{dlog}(z)=&\frac{1}{2}\Big({\text{dlog}W_1+\text{dlog}W_3}{-\text{dlog}W_2-\text{dlog}W_4-\text{dlog}W_{25}}\Big),\\
    \text{dlog}(1-z)=&{\frac{1}{2}}\Big({\text{dlog}W_5+\text{dlog}W_6}{-\text{dlog}W_2-\text{dlog}W_4}\Big)+\frac{1}{4}\Big({\text{dlog}W_{{30}}}{-\text{dlog}W_{25}}\Big),\\
    \text{dlog}(\bar{z})=&\frac{1}{2}\Big({\text{dlog}W_1+\text{dlog}W_3+\text{dlog}W_{25}}{-\text{dlog}W_2-\text{dlog}W_4}\Big),\\
    \text{dlog}(1-\bar{z})=&{\frac{1}{2}}\Big({\text{dlog}W_5+\text{dlog}W_6}{-\text{dlog}W_2-\text{dlog}W_4}\Big)+{\frac{1}{4}}\Big({\text{dlog}W_{25}}{-\text{dlog}W_{30}}\Big),\\
\end{aligned}
\end{equation}
where the letters $W_i$ are defined in section \ref{sec:alphabet}. With these relations we can rewrite $\mathcal S(f^{(2)})$ in terms of $W_i$'s, considering the symbol property that
\begin{equation}
    S[a_1,\dots,a_{i-1},a_i,a_{i+1},\dots,a_m]=\sum_j c_j S[a_1,\dots,a_{i-1},b_{ij},a_{i+1},\dots,a_m],
\end{equation}
if
\begin{equation}\label{eq:3.115.2022.05.30.12.56}
    \text{dlog}(a_i)=\sum_j c_j \text{dlog}(b_{ij}).
\end{equation}

After the variable replacement using \eqref{eq:symbol-f2-in-z} to \eqref{eq:3.115.2022.05.30.12.56}, $\mathcal S(f^{(2)})$ is written in a summation of $474$ terms of symbols formed by letters $W_1\sim W_6$, $W_{25}$ and $W_{30}$. This expression is identical to the symbol of $I_1^{(4)}$ calculated in the section \ref{subsec:symbol}.

\section{The symbology from limits of DCI integrals}\label{sec:DCI}

In this section, we move to the study of the alphabet as well as certain properties of the symbols from viewpoint of DCI integrals. We will first give a quick review of DCI integrals, focusing on those relevant for our studies. 


\subsection{Review of relevant DCI integrals}

Recall that for a $n$-point massless planar integral, we can first introduce dual points $\{x_i\}$ such that $x_{i+1}-x_i=p_i$ with $x_{n+1}=x_1$ to make the momentum conservation manifest, and then the integral only depends on planar variables, which are also the first entries:
\[
x_{i,j}^2:=(x_i-x_j)^2=(p_i+\cdots+p_{j-1})^2.
\]
It is convenient to further introduce momentum twistors $Z_i\in \mathbb P^3$ such that each dual point $x_i$ is associated with a bi-twistor $Z_{i-1}\wedge Z_{i}$ and planar variables become
\[
(x_i-x_j)^2=\frac{Z_{i-1}\wedge Z_{i}\wedge Z_{j-1}\wedge Z_{j}}{(Z_{i-1}\wedge Z_{i}\wedge I_\infty)(Z_{j-1}\wedge Z_{j}\wedge I_\infty)}=\frac{\langle i-1\,i\,j-1\,j\rangle}{\langle i-1\,i\,\infty\rangle\langle j-1\,j\,\infty\rangle},
\]
where $\langle ijkl\rangle:=\det(Z_iZ_jZ_kZ_l)$ and $I_\infty$ is the {\it infinity bi-twistor}. In terms of momentum twistors, the massless conditions $(x_{i+1}-x_i)^2=p_i^2=0$ are automatically solved, and conformal transformations of dual momenta become $SL(4)$ linear transformations of $\{Z_i\}$ together with rescaling $Z_i\to t_i Z_i$. We say that an integral is dual conformal invariant if it only depends on cross-ratios of planar variables (or four-brackets $\langle i-1\,i\,j-1\,j\rangle$) and do not depend on the infinity bitwistor $I_\infty$. 

Note that for a general planar integral with $m$ (possibly massive) legs, the kinematics can be described by a subset of $n\geq m$ dual points as above, by identifying each massive momentum with two (or more) massless legs~\cite{Chicherin:2020umh}. For example, a one-mass triangle kinematics depends on $3$ of $4$ dual points, {\it e.g.} $x_2, x_3, x_4$ (but not $x_1$), and a four-mass box kinematics depends on $4$ of $8$ dual points, {\it e.g.} $x_2, x_4, x_6, x_8$. Moreover, if we start with a DCI integral which has at least one dual point which is not null separated from adjacent ones, we can arrive at a non-DCI integral by simply sending it to infinity, or identifying this bi-twistor with $I_\infty$. As explained in~\cite{He:2021eec}, a general DCI kinematics with $n{-}2m$ massless legs and $m$ massive legs has dimension $3n{-}2m{-}15$~\footnote{Except for the special case $n=8, m=4$ which has dimension $2$ instead of $1$.}. By sending a point (between two massive legs) to infinity, this is exactly the dimension of a non-DCI kinematics with $m{-}1$ massive legs and $n{-}2m$ massless ones: we have $3$ degree of freedoms for each massless leg, and $4$ for each massive one, minus the dimension of Poincare group $-10$ and an overall scaling, $3\times (n{-}2m)+4 \times (m{-}1)-10-1$. 


For our purpose, we start with a DCI kinematics that depends on five generic dual points (fully massive pentagon) with $n=10, m=5$:
\begin{center}
\begin{tikzpicture}[scale=0.2]
\draw[black,thick](0,0)--(5,0)--(6.54,4.75)--(2.50,7.69)--(-1.54,4.75)--cycle;
\draw[black,thick](-1.5,-1)--(0,0)--(0.4,-1.5);
\draw[black,thick](6.5,-1)--(5,0)--(4.6,-1.5);
\draw[black,thick](7.94,4.25)--(6.54,4.75)--(7.94,6);
\draw[black,thick] (-3.04,4.25)--(-1.54,4.75)--(-3.04,6);
\draw[black,thick](1.6,9.19)--(2.50,7.69)--(3.4,9.19);
\filldraw[black] (1.6,9.19) node[anchor=south] {{$1$}};
\filldraw[black] (3.4,9.19) node[anchor=south] {{$10$}};
\filldraw[black] (-3.04,4.25) node[anchor=east] {{$3$}};
\filldraw[black] (-3.04,6) node[anchor=east] {{$2$}};
\filldraw[black] (-1.5,-1) node[anchor=east] {{$4$}};
\filldraw[black] (0.4,-1.5) node[anchor=north] {{$5$}};
\filldraw[black] (4.6,-1.5) node[anchor=north] {{$6$}};
\filldraw[black] (6.5,-1) node[anchor=west] {{$7$}};
\filldraw[black] (7.94,4.25) node[anchor=west] {{$8$}};
\filldraw[black] (7.94,6) node[anchor=west] {{$9$}};
\filldraw[black] (-1,6.3) node[anchor=south] {{$x_2$}};
\filldraw[black] (5.6,6.3) node[anchor=south] {{$x_{10}$}};
\filldraw[black] (-1,6.3) node[anchor=south] {{$x_2$}};
\filldraw[black] (5.6,6.3) node[anchor=south] {{$x_{10}$}};
\filldraw[black] (-2,1.25) node[anchor=east] {{$x_4$}};
\filldraw[black] (6.6,1.25) node[anchor=west] {{$x_{8}$}};
\filldraw[black] (2.5,0) node[anchor=north] {{$x_{6}$}};
\end{tikzpicture}
\end{center}
which depends on $3\times 10-2\times 5-15=5$ cross-ratios
\begin{equation}\label{allus}
\begin{aligned}
&u_1=\frac{x_{4,10}^2x_{6,8}^2}{x_{6,10}^2 x_{4,8}^2},\ u_2=\frac{x_{2,8}^2 x_{4,6}^2}{x_{2,6}^2 x_{4,8}^2},\ v_1=\frac{x_{8,10}^2x_{4,6}^2}{x_{4,8}^2 x_{6,10}^2},v_2=\frac{x_{2,4}^2 x_{6,8}^2}{x_{2,6}^2 x_{4,8}^2},\ u_3=\frac{x_{2,10}^2 x_{4,6}^2}{x_{2,6}^2 x_{4,10}^2}.
\end{aligned}
\end{equation}
At one-loop level, any integral with this pentagon kinematics can be reduced to five four-mass boxes 
\begin{equation}
I_{\text{4m}}(4,6,8,10),\,\,
I_{\text{4m}}(2,4,6,10),\,\,
I_{\text{4m}}(2,4,8,10),\,\,
I_{\text{4m}}(2,6,8,10),\,\,
I_{\text{4m}}(2,4,6,8),
\label{4ms}
\end{equation}
where each DCI four-mass box integral gives
\begin{equation*}
I_{\text{4m}}(a,b,c,d)=
    \begin{tikzpicture}[baseline={([yshift=-.5ex]current bounding box.center)},scale=0.15]
                \draw[black,thick] (0,5)--(-5,5)--(-5,0)--(0,0)--cycle;
                \draw[black,thick] (1.93,5.52)--(0,5)--(0.52,6.93);
                \draw[black,thick] (1.93,-0.52)--(0,0)--(0.52,-1.93);
                \draw[black,thick] (-6.93,5.52)--(-5,5)--(-5.52,6.93);
                \draw[black,thick] (-6.93,-0.52)--(-5,0)--(-5.52,-1.93);
 \node at (-2.5,6.5) {$x_a$};
\node at (2,2.5) {$x_d$};
\node at (-2.5,-1.5) {$x_c$};
\node at (-6.5,2.5) {$x_b$};
\end{tikzpicture}=\frac{1}{\Delta_{abcd}}\left(\operatorname{Li}_2(1-z_{abcd})-\operatorname{Li}_2(1-\bar z_{abcd})+\frac{1}{2}\log(v_{abcd})\log(\frac{z_{abcd}}{\bar z_{abcd}})\right).
        \label{4mbox}
\end{equation*}
We have seen that $I_{4m}$ only depends two DCI variables
\[
u_{abcd}=\displaystyle\frac{x_{a,b}^{2}x_{c,d}^{2}}{x_{a,c}^{2}x_{b,d}^{2}}=:z_{abcd}\bar z_{abcd},\quad  v_{abcd}=\displaystyle\frac{x_{a,d}^{2}x_{b,c}^{2}}{x_{a,c}^{2}x_{b,d}^{2}}=:(1-z_{abcd})(1-\bar z_{abcd}),
\]
and it involves a square root, \[\Delta_{abcd}:=\sqrt{(1-u_{abcd}-v_{abcd})^{2}-4u_{abcd}v_{abcd}}.\] Note that their $\{u,v\}$ are related to the five cross-ratios \eqref{allus} as
\[
\{u_1,v_1\},\{u_3,\frac{v_2}{u_1}\},\{\frac{v_1 v_2}{u_1 u_2},\frac{u_3}{u_2}\},\{\frac{v_1}{u_2},\frac{u_1 u_3}{u_2}\},\{u_2,v_2\}
\]
respectively. In order to arrive at non-DCI kinematics, one can send any one of the five dual points to infinity and identify the resulting massive corners with four massive legs of four-mass non-DCI kinematics. As shown in the following subsection, this limit induces a map between cross-ratios \eqref{allus} and kinematics variables $\{s,t,m_i^2\}_{i=1,\cdots,4}$, and reveals connections between DCI and non-DCI letters for one-loop integrals. In fact, this is nothing but the well-known example that in such limits, a DCI four-box becomes a (non-DCI) massive triangle integral (literally the box with a point at infinity). 


Similarly, we can move to two loops, and the prototype of DCI integrals are $10$-point double-box of the form:
\begin{center}
\begin{tikzpicture}[baseline={([yshift=-.5ex]current bounding box.center)},scale=0.25]
                \draw[black,thick] (0,5)--(-5,5)--(-5,0)--(0,0)--cycle;
                \draw[black,thick] (1.93,5.52)--(0,5)--(0.52,6.93);
                \draw[black,thick] (1.93,-0.52)--(0,0)--(0.52,-1.93);
                \draw[black,thick] (-6.52,6.93)--(-5,5)--(-3.48,6.93); 
                \draw[black,thick] (-5,0)--(-5,5)--(-10,5)--(-10,0)--cycle;
                \draw[black,thick] (-11.93,5.52)--(-10,5)--(-10.52,6.93);
                \draw[black,thick] (-11.93,-0.52)--(-10,0)--(-10.52,-1.93);
                \filldraw[black] (1.93,6) node[anchor=west] {{$8$}};
                \filldraw[black] (0.52,6.93) node[anchor=south] {{$9$}};
                \filldraw[black] (1.93,-1) node[anchor=west] {{$7$}};
                \filldraw[black] (0.52,-1.93) node[anchor=north] {{$6$}};
                \filldraw[black] (-11.93,6) node[anchor=east] {{$3$}};
                \filldraw[black] (-10.52,6.93) node[anchor=south] {{$2$}};
                \filldraw[black] (-11.93,-1) node[anchor=east] {{$4$}};
                \filldraw[black] (-10.52,-1.93) node[anchor=north] {{$5$}};
                \filldraw[black] (-6.52,6.93) node[anchor=south] {{$1$}};
                \filldraw[black] (-3.48,6.93) node[anchor=south] {{$10$}};
            \end{tikzpicture}

\end{center}
Note that this DCI integral has not been computed directly, though a similar $9$-point double box  (with leg $10$ removed) has been computed. However, alphabets of such integrals can be predicted by considering twistor geometries associated with leading singularities (or the Schubert problems) as first proposed by N. Arkani-Hamed and worked out for numerous one- and two-loop examples in~\cite{Yang:2022gko}. We will not review the details of such a method, but just to outline the basic idea as follows.

The Schubert problem concerns geometric configurations of intersecting lines in momentum twistor space, which are associated with maximal cuts, or leading singularities of (four-dimensional) integrals~\cite{Arkani-Hamed:2010pyv}. By representing each loop and external dual point as a line in twistor space, each cut propagator corresponds to intersecting two lines, and it is a beautiful problem for determining the intersections on internal or external lines. Whenever there are at least four points on such a line, the only DCI quantities one can write down are cross-ratios of such points. Remarkably, as shown in~\cite{Yang:2022gko}, such cross-ratios lead to symbol letters of DCI integrals! The classical example include $A_3, E_6$ alphabets for $n=6,7$ ~\cite{Golden:2013xva} and the $9+9$ algebraic letters for $n=8$ two-loop integrals~\cite{Zhang:2019vnm}.

It is straightforward to apply Schubert-problem-based method to $10$-point double box (similar computation has been done for $9$-point case in~\cite{Yang:2022gko}), as we will demonstrate soon. Here we first give the result for its leading singularity, which is proportional to (inverse of) $$\Delta_{2}{=}\sqrt{(1{-}u_1{-}u_2{+}u_1u_3)^2{-}4v_1 v_2}.$$ 
As we will see, by sending certain dual point $x_i$ to infinity, this gives remaining square roots for non-DCI integrals, in addition to those from one-loop leading singularities. 

\subsection{The alphabet explained by limits of DCI integrals}

\subsubsection{Even letters}


To reduce the DCI kinematics to non-DCI four-mass kinematics, we take the non-DCI limit $x_2\to\infty$ in the DCI fully massive pentagon, then we get a non-DCI box
\begin{center}
\begin{tikzpicture}[baseline={([yshift=-.5ex]current bounding box.center)},scale=0.2]
\draw[black,thick](0,0)--(5,0)--(6.54,4.75)--(2.50,7.69)--(-1.54,4.75)--cycle;
\draw[black,thick](-1.5,-1)--(0,0)--(0.4,-1.5);
\draw[black,thick](6.5,-1)--(5,0)--(4.6,-1.5);
\draw[black,thick](7.94,4.25)--(6.54,4.75)--(7.94,6);
\draw[black,thick] (-3.04,4.25)--(-1.54,4.75)--(-3.04,6);
\draw[black,thick](1.6,9.19)--(2.50,7.69)--(3.4,9.19);
\node at (5.5,7.5) {$x_{10}$};
\node at (7,1.5) {$x_8$};
\node at (2.5,-1.5) {$x_6$};
\node at (-2,1.5) {$x_4$};
\node at (-0.5,7.5) {$x_2$};
\end{tikzpicture}
$\longrightarrow$
\begin{tikzpicture}[baseline={([yshift=-.5ex]current bounding box.center)},scale=0.2]
                \draw[black,thick] (0,5)--(-5,5)--(-5,0)--(0,0)--cycle;
                \draw[black,thick] (1.93,5.52)--(0,5)--(0.52,6.93);
                \draw[black,thick] (1.93,-0.52)--(0,0)--(0.52,-1.93);
                \draw[black,thick] (-6.93,5.52)--(-5,5)--(-5.52,6.93);
                \draw[black,thick] (-6.93,-0.52)--(-5,0)--(-5.52,-1.93);
\filldraw[black] (-7.5,-2.5) node {{$(P_2)$}};
\filldraw[black] (-7.5,7.5) node {{$(P_1)$}};
\filldraw[black] (2.5,-2.5) node {{$(P_3)$}};
\filldraw[black] (2.5,7.5) node {{$(P_4)$}};
            \node at (-2.5,6.5) {$x_{10}$};
\node at (2,2.5) {$x_8$};
\node at (-2.5,-1.5) {$x_6$};
\node at (-6.5,2.5) {$x_4$};
            \end{tikzpicture},
\end{center}
and we can represent massive legs by $P_1=x_{10,4}$, $P_2=x_{4,6}$, $P_3=x_{6,8}$ and $P_4=x_{8,10}$. 
In this limit, finite planar variables become
\begin{equation}
\begin{aligned}
&x_{6,10}^2\to (P_1+P_2)^2=s,\quad x_{4,8}^2\to (P_2+P_3)^2=t,\\
&x_{4,10}^2\to P_1^2=m_1^2,\quad x_{4,6}^2\to P_2^2=m_2^2,\quad x_{6,8}^2\to P_3^2=m_3^2,\quad x_{8,10}^2\to P_4^2=m_4^2,
\end{aligned}
\end{equation}
and other infinite $x_{2,*}^2$ cancel as $x_{2,i}^2/x_{2,j}^2\to 1$ in cross-ratios $\{u_{abcd},v_{abcd}\}$. For example, 
\begin{equation}
u_{24810}\to \frac{x_{8,10}^2}{x_{4,10}^2}\to \frac{m_4^2}{m_1^2},\quad 
v_{24810}\to \frac{x_{4,8}^2}{x_{4,10}^2}\to \frac{t}{m_1^2},
\end{equation}
and we can also see that
\begin{equation}
\Delta_{24810}\to \frac{1}{m_1^2}\sqrt{(m_1^2-t-m_4^2)^2-4tm_4^2}=\frac{r_3}{m_1^2},
\end{equation}
so $\Delta_{24810}$ becomes $r_3$ up to a factor $m_1^2$! Similarly, the square roots of five four-mass boxes in eq.\eqref{4ms} give $\{r_1,r_2,r_3,r_4,r_5\}$ (defined in eq.\eqref{defi:r1-5}) in this limit respectively.

Note that we can also set another $x_i$ to $\infty$ with certain identification of massive legs, then the same four-mass square root may correspond to another one-loop square root. For example, sending $x_8\to \infty$, $\Delta_{24810}$ gives $r_5$ by identifying $P_1=x_{6,10}$, $P_2=x_{10,2}$, $P_3=x_{2,4}$ and $P_4=x_{4,6}$. In addition, under any identification $\Delta_{24810}$ always gives $r_1$ when sending $x_6\to\infty$. The same four-mass square root may also correspond to different forms of the same square root under different non-DCI limits. For example, if we take $x_4\to\infty$ and set $P_1=x_{10,2},P_2=x_{2,6},P_3=x_{6,8},P_4=x_{8,10}$, then
\begin{equation}
\Delta_{24810}\to \frac{1}{x_{2,8}^2}\sqrt{(x_{2,8}^2-x_{2,10}^2-x_{8,10}^2)^2-x_{2,10}x_{8,10}}
\to \frac{1}{t}\sqrt{(t-m_1^2-m_4^2)^2-4 m_1^2 m_4^2}=\frac{r_3}{t},
\end{equation}
Two limits $x_2\to\infty$ and $x_4\to \infty$ coincidentally give the same square root $r_3$ since the four-mass box $I_{\text{4m}}(2,4,8,10)$ becomes the same triangle
\[
    \begin{tikzpicture}[baseline={([yshift=-.5ex]current bounding box.center)},scale=0.25]
                \draw[black,thick] (0,5)--(-5,5)--(-5,0)--(0,0)--cycle;
                \draw[black,thick] (1.93,5.52)--(0,5)--(0.52,6.93);
                \draw[black,thick] (1.93,-0.52)--(0,0)--(0.52,-1.93);
                \draw[black,thick] (-6.93,5.52)--(-5,5)--(-5.52,6.93);
                \draw[black,thick] (-6.93,-0.52)--(-5,0)--(-5.52,-1.93);
 \node at (-2.5,6.5) {$x_{10}$};
\node at (2,2.5) {$x_{8}$};
\node at (-2.5,-1.5) {$x_4$};
\node at (-6.5,2.5) {$x_{2}$};
\end{tikzpicture}
\longrightarrow 
\begin{tikzpicture}[baseline={([yshift=-.5ex]current bounding box.center)},scale=0.2]
                \draw[black] (1.5,5)--(-5.5,5)--(-2,0)--cycle;
                \draw[black,thick] (3,6)--(1.5,5);
                \draw[black,thick] (-7,6)--(-5.5,5);
                \draw[black,thick] (-3.5,-2)--(-2,0)--(-0.5,-2);
\filldraw[black] (-4,-3.5) node {{$(P_2)$}};
\filldraw[black] (-8,7.5) node {{$(P_1)$}};
\filldraw[black] (0.5,-3.5) node {{$(P_3)$}};
\filldraw[black] (4,7.5) node {{$(P_4)$}};
            \end{tikzpicture}.
\]
In a word, five four-mass square roots always give the set $\{r_1,\dots,r_5\}$, no matter how we take the limit $x_i\to \infty$ and how we identify external massive legs.

For two-loop cases, we again send $x_2\to\infty$, then
\begin{center}
\begin{tikzpicture}[baseline={([yshift=-.5ex]current bounding box.center)},scale=0.25]
                \draw[black,thick] (0,5)--(-5,5)--(-5,0)--(0,0)--cycle;
                \draw[black,thick] (1.93,5.52)--(0,5)--(0.52,6.93);
                \draw[black,thick] (1.93,-0.52)--(0,0)--(0.52,-1.93);
                \draw[black,thick] (-6.52,6.93)--(-5,5)--(-3.48,6.93); 
                \draw[black,thick] (-5,0)--(-5,5)--(-10,5)--(-10,0)--cycle;
                \draw[black,thick] (-11.93,5.52)--(-10,5)--(-10.52,6.93);
                \draw[black,thick] (-11.93,-0.52)--(-10,0)--(-10.52,-1.93);
                \filldraw[black] (1.93,6) node[anchor=west] {{$8$}};
                \filldraw[black] (0.52,6.93) node[anchor=south] {{$9$}};
                \filldraw[black] (1.93,-1) node[anchor=west] {{$7$}};
                \filldraw[black] (0.52,-1.93) node[anchor=north] {{$6$}};
                \filldraw[black] (-11.93,6) node[anchor=east] {{$3$}};
                \filldraw[black] (-10.52,6.93) node[anchor=south] {{$2$}};
                \filldraw[black] (-11.93,-1) node[anchor=east] {{$4$}};
                \filldraw[black] (-10.52,-1.93) node[anchor=north] {{$5$}};
                \filldraw[black] (-6.52,6.93) node[anchor=south] {{$1$}};
                \filldraw[black] (-3.48,6.93) node[anchor=south] {{$10$}};
            \end{tikzpicture}
$\longrightarrow$
\begin{tikzpicture}[baseline={([yshift=-.5ex]current bounding box.center)},scale=0.25]
                \draw[black,thick] (0,5)--(-5,5)--(-5,0)--(0,0)--cycle;
                \draw[black,thick] (1.93,5.52)--(0,5)--(0.52,6.93);
                \draw[black,thick] (1.93,-0.52)--(0,0)--(0.52,-1.93);
                \draw[black,thick] (-6.52,6.93)--(-5,5)--(-3.48,6.93); 
                \draw[black,thick] (-5,5)--(-9,2.5)--(-5,0);
                \draw[black,thick] (-10,3.5)--(-9,2.5)--(-10,1.5);
                \filldraw[black] (1.93,6) node[anchor=west] {{$8$}};
                \filldraw[black] (0.52,6.93) node[anchor=south] {{$9$}};
                \filldraw[black] (1.93,-1) node[anchor=west] {{$7$}};
                \filldraw[black] (0.52,-1.93) node[anchor=north] {{$6$}};
                \filldraw[black] (-6.52,6.93) node[anchor=south] {{$3$}};
                \filldraw[black] (-3.48,6.93) node[anchor=south] {{$10$}};
\filldraw[black] (-10,3.5) node[anchor=east] {{$4$}};
\filldraw[black] (-10,1.5) node[anchor=east] {{$5$}};
            \end{tikzpicture}.
\end{center}
 Now we have different choices to label the massive corners as the non-DCI massive legs. For example, if we still identify $P_1=x_{10,4}$, $P_2=x_{4,6}$, $P_3=x_{6,8}$ and $P_4=x_{8,10}$, then $\Delta_2\to \frac{1}{st}r_9$. Similarly, if we identify $P_3=x_{10,4}$, $P_4=x_{4,6}$, $P_1=x_{6,8}$ and $P_2=x_{8,10}$, $\Delta_2\to \frac{1}{st}r_{11}$.

Here we list all the proper limits and corresponding identifications to get the other square roots:
\begin{itemize}
\item [$r_6$]: sending $x_6\to\infty$ with the identification $P_1=x_{10,2}$, $P_2=x_{2,4}$, $P_3=x_{4,8}$ and $P_4=x_{8,10}$,
\item [$r_7$]: sending $x_6\to\infty$ with the identification $P_4=x_{10,2}$, $P_1=x_{2,4}$, $P_2=x_{4,8}$ and $P_3=x_{8,10}$; 
\item [$r_8$]: sending $x_8\to\infty$ with the identification $P_2=x_{10,2}$, $P_3=x_{2,4}$, $P_4=x_{4,6}$ and $P_1=x_{6,10}$,
\item [$r_{10}$]: sending $x_8\to\infty$ with the identification  $P_4=x_{10,2}$, $P_1=x_{2,4}$, $P_2=x_{4,6}$ and $P_3=x_{6,10}$;
\end{itemize}
Note that we can also reproduce the six two-loop roots starting from other double-box integrals with the same topology but the external legs cyclically rotated.

A nice consequence of such identifications is that the roots always take the form $r_i^2:=A_i^2-4 B_i$, where $A_i$ and $B_i$ are polynomials in kinematic variables. We we record them as
\begin{equation}
\begin{aligned}
&r_1^2=(st-m_1^2 m_3^2-m_2^2 m_4^2)^2-4 m_1^2 m_2^2 m_3^2 m _4^2,\\
&r_2^2=(s-m_1^2-m_2^2)^2-4 m_1^2 m_2^2, \quad 
r_3^2=(t-m_1^2-m_4^2)^2-4 m_1^2 m_4^2,\\
&r_4^2=(s-m_3^2-m_4^2)^2-4 m_3^2 m_4^2, \quad 
r_5^2=(t-m_2^2-m_3^2)^2-4 m_2^2 m_3^2,\\
&r_6^2=(s+t-m_1^2-m_3^2)^2-4 m_2^2 m_4^2,\quad r_7^2=(s+t-m_2^2-m_4^2)^2-4 m_1^2 m_3^2,\\
&r_8^2=((m_1^2-m_2^2) m_4^2+ s (t-m_1^2))^2-4 s m_1^2 m_3^2 m_4^2,\\
&r_9^2=((m_2^2-m_1^2) m_3^2+ s (t-m_2^2))^2-4 s m_2^2 m_3^2 m_4^2,\\
&r_{10}^2=((m_3^2-m_4^2) m_2^2+ s (t-m_3^2))^2-4 s m_1^2 m_2^2 m_3^2,\\
&r_{11}^2=((m_4^2-m_3^2) m_1^2+ s (t-m_4^2))^2-4 s m_1^2 m_2^2 m_4^2,
\end{aligned}
\end{equation}
where $r_2, \cdots, r_5$ form a cyclic orbit, so do $r_6, r_7$ and $r_8, \cdots, r_{11}$. We can write ``one-loop" square roots $r_i^2$ for $i=1,\dots,5$ as $A_i^2-4 B_i$ in different ways, e.g.
\begin{equation}
r_3^2=(t-m_1^2-m_4^2)^2-4 m_1^2 m_4^2=(m_1^2-t-m_4^2)^2-4tm_4^2,
\end{equation}
which reflect the fact that there are different choices to take limits for the one-loop case. Note that no matter how we write it, $B_i$ is always a monomial which is {\it positive} when all kinematic variables are positive. Furthermore, since the DCI square root is {\it positive} for any positive DCI kinematic point, these square roots stay positive in non-DCI limit for such positive regions.

We have one more comment on the $W_{18}$ and its positivity. Note that there exists a DCI letter
\begin{align}
 \tilde W&=u_1^2 u_2 u_3-u_1^2 u_2-u_1^2 u_3^2-u_1^2 u_3 v_2+u_1^2 u_3-u_1 u_2^2-u_1 u_2 u_3 v_1+u_1 u_2 u_3+u_1 u_2 v_1 \nonumber\\
 &+u_1 u_2 v_2+u_1 u_2+u_1 u_3 v_1 v_2+u_1 u_3 v_1+u_1 u_3 v_2-u_1 u_3+u_1 v_1 v_2-u_1 v_1+u_2 v_1 v_2\nonumber\\
 &-u_2 v_2-v_1^2 v_2-v_1 v_2^2+v_1 v_2   
\end{align}
whose non-DCI limits are always proportional to $W_{18}$, independent of the dual point $x_i$ we send to infinity. For example, the DCI letter becomes
\begin{equation}
\tilde W\to -\frac{m_2^2m_3^2}{s^3t^2}W_{18},\ \tilde W\to -\frac{m_3^2}{m_2^4s^2}W_{18},
\end{equation}
when sending $x_2\to\infty$ or $x_4\to\infty$ respectively, and similar for the other $3$ non-DCI limits. The crucial point is that $\tilde W$ is positive definite in the positive region $G_+(4,10)$, as can be checked by any positive parameterization~\cite{postnikov2006total}. It is believed that $\tilde W$ is a letter of $10$-point double-box integral.

\subsubsection{Odd letters}

Next we study algebraic or odd letters, {\it i.e.} those that involve square roots, and their $\log$ flip signs when we take flip the sign of a square root. The discussion above directly motivates us to relate odd letters that only involve one square root, {\it i.e. } $W_{19},W_{20},\dots,W_{34}$, to similar odd letters for DCI integrals. For one-loop case, it is well known that odd letters of four-mass box (and higher-loop generalizations such as ladder integrals) take the form $z_i/\bar{z}_i$ and $(1-z_i)/(1-\bar{z}_i)$. In the  ``non-DCI" limit $x_2\to\infty$ with $P_1=x_{10,4}$, $P_2=x_{4,6}$, $P_3=x_{6,8}$ and $P_4=x_{8,10}$, the five $z$'s obtained from limits of one-loop case take the form
\begin{equation}\label{z1-5}
\begin{aligned}
&z_1 = \frac 12+\frac{m_1^2 m_3^2-m_2^2 m_4^2-r_1}{2 s t},\quad z_2 =\frac 12+ \frac{m_2^2-s+r_2}{2 m_1^2},\quad z_3 = \frac12+\frac{m_4^2-t+r_3}{2 m_1^2},\\
&z_4 = \frac{1}{2}+\frac{-m_3^2+m_4^2-r_4}{2 s},\quad z_5 =\frac12+ \frac{m_2^2-m_3^2-r_5}{2 t},
\end{aligned}
\end{equation}
and $\bar z_i:=z_i(r_i\to -r_i)$. Remarkably, we find that the $10$ odd letters from four-mass boxes, $z_i/\bar{z}_i$ and $(1-z_i)/(1-\bar{z}_i)$ for $i=1,\cdots, 5$ are nothing but (multiplicative combinations of) $W_{21},\dots, W_{30}$; precisely we have
\begin{equation}
\begin{aligned}
&W_{21}=\frac{z_2}{\bar z_2}, W_{22}=\frac{1-z_4}{1-\bar z_4}\frac{\bar z_4}{ z_4}, W_{23}=\frac{1-z_5}{1-\bar z_5}\frac{\bar z_5}{ z_5}, W_{24}=\frac{\bar z_3}{z_3},
W_{25}=\frac{1-z_1}{1-\bar z_1}\frac{\bar z_1}{ z_1}, W_{26}=\biggl(\frac{1-z_2}{1-\bar z_2}\biggr)^2\frac{\bar z_2}{ z_2},\\
&
W_{27}=\frac{1-z_4}{1-\bar z_4}\frac{z_4}{\bar z_4},\,
W_{28}=\frac{1-\bar z_5}{1-z_5}\frac{\bar z_5}{z_5},\,
W_{29}=\biggl(\frac{1-\bar z_3}{1- z_3}\biggr)^2\frac{z_3}{\bar  z_3},\,
W_{30}=\frac{1-\bar z_1}{1- z_1}\frac{\bar z_1}{z_1}.
\end{aligned}
\end{equation}
For the remaining $6$ odd letters with single square roots obtained from two-loop diagrams, they cannot be interpreted as odd letters of a four-mass box, but it is straightforward to introduce analogous $z$'s
\begin{equation}
\begin{aligned}
&z_6=\frac{1}{2}+\frac{s-m_3^2+r_6}{2 \left(t-m_1^2\right)},
\quad z_7=\frac{1}{2}+\frac{t-m_4^2+r_7}{2 \left(s-m_2^2\right)},\\
&z_8=\frac{1}{2}+\frac{(m_1^2-m_2^2) m_4^2+r_8}{2 s \left(t-m_1^2\right)},\,\,
z_9=\frac{1}{2}+\frac{(m_2^2-m_1^2) m_3^2+r_9}{2 s \left(t-m_2^2\right)},\\
&z_{10}=\frac{1}{2}+\frac{(m_3^2-m_4^2) m_2^2+r_{10}}{2 s \left(t-m_3^2\right)},\,
z_{11}=\frac{1}{2}+\frac{(m_4^2-m_3^2) m_1^2+r_{11}}{2 s \left(t-m_4^2\right)}
\end{aligned}   
\end{equation}
which nicely simplify them as (note there is no letters of the form $(1-z)/(1-\bar{z})$):
\[
W_{19}=\frac{\bar z_6}{z_6},\,
W_{20}=\frac{\bar z_7}{z_7},\,
W_{31}=\biggl(\frac{z_{10}}{\bar z_{10}}\biggr)^2,\,
W_{32}=\biggl(\frac{\bar z_{11}}{z_{11}}\biggr)^{2},\,
W_{33}=\biggl(\frac{z_{8}}{\bar z_{8}}\biggr)^2,\,
W_{34}=\biggl(\frac{\bar z_{9}}{z_{9}}\biggr)^{2}.
\]
Thus all odd letters with single square root can be explained from those for DCI integrals. 

Now we turn to the other $34$ odd letters $W_{35},\dots,W_{68}$ involving two square roots, which take the form
\[
\frac{a+r_ir_j}{a-r_ir_j}
\]
with $a$ being generally involved polynomials in kinematic variables, and we call them mixed algebraic letters to distinguish them from those letters with only one square root. Mixed algebraic letters only show up as two-loop letters. We can generate them from the approach of Schubert problems and the alphabet of  $10$-point double-box integral above. 

To be explicit, we consider $5$ lines, $(12)$, $(34)$, $(56)$, $(78)$, $(9\ 10)$ in momentum twistor space, which correspond to dual points $\{x_2,x_4,x_6,x_8,x_{10}\}$ respectively. Following the procedure in \cite{Yang:2022gko}, for each DCI four-mass-box $I_{4m}(i,j,k,l)$, two solutions $(AB)^\pm_{i,j,k,l}$ from its maximal cut  produce $8$ intersections on its $4$ external lines, which will be denoted as $\{\alpha_{i,j,k,l}^\pm,\beta_{i,j,k,l}^\pm,\gamma_{i,j,k,l}^\pm,\delta_{i,j,k,l}^\pm\}$ on $\{(i{-}1i),(j{-}1 j),(k{-}1k),(l{-}1l)\}$ respectively. Furthermore, after cutting all propagators of the $10$-point double-box integral, {\it i.e.} looking for two lines $(CD)$ and $(EF)$ such that $(CD)$ intersects with $(12)$, $(34)$, $(56)$ $(EF)$, and  $(EF)$ intersects with $(CD)$, $(56)$, $(78)$, $(9\ 10)$, we have two pairs of solutions $\{(CD)_i,(EF)_i\}_{i=\pm}$, and each external line contains two new intersections, which we denote them as $\{\epsilon_{k}^\pm\}_{k=2,4,6,8,10}$ on $x_k$. Note that no matter $i=+$ or $i=-$, three lines $(CD)_i$ $(EF)_i$ and $(56)$ share a same intersection $\epsilon_6^i$, thus $(56)$ has only two new intersections on it as well. 
\begin{center}
\begin{tikzpicture}[scale=0.72]
\draw[black,ultra thick](-4,2)--(-4,-2);
\draw[black,ultra thick](-2,2)--(-2,-2);
\draw[black,ultra thick](0,2)--(0,-2);
\draw[black,ultra thick](2,2)--(2,-2);
\draw[blue,thick](-4.5,1)--(2.5,1);
\draw[blue,thick](-4.5,-1)--(2.5,-1);
\filldraw[blue] (2.5,1) node[anchor=west] {{$(AB)^+_{i,j,k,l}$}};
\filldraw[blue] (2.5,-1) node[anchor=west] {{$(AB)^+_{i,j,k,l}$}};
\filldraw[black] (-4,2) node[anchor=south] {{$i{-}1$}};
\filldraw[black] (-4,-2) node[anchor=north] {{$i$}};
\filldraw[black] (-2,2) node[anchor=south] {{$j{-}1$}};
\filldraw[black] (-2,-2) node[anchor=north] {{$j$}};
\filldraw[black] (0,2) node[anchor=south] {{$k{-}1$}};
\filldraw[black] (0,-2) node[anchor=north] {{$k$}};
\filldraw[black] (2,2) node[anchor=south] {{$l{-}1$}};
\filldraw[black] (2,-2) node[anchor=north] {{$l$}};
\filldraw[blue]  (-4,1) circle [radius=2pt];
\filldraw[blue]  (-4,-1) circle [radius=2pt];
\filldraw[blue]  (-2,1) circle [radius=2pt];
\filldraw[blue]  (-2,-1) circle [radius=2pt];
\filldraw[blue]  (0,1) circle [radius=2pt];
\filldraw[blue]  (0,-1) circle [radius=2pt];
\filldraw[blue]  (2,1) circle [radius=2pt];
\filldraw[blue]  (2,-1) circle [radius=2pt];
\filldraw[blue] (-4,1) node[anchor=north west] {{$\alpha_{i,j,k,l}^+$}};
\filldraw[blue] (-4,-1) node[anchor=north west] {{$\alpha_{i,j,k,l}^-$}};
\filldraw[blue] (-2,1) node[anchor=north west] {{$\beta_{i,j,k,l}^+$}};
\filldraw[blue] (-2,-1) node[anchor=north west] {{$\beta_{i,j,k,l}^-$}};
\filldraw[blue] (0,1) node[anchor=north west] {{$\gamma_{i,j,k,l}^+$}};
\filldraw[blue] (0,-1) node[anchor=north west] {{$\gamma_{i,j,k,l}^-$}};
\filldraw[blue] (2,1) node[anchor=north west] {{$\delta_{i,j,k,l}^+$}};
\filldraw[blue] (2,-1) node[anchor=north west] {{$\delta_{i,j,k,l}^-$}};
\end{tikzpicture}
        \begin{tikzpicture}[scale=0.72]
                \draw[black,ultra thick] (-3,1)--(3,1);
                 \draw[black,ultra thick] (-3.6957,1.9535)--(-1.2369,3.4516);
                  \draw[black,ultra thick] (3.6957,1.9535)--(1.2369,3.4516);
                   \draw[black,ultra thick] (3.6638,3.1899)--(1.4426,4.5521);
                   \draw[black,ultra thick] (-3.6638,3.1899)--(-1.2935,4.6776);
\draw [purple,thick](-3.9774,4.5424) -- (-0.6694,0.0994);
\draw [purple,thick](-2.3401,0.2458) -- (3.0955,4.3863);
\draw [purple,thick](3.9774,4.5424) -- (0.6694,0.0994);
\draw [purple,thick](2.3401,0.2458) -- (-3.0955,4.3863);
\node at (-1.3046,4.6518) {2};
\node at (1.4791,4.5484) {10};
\filldraw[purple]  (-1.3364,1.001) circle [radius=2pt];
\filldraw[purple]  (1.3427,0.9717) circle [radius=2pt];
\filldraw[purple]  (3.1939,3.4457) circle [radius=2pt];
\filldraw[purple]  (2.4755,3.89) circle [radius=2pt];
\filldraw[purple]  (2.5511,2.6422) circle [radius=2pt];
\filldraw[purple]  (1.5869,3.1999) circle [radius=2pt];
\filldraw[purple]  (-2.4591,3.9184) circle [radius=2pt];
\filldraw[purple]  (-3.187,3.4552) circle [radius=2pt];
\filldraw[purple]  (-1.58,3.1905) circle [radius=2pt];
\filldraw[purple]  (-2.5537,2.5949) circle [radius=2pt];
\node at (-4.0323,3.2628) {1};
\node at (4.088,3.2533) {9};
\node at (-3.9381,1.8448) {3};
\node at (-0.9884,3.5842) {4};
\node at (0.9117,3.6503) {8};
\node at (4.0124,1.8164) {7};
\node at (-3.2194,1.0223) {5};
\node at (3.2372,0.994) {6};
\node [purple] at (-4.3727,4.7564) {$(CD)_+$};
\node [purple] at (-2.9329,4.7469) {$(CD)_-$};
\node [purple] at (4.3727,4.7564) {$(EF)_-$};
\node [purple] at (2.9329,4.7469) {$(EF)_+$};
\node [purple] at (-1.4234,0.5995) {$\epsilon_6^+$};
\node [purple] at (2.4364,3.5697) {$\epsilon_{10}^+$};
\node [purple] at (1.4793,2.8074) {$\epsilon_8^+$};
\node [purple] at (-2.6116,2.1044) {$\epsilon_4^+$};
\node [purple] at (-3.2383,3.0276) {$\epsilon_2^+$};
\node [purple] at (1.4234,0.5995) {$\epsilon_6^-$};
\node [purple] at (3.2485,3.0191) {$\epsilon_{10}^-$};
\node [purple] at (2.6261,2.2427) {$\epsilon_8^-$};
\node [purple] at (-1.0498,3.2166) {$\epsilon_4^-$};
\node [purple] at (-1.9137,3.8858) {$\epsilon_2^-$};
\end{tikzpicture}
\end{center}

Similar to the $9$-point case in \cite{Yang:2022gko}, now we can construct possible DCI letters for $10$-point two-loop double-box integral from the intersections on external lines, once a line has four distinct points on it. The upshot is that all $34$ mixed algebraic letters here are non-DCI limits of certain DCI letters constructed in this approach!

For instance, to generate $W_{35}$, which involves two one-loop square roots $r_2$ and $r_4$. We consider four distinct intersections
\begin{equation}
\{X_1=\alpha_{2,4,6,10}^+,X_2=\alpha_{2,4,6,10}^-,X_3=\alpha_{2,6,8,10}^+,X_4=\alpha_{2,6,8,10}^-\}
\end{equation}
on the line $(12)$. Since each $X_i$ is an individual momentum twistor, we can construct the following cross-ratio
\[\frac{(X_1,X_3)(X_2,X_4)}{(X_1,X_4)(X_2,X_3)}:=\frac{\langle X_1X_3I\rangle\langle X_2X_4I\rangle}{\langle X_1X_4I\rangle\langle X_2X_3I\rangle}\]
from this configuration. Here $I$ is a reference line (bitwistor) in momentum twistor space that does not intersect with $(12)$.  It can be directly checked that such a cross-ratio is DCI and actually independent of $I$. Moreover, after taking its non-DCI limit $x_2\to\infty$ and identifying $P_1=x_{10,4}$, $P_2=x_{4,6}$, $P_3=x_{6,8}$, $P_4=x_{8,10}$, this cross-ratio yields the mixed algebraic letter $W_{35}$! Furthermore, $W_{37}$, involving one one-loop square root $r_2$ and one two-loop square root $r_6$, can also be constructed from this approach. In this case we take four points on $(56)$ as
\begin{equation}
\{X_1=\epsilon_6^+,X_2=\epsilon_6^-,X_3=\gamma_{2,4,6,10}^+,X_4=\gamma_{2,4,6,10}^-\},
\end{equation}
and consider $\frac{(X_1,X_3)(X_2,X_4)}{(X_1,X_4)(X_2,X_3)}$ again. Non-DCI limit $x_6\to\infty$ of this letter with the identification $P_1=x_{10,2}$, $P_2=x_{2,4}$, $P_3=x_{4,8}$ and $P_4=x_{8,10}$ then gives $W_{37}$ as we want.

In the rest of this subsection, we list all the configurations we use to construct $34$ mixed algebraic letters. We first classify $34$ mixed algebraic letters according to different square roots, {\it e.g.} letters with only one-loop square roots and letters with various two-loop roots. For each group we fix a non-DCI limit and the corresponding identification of external legs, and present one possible choice $\{X_1,X_2,X_3,X_4\}$ to construct each individual letter. We stick to the cross-ratio $\frac{(X_1,X_3)(X_2,X_4)}{(X_1,X_4)(X_2,X_3)}$ from each configuration and consider its non-DCI limit in each case. Note that the configurations we present here are not unique and these mixed algebraic letters can be generated by many different choices.
\begin{itemize}

    \item [1.] letters with only one-loop square roots $r_1\sim r_5$: This group of letters consists of $W_{35}$, $W_{36}$, $W_{38}$, $W_{40}$, $W_{42}$ and $W_{54}$ with only triangle roots, and $W_{46}$, $W_{51}$, $W_{59}$, $W_{62}$ with $r_1$. To generate them, we keep the non-DCI limit $x_2\to\infty$ and the identification $P_1=x_{10,4}$, $P_2=x_{4,6}$, $P_3=x_{6,8}$, $P_4=x_{8,10}$.  As we have pointed out, five square roots of DCI four-mass boxes in \eqref{4mbox} give $\{r_1,r_2,r_3,r_4,r_5\}$ respectively. Then the configurations we need are 
    \begin{align*}
        &W_{35}:\{\alpha_{2,4,6,10}^+,\alpha_{2,4,6,10}^-,\alpha_{2,6,8,10}^+,\alpha_{2,6,8,10}^-\},W_{36}:\{\alpha_{2,4,6,10}^+,\alpha_{2,4,6,10}^-,\alpha_{2,4,6,8}^+,\alpha_{2,4,6,8}^-\}\\
        &W_{38}:\{\alpha_{2,4,6,10}^+,\alpha_{2,4,6,10}^-,\alpha_{2,4,8,10}^+,\alpha_{2,4,8,10}^-\},W_{40}:\{\alpha_{2,4,8,10}^+,\alpha_{2,4,8,10}^-,\alpha_{2,6,8,10}^+,\alpha_{2,6,8,10}^-\}\\
        &W_{42}:\{\alpha_{2,6,8,10}^+,\alpha_{2,6,8,10}^-,\alpha_{2,4,6,8}^+,\alpha_{2,4,6,8}^-\},\ \ W_{54}:\{\alpha_{2,4,8,10}^+,\alpha_{2,4,8,10}^-,\alpha_{2,4,6,8}^+,\alpha_{2,4,6,8}^-\},
    \end{align*}
    for those with only triangle square roots, and
    \begin{align*}
        &W_{46}:\{\delta_{4,6,8,10}^+,\delta_{4,6,8,10}^-,\delta_{2,4,6,10}^+,\delta_{2,4,6,10}^-\},W_{51}:\{\delta_{4,6,8,10}^+,\delta_{4,6,8,10}^-,\delta_{2,6,8,10}^+,\delta_{2,6,8,10}^-\}\\
        &W_{59}:\{\gamma_{4,6,8,10}^+,\gamma_{4,6,8,10}^-,\gamma_{2,4,6,8}^+,\gamma_{2,4,6,8}^-\}, W_{62}:\{\gamma_{4,6,8,10}^+,\gamma_{4,6,8,10}^-,\gamma_{2,4,8,10}^+,\gamma_{2,4,8,10}^-\}
    \end{align*}
    for those with one triangle square root $r_2\sim r_5$ and one $r_1$.

    \item [2.] letters with $r_6$ or $r_7$: letters involving $r_6$ consists of $W_{37}$, $W_{41}$, $W_{55}$, $W_{56}$, and similarly those with $r_7$ are $W_{39}$, $W_{43}$, $W_{57}$ and $W_{58}$. Sending $x_6\to\infty$ with the identification $P_1=x_{10,2}$, $P_2=x_{2,4}$, $P_3=x_{4,8}$ and $P_4=x_{8,10}$ (note that under this limit five square roots of DCI four-mass boxes \eqref{4mbox} give $\{r_4,r_2,r_1,r_3,r_5\}$ respectively), we can then recover the four letters from $(56)$ as
    \begin{align*}
        &W_{37}:\{\gamma_{2,4,6,10}^+,\gamma_{2,4,6,10}^-,\epsilon_6^+,\epsilon_6^-\}, W_{41}:\{\beta_{4,6,8,10}^+,\beta_{4,6,8,10}^-,\epsilon_6^+,\epsilon_6^-\},\\
        &W_{55}:\{\gamma_{2,4,6,8}^+,\gamma_{2,4,6,8}^-,\epsilon_6^+,\epsilon_6^-\}, W_{56}:\{\beta_{2,6,8,10}^+,\beta_{2,6,8,10}^-,\epsilon_6^+,\epsilon_6^-\}.
    \end{align*}
On the other hand, if we keep $x_6\to\infty$ but change the identification to $P_4=x_{10,2}$, $P_1=x_{2,4}$, $P_2=x_{4,8}$ and $P_3=x_{8,10}$ (five square roots of DCI four-mass boxes \eqref{4mbox} give $\{r_5,r_3,r_1,r_4,r_2\}$ respectively), we will get the four letters with $r_7$ from the four configurations instead. 
    \item [3.] letters with $r_8$ or $r_{10}$: letters involving $r_8$ consists of $W_{45}$, $W_{52}$, $W_{63}$ $W_{65}$. Sending $x_8\to\infty$ with the identification $P_2=x_{10,2}$, $P_3=x_{2,4}$, $P_4=x_{4,6}$ and $P_1=x_{6,10}$ (five square roots of DCI four-mass boxes \eqref{4mbox} give $\{r_3,r_1,r_5,r_2,r_4\}$ respectively), we can then recover these four letters from 
    \begin{align*}
        W_{45}: \{\gamma_{2,6,8,10}^+,\gamma_{2,6,8,10}^-,\epsilon_8^+,\epsilon_8^-\}, W_{52}: \{\delta_{2,4,6,8}^+,\delta_{2,4,6,8}^-,\epsilon_8^+,\epsilon_8^-\},\\
        W_{63}: \{\gamma_{4,6,8,10}^+,\gamma_{4,6,8,10}^-,\epsilon_8^+,\epsilon_8^-\}, W_{65}: \{\delta_{2,4,6,10}^+,\delta_{2,4,6,10}^-,\epsilon_{10}^+,\epsilon_{10}^-\}.
    \end{align*}

On the other hand, if we keep $x_8\to\infty$ but change the identification to $P_4=x_{10,2}$, $P_1=x_{2,4}$, $P_2=x_{4,6}$ and $P_3=x_{6,10}$ (five square roots of DCI four-mass boxes \eqref{4mbox} give $\{r_5,r_1,r_3,r_4,r_2\}$ respectively), we will get the four letters $W_{47}$, $W_{50}$, $W_{61}$ and $W_{67}$ with $r_{10}$ from the four configurations instead.
    \item [4.] letters with $r_9$ or $r_{11}$: letters involving $r_9$ consists of $W_{44}$, $W_{53}$, $W_{60}$ and $W_{66}$. Sending $x_2\to\infty$ with the identification $P_1=x_{10,4}$, $P_2=x_{4,6}$, $P_3=x_{6,8}$, $P_4=x_{8,10}$ (same limit and identification as that for the first group), we can then recover these four letters from 
    \begin{align*}
        W_{44}: \{\alpha_{2,4,6,10}^+,\alpha_{2,4,6,10}^-,\epsilon_2^+,\epsilon_2^-\}, W_{53}: \{\alpha_{2,6,8,10}^+,\alpha_{2,6,8,10}^-,\epsilon_2^+,\epsilon_2^-\},\\
        W_{60}: \{\alpha_{2,4,6,8}^+,\alpha_{2,4,6,8}^-,\epsilon_2^+,\epsilon_2^-\}, W_{66}: \{\delta_{4,6,8,10}^+,\delta_{4,6,8,10}^-,\epsilon_{10}^+,\epsilon_{10}^-\}.
    \end{align*}

On the other hand, if we change the identification to $P_3=x_{10,4}$, $P_4=x_{4,6}$, $P_1=x_{6,8}$ and $P_2=x_{8,10}$ (five square roots of DCI four-mass boxes \eqref{4mbox} give $\{r_1,r_4,r_5,r_2,r_3\}$), we recover letters  $W_{48}$, $W_{49}$, $W_{64}$ and $W_{68}$ with $r_{11}$.
\end{itemize}
Note that the appearance of algebraic letters with two-loop square roots has certain pattern. For instance, on the $17$th row of the DE ($I_{17}=r_6 G_{0,1,1,0,1,1,1,0,0}$, which is finite), there are five off-diagonal elements which are just the five odd letters $\{W_{19},W_{37},W_{41},W_{55},W_{56}\}$ with $r_6$. Furthermore, these are exactly the last entries of $I_{17}$ at ${\cal O}(\epsilon)$! Similar phenomenon applies to double-triangle $I_{32}$ with $r_7$ and box-triangles $I_5$, $I_6$, $I_7$, $I_8$ with $r_8,\cdots, r_{11}$, respectively: at leading order, there are exactly $5$ last entries, namely those $5$ odd letters containing $r_i$, for each of these integrals.

Last but not least, we note that the product of numerator and denominator of each mixed algebraic letter is always proportional to the last rational letter, which was shown to be the Gram determinant, and other factors of this product are just kinematic variables $m_1^2, \dots, m_4^2, s, t$. For example, the product of numerator and denominator of $W_{36}$ 
is
\begin{equation}
(f_{36}+r_2r_5)(f_{36}-r_2r_5)=4 m_2^2 W_{18}.
\end{equation}
Therefore, any mixed odd letters are positive in the positive region.

\subsection{Properties of the symbols}

In this section, we investigate some analytic structures of these integrals from their symbols, which can be nicely understood from DCI integrals.  

From canonical differential equations $dI_i=\epsilon\sum_j d\log A_{ij}\,I_j$, we can construct symbols of integrals in the basis by
\begin{equation}
    I^{(w)}_{a}=\sum_{i_w} I^{(w-1)}_{i_w}\otimes  A_{a i_{w}}=\sum_{i_1,\dots,i_{w}}I_{i_1}^{(0)} A_{i_2i_1}\otimes  A_{i_3 i_2}\otimes \cdots \otimes  A_{i_wi_{w-1}} \otimes  A_{a i_{w}},
\end{equation}
where $I_k^{(w)}$ is the series coefficient of $I_k(\epsilon)=\sum_{w\geq 0} I_k^{(w)} \epsilon^{w-4}$. One can easily check the solution (\theequation) by series expansion of the canonical differential equations.

The starting point is to understand branch points, or the first entries of the symbol, and corresponding physical discontinuities. For any planar integral, branch points must correspond to {\it planar variables} $s_{i,\dots,k}:=(p_{i}+p_{i+1}+\dots+p_k)^2=0$. In our cases, this means that first entries of $I_a$ should be
\[
\{s,t,m_1^2,m_2^2,m_3^2,m_4^2\},
\]
which can be directly verified from $\sum_a A_{ai} I^{(0)}_i$. The corresponding physical discontinuity for the channel $x=0$ is 
\begin{equation}
\operatorname{Disc}_{x=0}(I^{(w)}_{a})=\sum_{i_1,\dots,i_{w}}I_{i_1}^{(0)}\operatorname{Disc}_{x=0}(\log A_{i_2i_1}) A_{i_3 i_2}\otimes \cdots \otimes A_{a i_{w}}.
\end{equation}

Then we consider double discontinuities. The corresponding physical constrains are Steinmann relations, which say that the double discontinuities taken in overlapping channels of any planar integrals should vanish. In our cases, the only overlapping channels are $s=0$ and $t=0$. We consider the combination
\begin{equation}
X_{ac}=\sum_b A_{bc}\otimes A_{ab},
\end{equation}
such that $I_i^{(w)}$ can be written as $\sum_{p,q} X_{p,q}\otimes Y_{i,p,q}$ where $Y_{i,p,q}$ are symbols of  weight-$(w-2)$, then
\begin{equation}
    \operatorname{Disc}_{s=0}\operatorname{Disc}_{t=0}(I_i^{(w)})=\sum_{p,q} \operatorname{Disc}_{s=0}\operatorname{Disc}_{t=0}(X_{p,q}) Y_{i,p,q}.
\end{equation}
The double discontinuity of $X_{ac}$ is just 
\begin{equation}
\operatorname{Disc}_{s=0}\operatorname{Disc}_{t=0}X_{ac}=\sum_b \operatorname{Disc}_{s=0}(\log A_{ab})\operatorname{Disc}_{t=0}(\log A_{bc}),
\end{equation}
the product of two discontinuity matrix. We check that both $\operatorname{Disc}_{s=0}\operatorname{Disc}_{t=0}X_{ac}$ and \\ $\operatorname{Disc}_{t=0}\operatorname{Disc}_{s=0}X_{ac}$ vanish. 

Steinmann relations constrain first-two entries of the symbol of $I^{(w)}_a$, we could further consider the same constrains on any adjacent entries of the symbol, which are known as the extended Steinmann relations. For the symbol of $I^{(w)}_a$, these constrains are
\begin{equation}
\sum_{i_1,\dots,\widehat{i_k}\dots,i_{w}}\hspace{-2ex}I_{i_1}^{(0)} A_{i_2i_1}\otimes \cdots \otimes \widehat{A_{i_{k}i_{k-1}}}\otimes \widehat{A_{i_{k+1}i_{k}}}\otimes \cdots\otimes  A_{a i_{w}} \operatorname{Disc}_{s=0}\operatorname{Disc}_{t=0}X_{i_{k+1},i_{k-1}}=0
\end{equation}
and 
\begin{equation}
\sum_{i_1,\dots,\widehat{i_k}\dots,i_{w}}\hspace{-2ex}I_{i_1}^{(0)} A_{i_2i_1}\otimes \cdots \otimes \widehat{A_{i_{k}i_{k-1}}}\otimes \widehat{A_{i_{k+1}i_{k}}}\otimes \cdots\otimes  A_{a i_{w}} \operatorname{Disc}_{t=0}\operatorname{Disc}_{s=0}X_{i_{k+1},i_{k-1}}=0,
\end{equation}
for any $k$ and the only overlapping channels $s=0$ and $t=0$. Therefore, Steinmann relations for $X$ also guarantee the extended Steinmann relations of $I^{(w)}_a$.

In \cite{He:2021mme}, it's conjectured from first entry condition and Steinmann relations that the first-two entries of DCI integrals can only be linear combination of one-loop box functions and some trivial $\log \log$ functions. For any order of $\epsilon$ of $I^{(w)}_a$, we also find that the first-two entries of $I_i$ can only be linear combinations of 
\begin{equation}
\{\log(s)^2,\log(t)^2,\log(m_i^2)\log(s),\log(m_i^2)\log(t),\log(m_i^2)\log(m_j^2),F(z_k,\bar z_k)\}_{i,j=1,\dots,4,k=1,\dots 5},
\end{equation}
where $F(z_k,\bar z_k)$ are (normalized) four-mass box function whose symbol is 
\begin{equation}
\mathcal S[F(z,\bar z)]:=((1-z) (1-\bar z))\otimes \frac{z}{\bar z}-(z \bar z)\otimes \frac{1-z}{1-\bar z}.
\end{equation}
Therefore, the first two entries are also box functions (after taking one dual point to infinity) except $\log \log$ functions.

\section{Summary and Outlook}\label{sec:summary}
In this paper, we provide insights for multi-loop Feynman integrals and their symbology from the viewpoint of dual conformal symmetry. The symbols of UT integrals, in principle, can be derived from the differential equations of the UT integral basis. As shown in this paper, the symbols of some integrals can be very complicated. Also, the symbols are with many "mysterious" structures. We also studied the symbology form DCI point of view. By sending certain dual point to infinity in DCI integrals, non-DCI finite integrals are acquired. Meanwhile, the corresponding symbol structures can also be analyzed. This gives us an insight about some properties of the symbols, before we comply the canonical differential equation computation, in future studies for new Feynman integrals.

To explain the above specifically, in this paper, we studied a cutting-edge example, the two-loop double box diagram with $4$ different external masses. We determine its UT integral basis mainly using loop-by-loop analysis of leading singularity in Baikov representation. The analytic derivation of corresponding differential equations is nontrivial and we use the package {\sc FiniteFlow} based on the finite-field method for this computation.

From the differential equations, we derived the symbols of the UT integrals from order $\epsilon^{-4}$ to $\epsilon^0$. The symbols consist of $68$ letters. Among the letters, there are 18 even letters being rational functions, $16$ odd letters containing one square root each, and $34$ odd letters containing $2$ square roots each. The symbols for the $74$ integrals are derived using the differential equations. We found their properties that: 1) The letters that can appear at the first and the second entries of the symbols are limited. 2) A large number of the possible letter pairs that can appear at two adjacent entries are forbidden, read from the letter structure of differential equations. 

What we find particularly remarkable is that all symbol letters of these integrals have clear origin from symbol letters of related dual conformal integrals, which have been studied extensively and exhibit rich mathematical structures. Both even and odd letters are nicely obtained by taking a dual point to infinity, of the letters for one- and two-loop DCI integrals with generic pentagon kinematics. In particular, not only do the square roots correspond to one- and two-loop leading singularities, but we determine all complicated mixed odd letters from associated twistor geometries, which are also ``last entries" for corresponding finite, two-loop integrals. We find it satisfying that one can obtain all letters by considering mathematical structure of DCI integrals without actually computing them, and we leave a more systematic study of this phenomenon to a future work. Last but not least, we find that important properties of DCI integrals nicely carry over to UT integrals we consider in this paper: the first two entries come from one-loop DCI box functions, and extended Steinmann relations further constrain any two adjacent entries to any order in $\epsilon$. It would be highly desirable to understand further how structures of their symbols follow from properties of these DCI integrals. 

\section*{Acknowledgement} 
We acknowledge Xiaodi Li for his participation and effort in the early stage of this work.  We also thank Christoph Dlapa, Johannes Henn, Gang Yang and Lilin Yang for enlightening discussions. SH is supported by National Natural Science Foundation of China under Grant No. 11935013, 11947301, 12047502, 12047503 and the Key Research Program of the Chinese Academy of Sciences, Grant NO. XDPB15. YZ is supported from the NSF of China through Grant No. 11947301, 12047502, 12075234, and the Key Research Program of the Chinese Academy of Sciences, Grant NO. XDPB15. YZ also acknowledges the Institute of Theoretical Physics, Chinese Academy
of Sciences, for the hospitality through the “Peng Huanwu visiting professor program”.

\appendix

\section{Diagram of the two-loop master integrals}

\label{appendix:2l4pMIs}
\begin{figure}[H] 
    \centering    
    
    \begin{minipage}{0.18\linewidth}
        \centering
        \includegraphics[width=2.3cm,height=1.6cm]{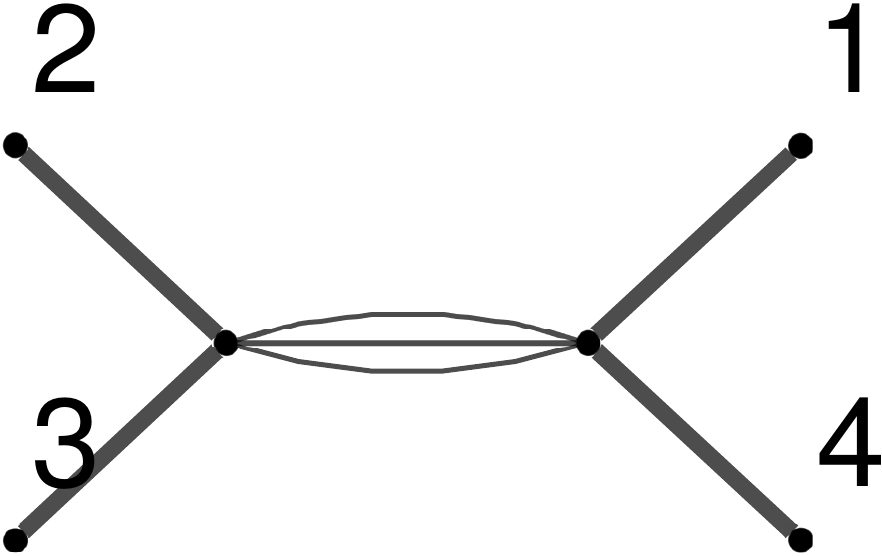}  
        \caption*{$\{2,5,7\}$ \\1 MI}  
        \label{5}  
    \end{minipage}
    \begin{minipage}{0.18\linewidth}
        \centering
        \includegraphics[width=2.3cm,height=1.6cm]{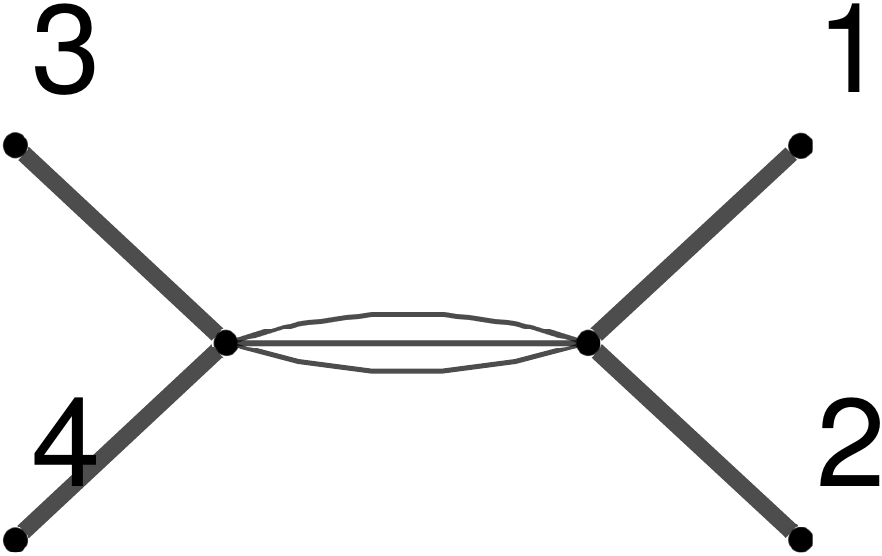}  
        \caption*{$\{1,4,7\}$ \\1 MI}  
        \label{5}  
    \end{minipage}
    
\caption{Diagrams for the two-loop master integrals with four external
massive legs}
\label{fig:FeynmanDiagrams}

\end{figure}

In this appendix, we list the diagrams for the two-loop Feynman integrals with four external massive legs of every sector with master integrals. They are shown in Fig.\ref{fig:FeynmanDiagrams}.

\begin{figure}[H]
    \centering  
    \begin{minipage}{0.18\linewidth}
        \centering
        \includegraphics[width=2.3cm,height=1.85cm]{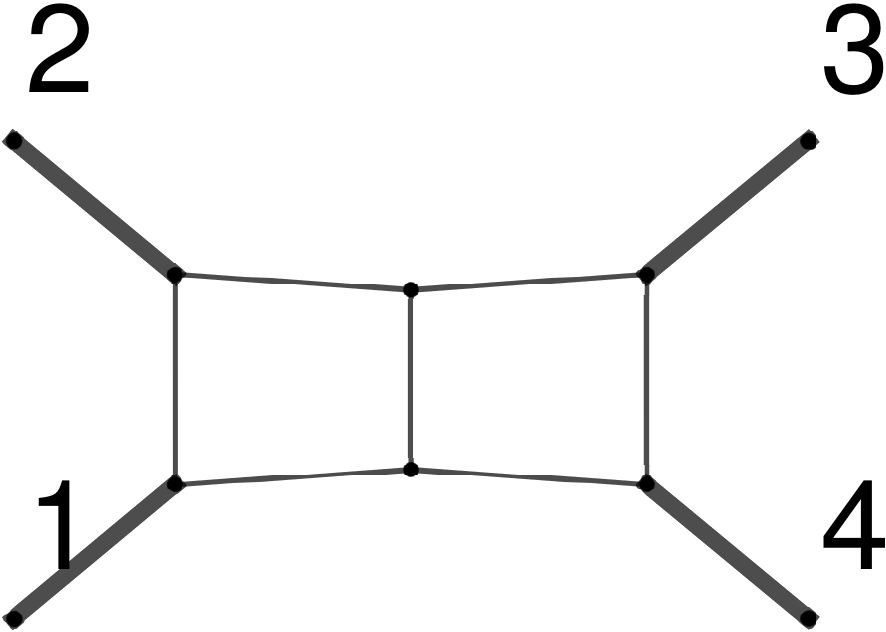}
        \caption*{$\{1,2,3,4,5,6,7\}$ \\4 MIs}  
        \label{1}  
    \end{minipage}
    \begin{minipage}{0.18\linewidth}
        \centering
        \includegraphics[width=2.3cm,height=1.85cm]{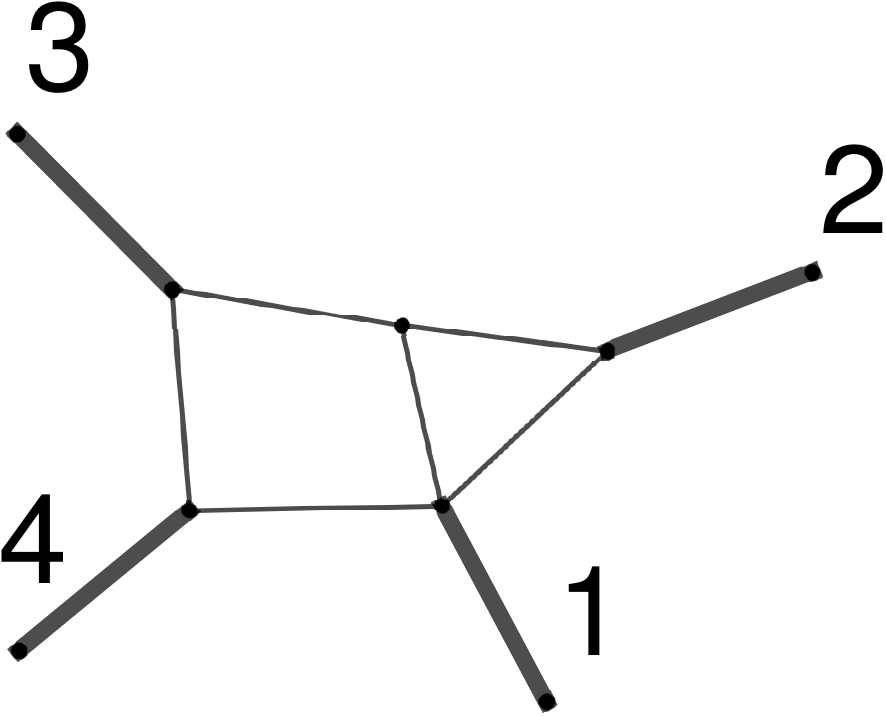} 
        \caption*{$\{2,3,4,5,6,7\}$ \\1 MI}  
        \label{2}  
    \end{minipage}
    \begin{minipage}{0.18\linewidth}
        \centering
        \includegraphics[width=2.3cm,height=1.85cm]{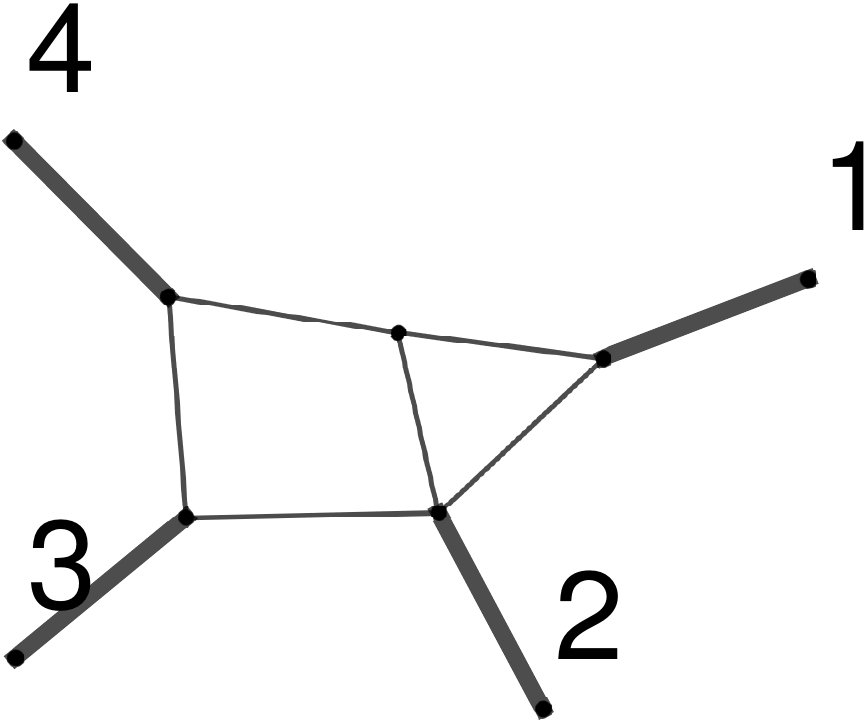} 
        \caption*{$\{1,2,4,5,6,7\}$ \\1 MI}  
        \label{3}  
    \end{minipage}
    \begin{minipage}{0.18\linewidth}
        \centering
        \includegraphics[width=2.3cm,height=1.85cm]{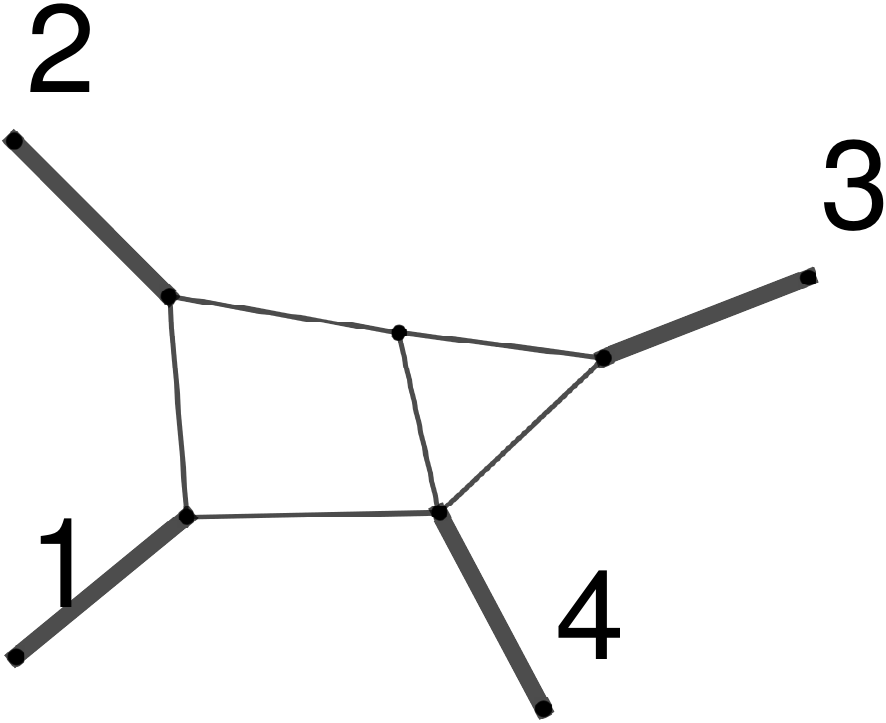} 
        \caption*{$\{1,2,3,4,5,7\}$ \\1 MI}  
        \label{5}  
    \end{minipage}
    \begin{minipage}{0.18\linewidth}
        \centering
        \includegraphics[width=2.3cm,height=1.85cm]{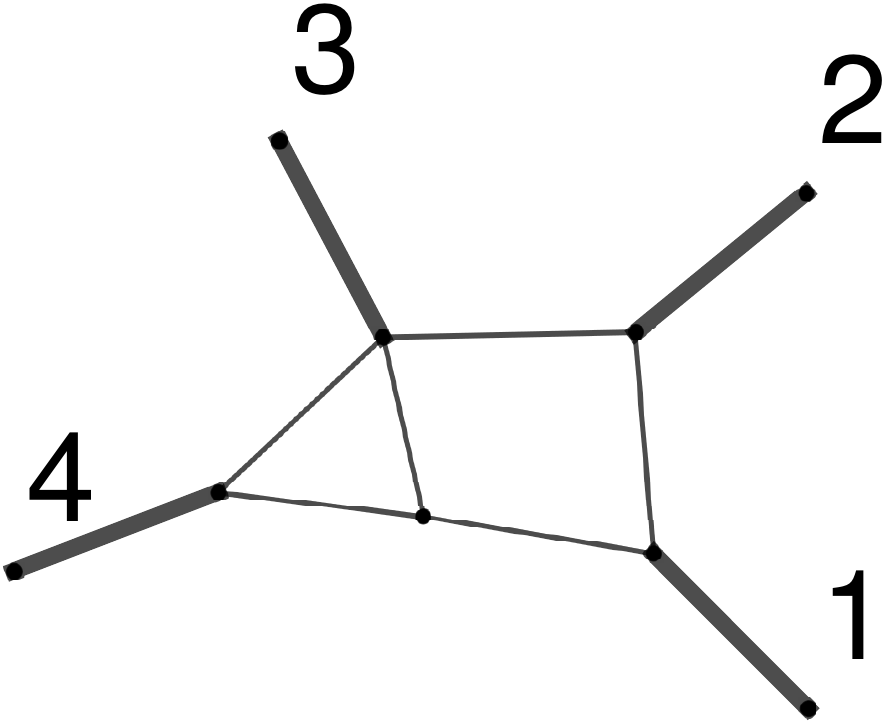} 
        \caption*{$\{1,2,3,5,6,7\}$ \\1 MI}  
        \label{4}  
    \end{minipage}
\end{figure}

\begin{figure}[H] 
    \centering   

    \begin{minipage}{0.18\linewidth}
        \centering
        \includegraphics[width=1.85cm,height=1.85cm]{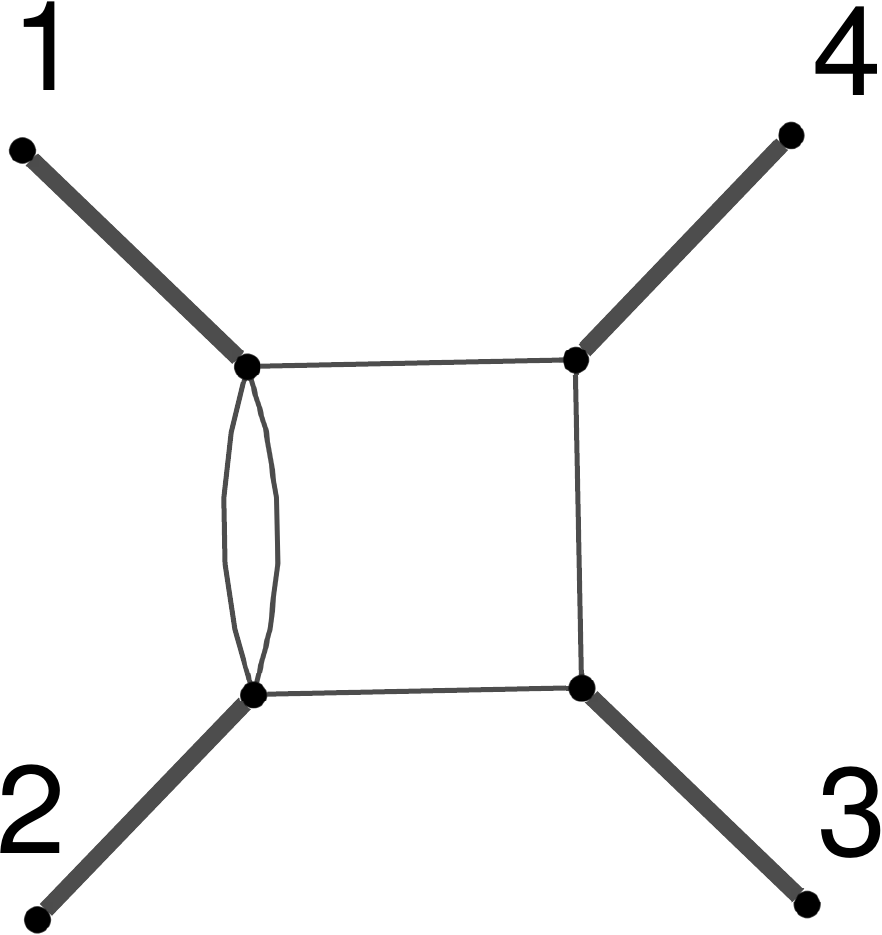}  
        \caption*{$\{2,4,5,6,7\}$ \\2 MIs}  
        \label{4}  
    \end{minipage}
    \begin{minipage}{0.18\linewidth}
        \centering
        \includegraphics[width=1.85cm,height=1.85cm]{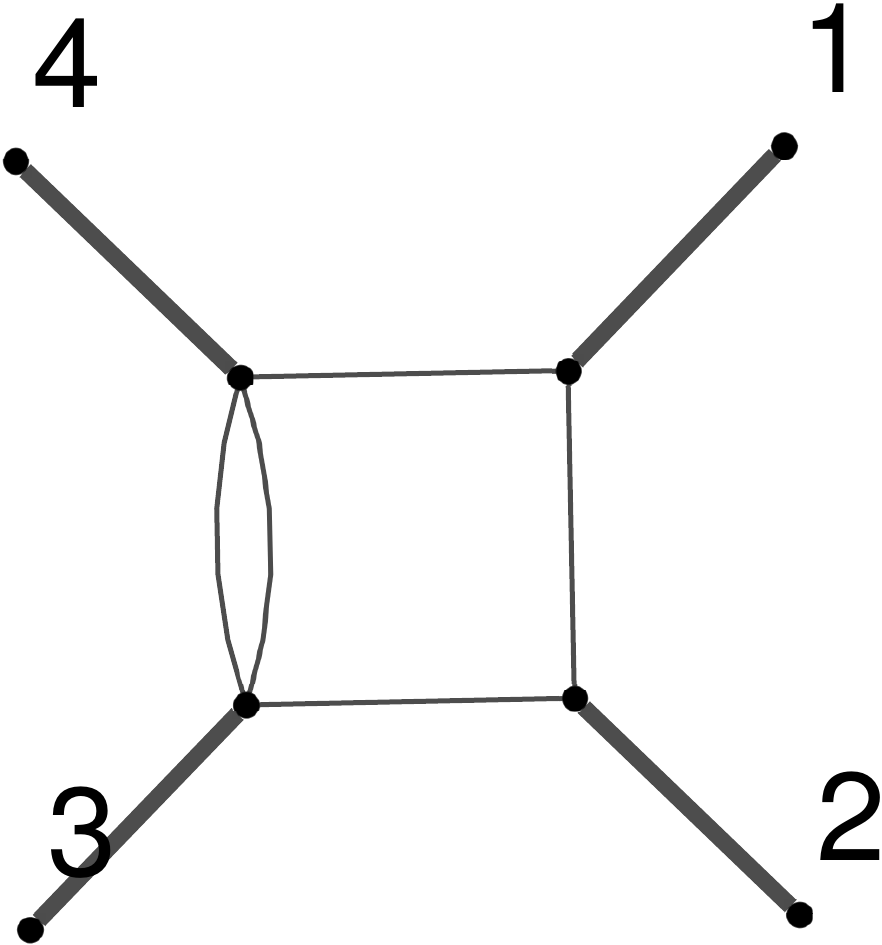}  
        \caption*{$\{1,2,3,5,7\}$ \\2 MIs}  
        \label{4}  
    \end{minipage}
    \begin{minipage}{0.18\linewidth}
        \centering
        \includegraphics[width=2.3cm,height=1.85cm]{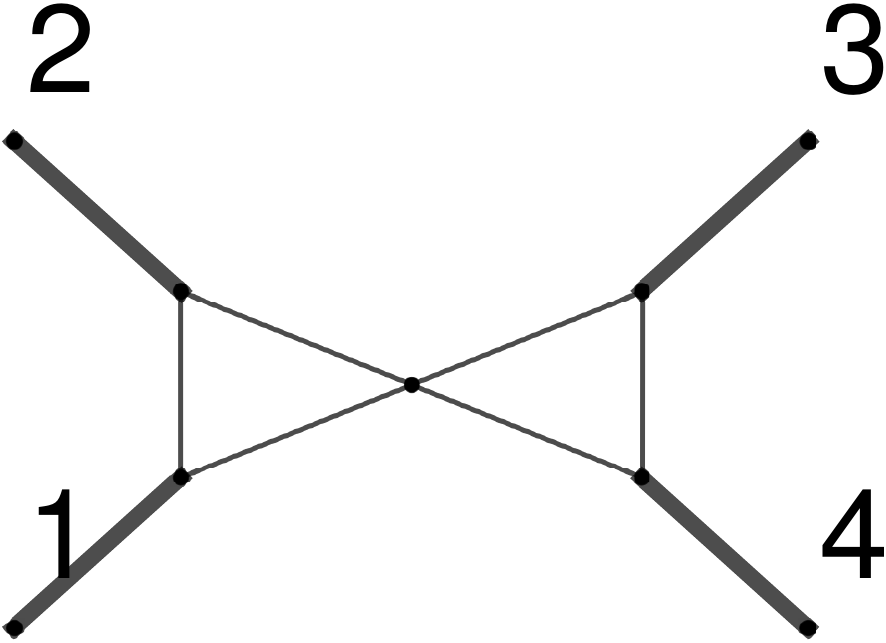} 
        \caption*{$\{1,2,3,4,5,6\}$ \\1 MI}  
        \label{6}  
    \end{minipage}
    
\end{figure}

\begin{figure}[H] 
    \centering
    
    \begin{minipage}{0.18\linewidth}
        \centering
        \includegraphics[width=2.3cm,height=1.85cm]{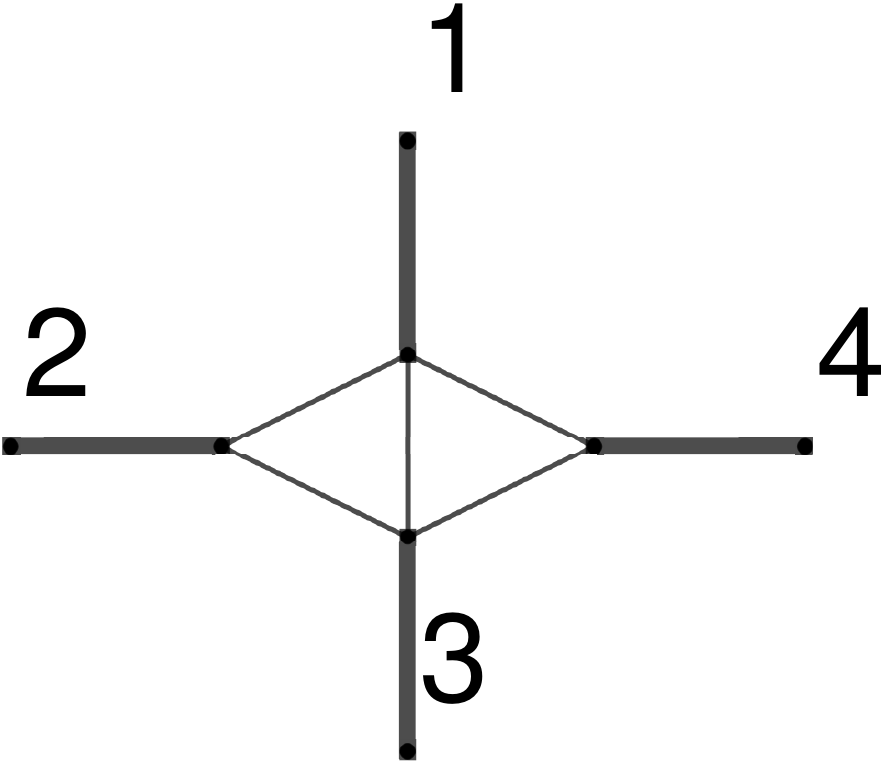}  
        \caption*{$\{2,3,5,6,7\}$ \\7 MIs}  
        \label{5}  
    \end{minipage}
    \begin{minipage}{0.18\linewidth}
        \centering
        \includegraphics[width=2.3cm,height=1.85cm]{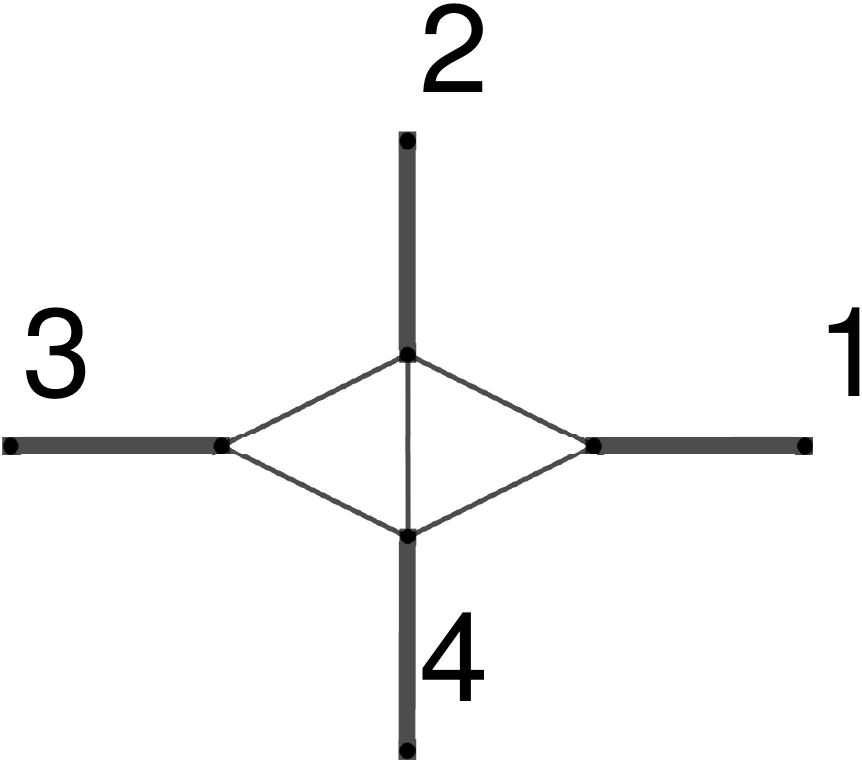}  
        \caption*{$\{1,2,4,5,7\}$ \\7 MIs}  
        \label{5}  
    \end{minipage}
     \begin{minipage}{0.18\linewidth}
        \centering
        \includegraphics[width=2.3cm,height=1.85cm]{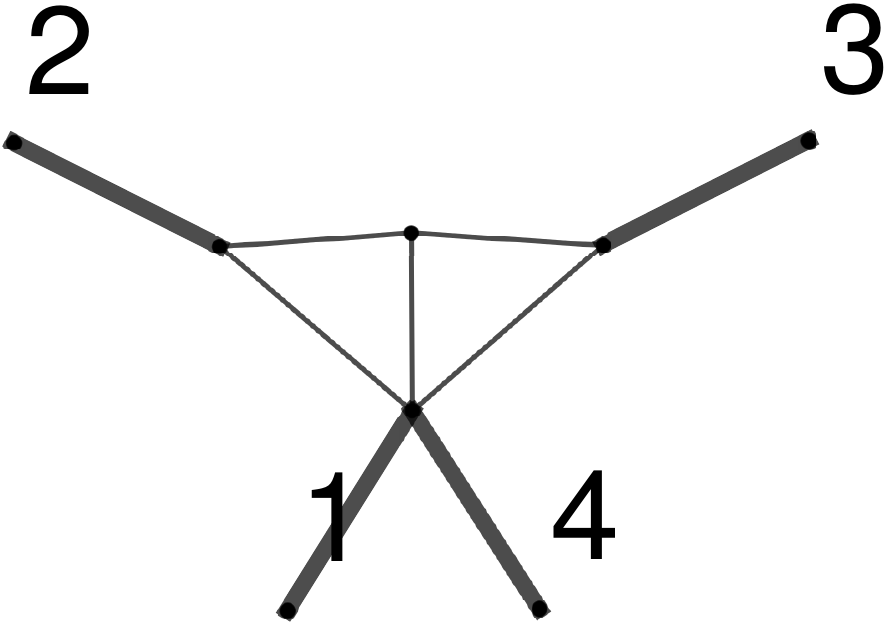}  
        \caption*{$\{2,3,4,5,7\}$ \\1 MI}  
        \label{4}  
    \end{minipage}
    \begin{minipage}{0.18\linewidth}
        \centering
        \includegraphics[width=2.3cm,height=1.85cm]{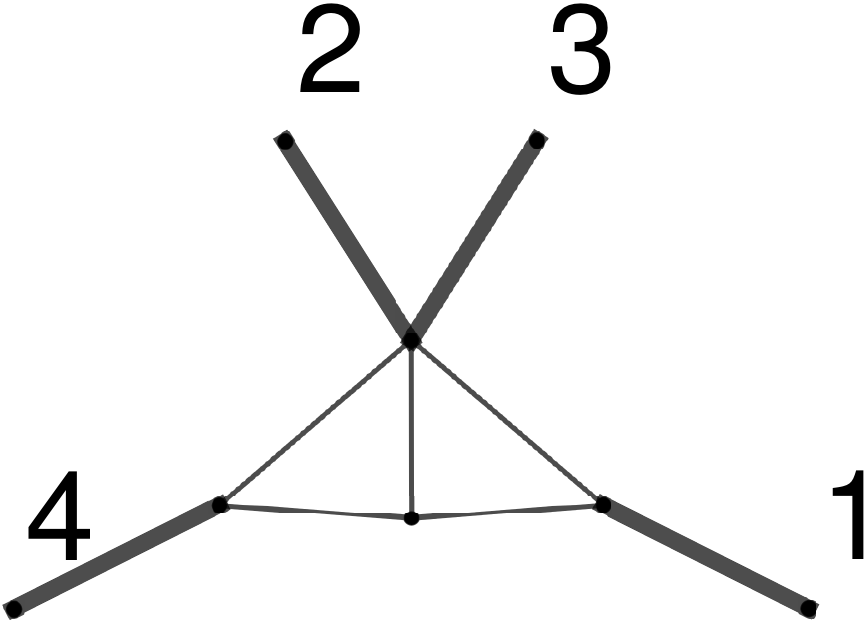}  
        \caption*{$\{1,2,5,6,7\}$ \\1 MI}  
        \label{6}  
    \end{minipage}
\end{figure}

\begin{figure}[H] 
    \centering    
    \begin{minipage}{0.18\linewidth}
        \centering
        \includegraphics[width=2.3cm,height=1.85cm]{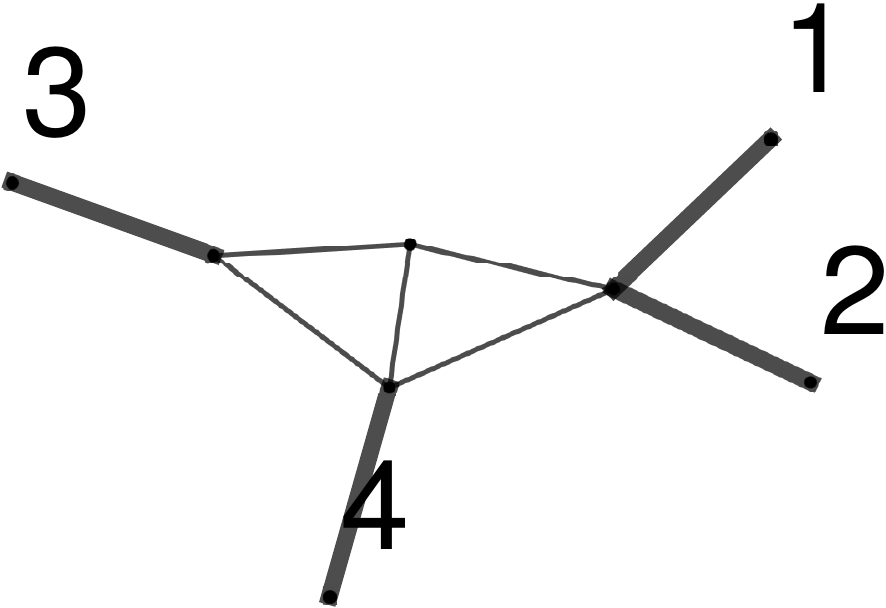}  
        \caption*{$\{1,3,4,5,7\}$ \\1 MI}  
        \label{4}  
    \end{minipage}
     \begin{minipage}{0.18\linewidth}
        \centering
        \includegraphics[width=2.3cm,height=1.85cm]{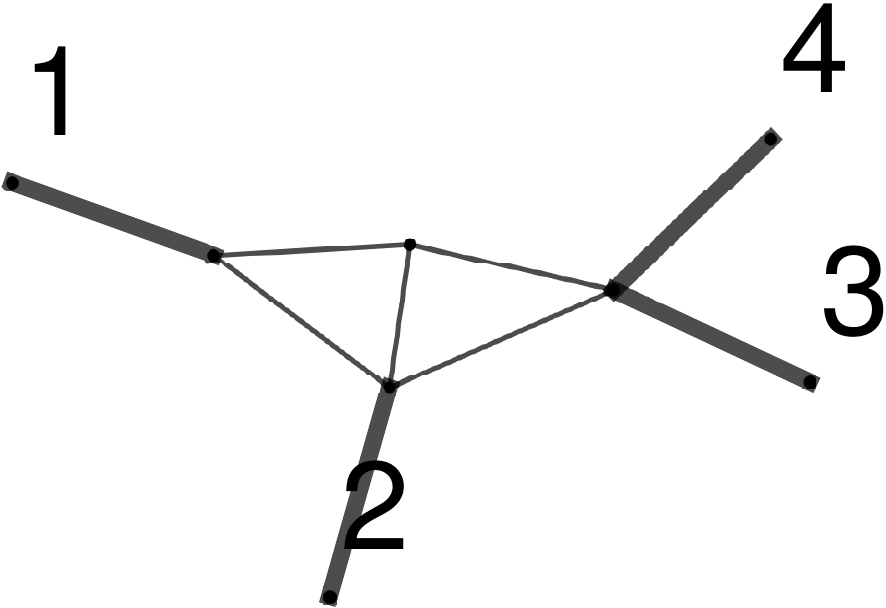}  
        \caption*{$\{1,2,4,6,7\}$ \\1 MI}  
        \label{4}  
    \end{minipage}
    \begin{minipage}{0.18\linewidth}
        \centering
        \includegraphics[width=2.3cm,height=1.85cm]{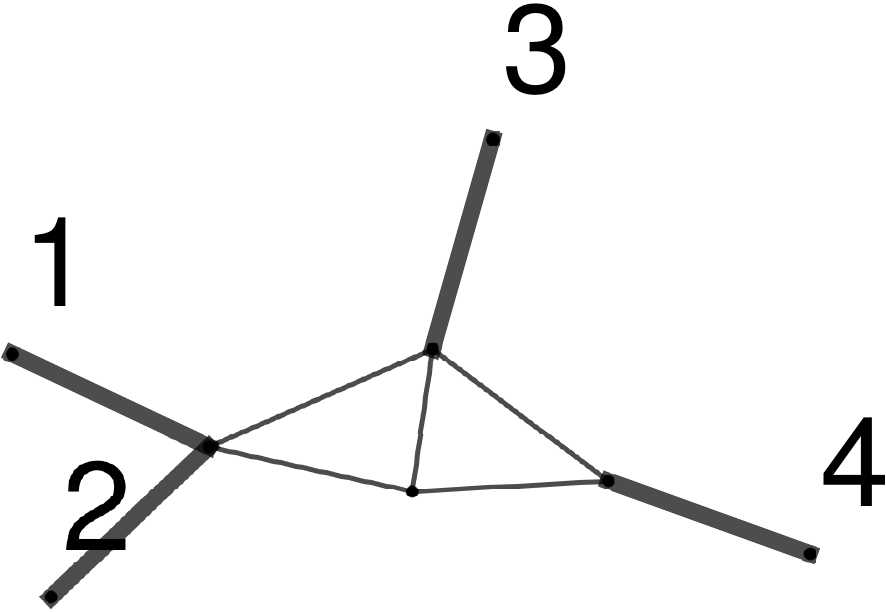}  
        \caption*{$\{1,3,5,6,7\}$ \\1 MI}  
        \label{6}  
    \end{minipage}
    \begin{minipage}{0.18\linewidth}
        \centering
        \includegraphics[width=2.3cm,height=1.85cm]{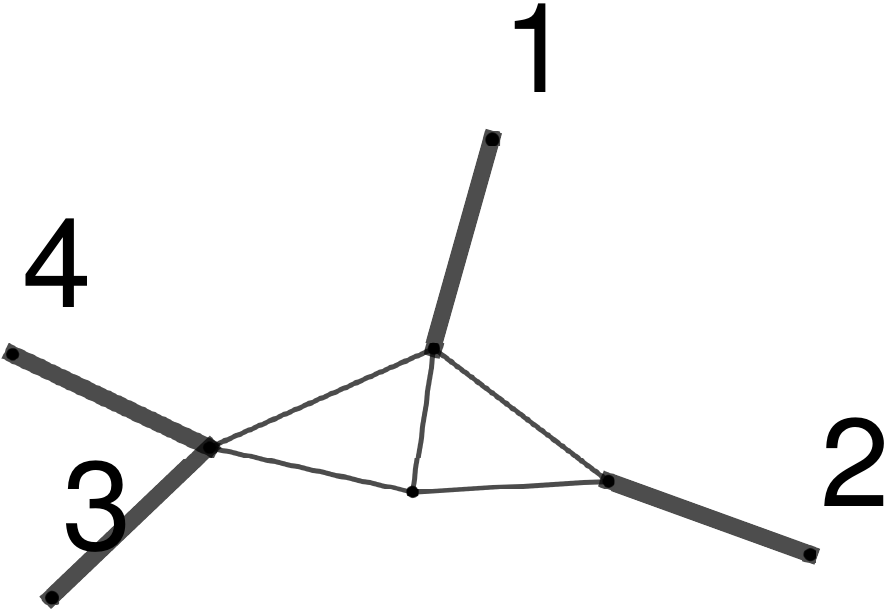}  
        \caption*{$\{2,3,4,6,7\}$ \\1 MI}  
        \label{6}  
    \end{minipage}
\end{figure}

\begin{figure}[H] 
    \centering    
    
    \begin{minipage}{0.18\linewidth}
        \centering
        \includegraphics[width=2.3cm,height=1.85cm]{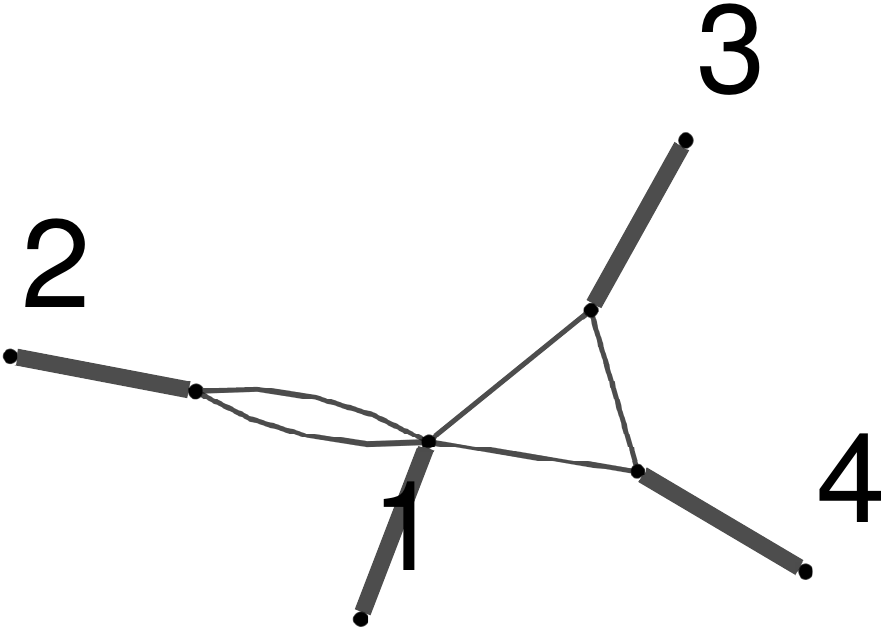}  
        \caption*{$\{2,3,4,5,6\}$ \\1 MI}  
        \label{5}  
    \end{minipage}
    \begin{minipage}{0.18\linewidth}
        \centering
        \includegraphics[width=2.3cm,height=1.85cm]{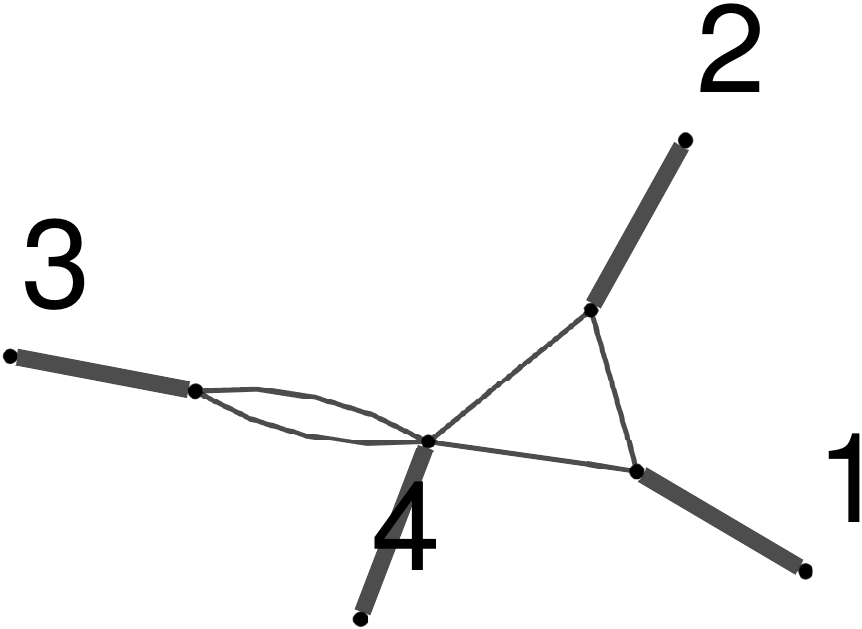}  
        \caption*{$\{1,2,3,4,5\}$ \\1 MI}  
        \label{4}  
    \end{minipage}
    \begin{minipage}{0.18\linewidth}
        \centering
        \includegraphics[width=2.3cm,height=1.85cm]{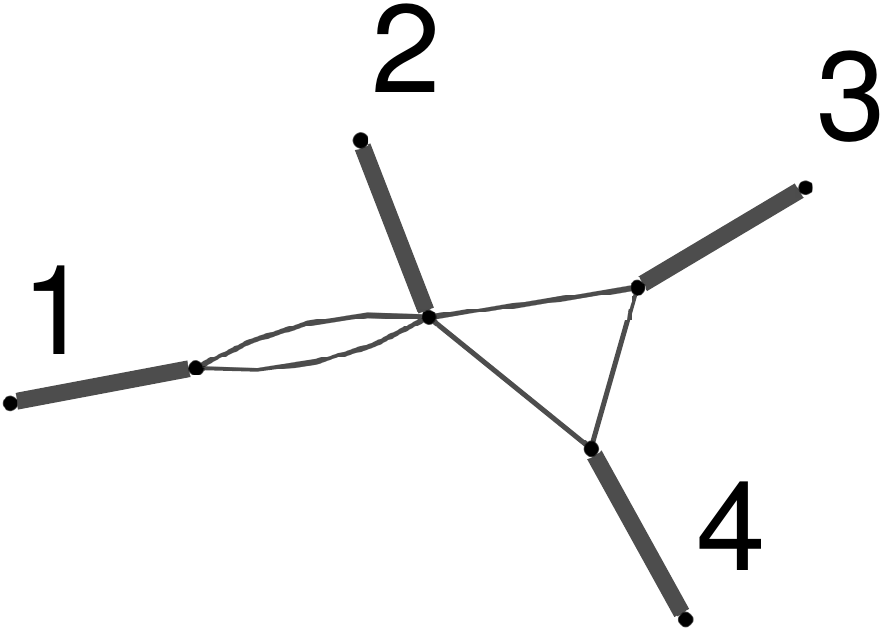}  
        \caption*{$\{1,2,4,5,6\}$ \\1 MI}  
        \label{6}  
    \end{minipage}
    \begin{minipage}{0.18\linewidth}
        \centering
        \includegraphics[width=2.3cm,height=1.85cm]{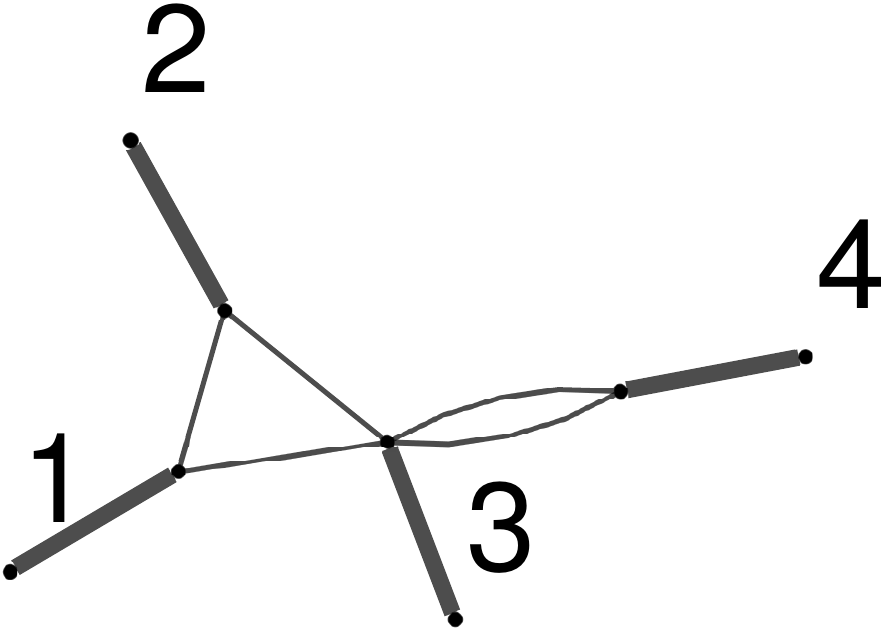}  
        \caption*{$\{1,2,3,5,6\}$ \\1 MI}  
        \label{5}  
    \end{minipage}
\end{figure}

\begin{figure}[H] 
    \centering
    \begin{minipage}{0.18\linewidth}
        \centering
        \includegraphics[width=2.3cm,height=1.85cm]{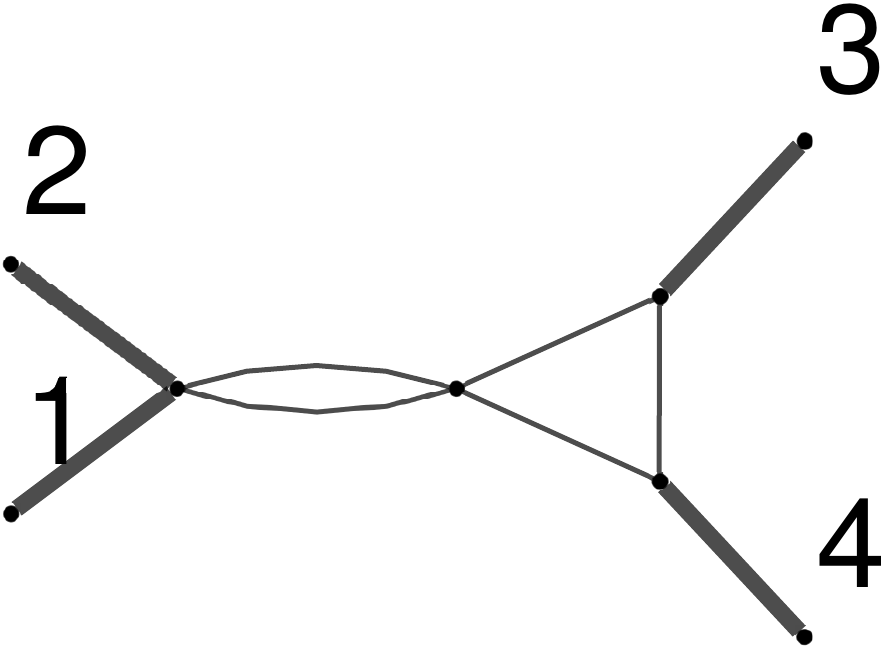}  
        \caption*{$\{1,3,4,5,6\}$ \\1 MI}  
        \label{5}  
    \end{minipage}
    \begin{minipage}{0.18\linewidth}
        \centering
        \includegraphics[width=2.3cm,height=1.85cm]{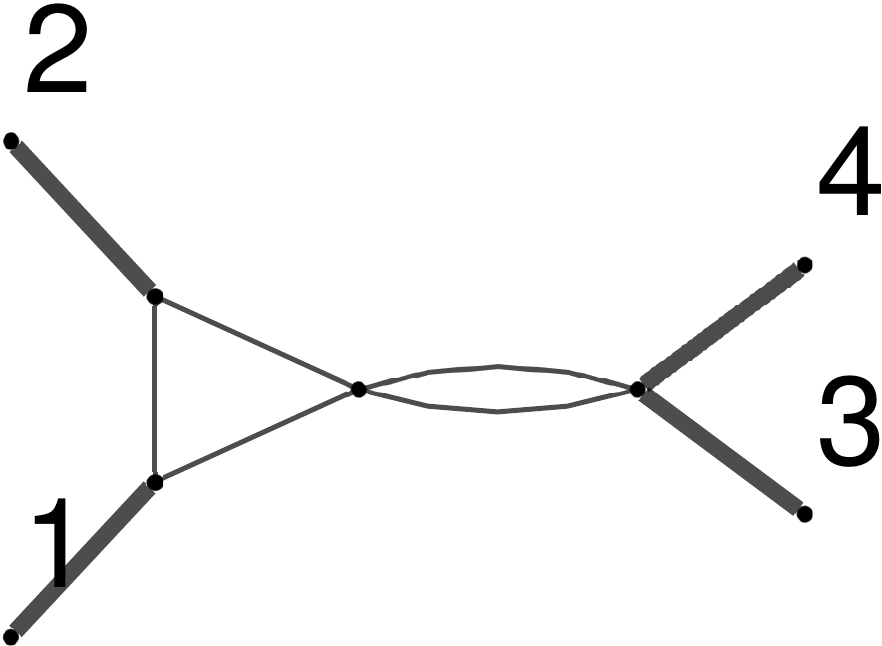}  
        \caption*{$\{1,2,3,4,6\}$ \\1 MI}  
        \label{6}  
    \end{minipage}
    
\end{figure}

\begin{figure}[H] 
    \centering   
    
    \begin{minipage}{0.18\linewidth}
        \centering
        \includegraphics[width=2.1cm,height=2.1cm]{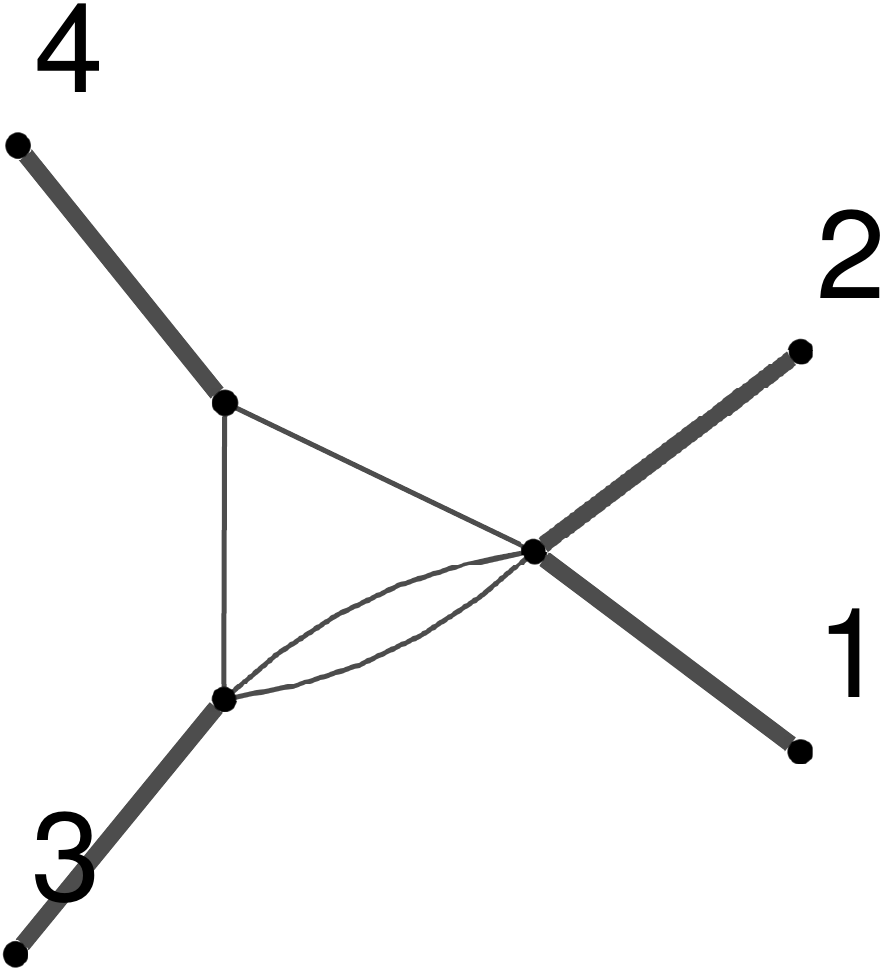}  
        \caption*{$\{3,5,6,7\}$ \\2 MIs}  
        \label{5}  
    \end{minipage}
    \begin{minipage}{0.18\linewidth}
        \centering
        \includegraphics[width=2.1cm,height=2.1cm]{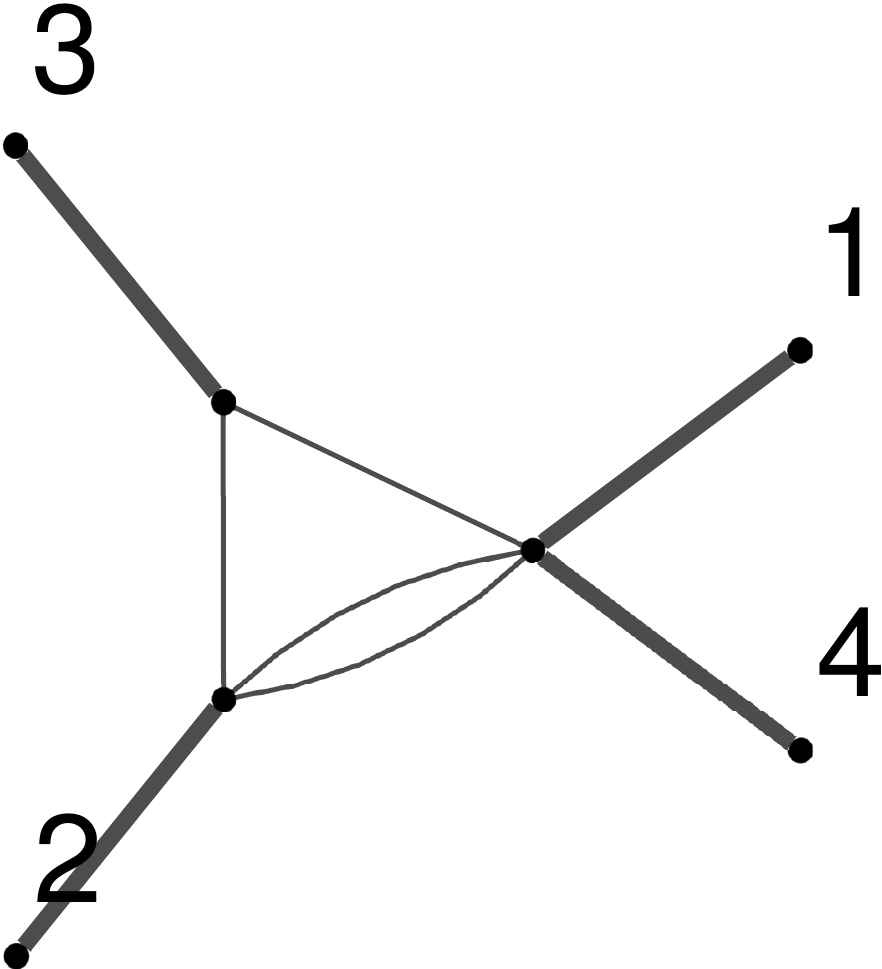}  
        \caption*{$\{2,4,5,7\}$ \\2 MIs}  
        \label{5}  
    \end{minipage}
    \begin{minipage}{0.18\linewidth}
        \centering
        \includegraphics[width=2.1cm,height=2.1cm]{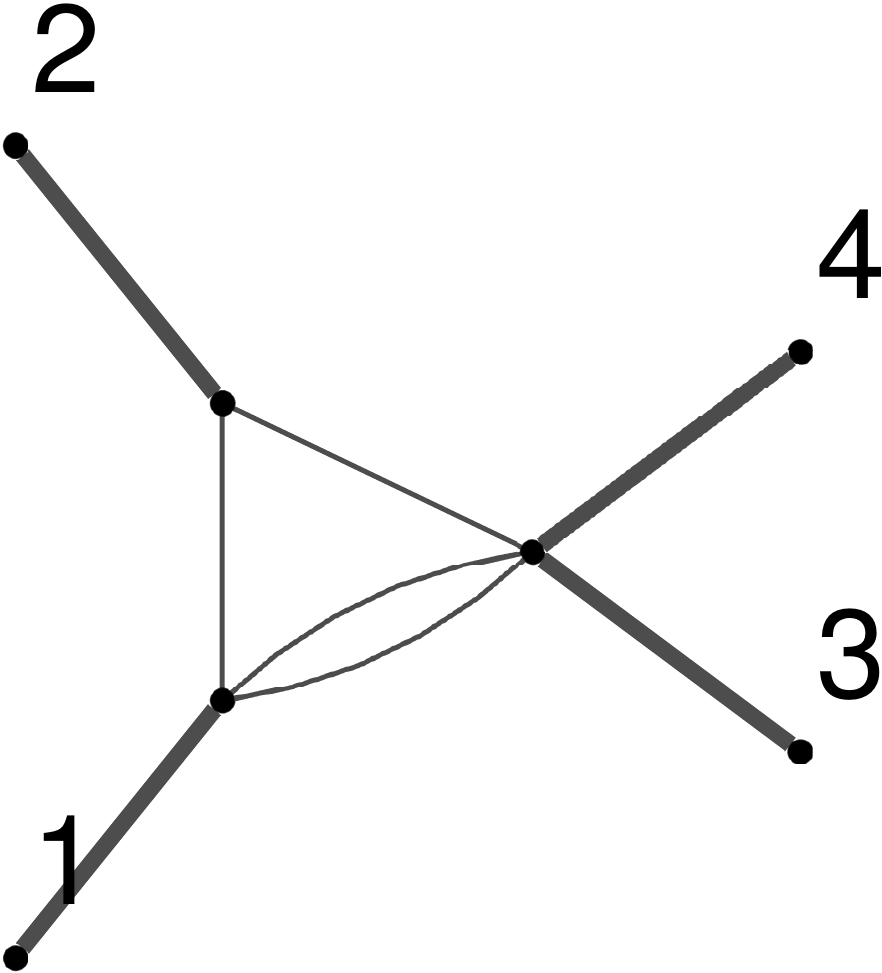}  
        \caption*{$\{1,3,5,6,7\}$ \\2 MIs}  
        \label{6}  
    \end{minipage}
    \begin{minipage}{0.18\linewidth}
        \centering
        \includegraphics[width=2.1cm,height=2.1cm]{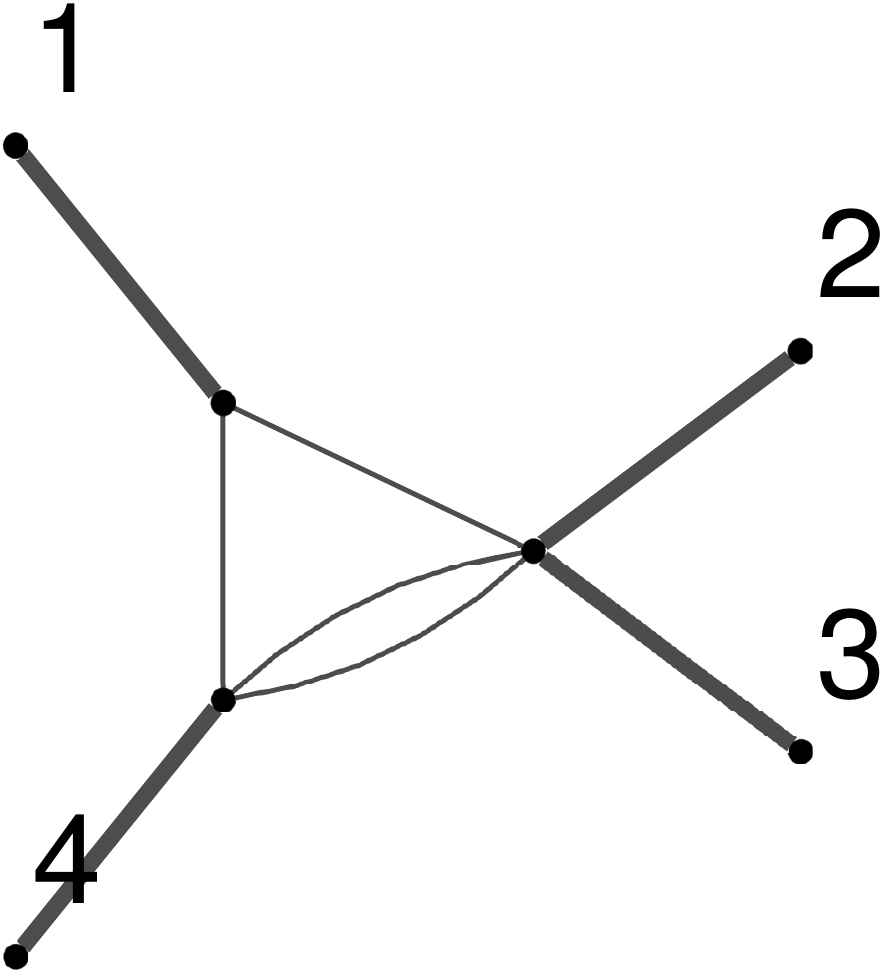}  
        \caption*{$\{1,2,5,7\}$ \\2 MIs}  
        \label{4}  
    \end{minipage}
    \begin{minipage}{0.18\linewidth}
        \centering
        \includegraphics[width=2.1cm,height=2.1cm]{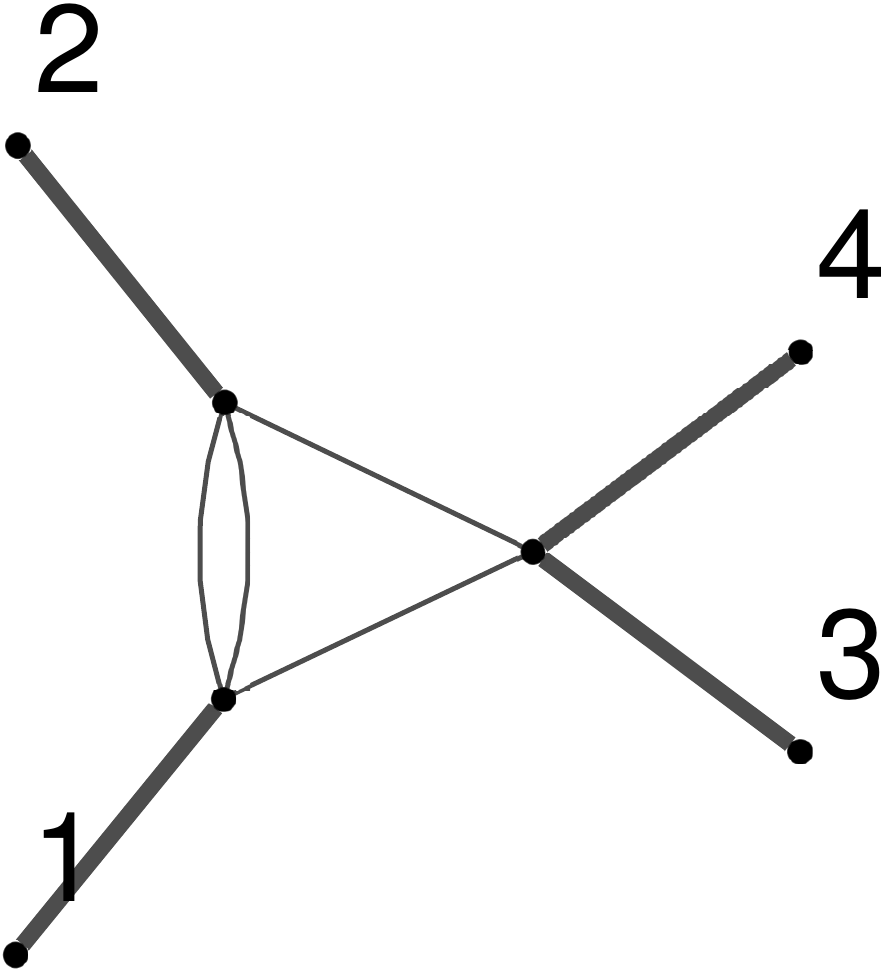}  
        \caption*{$\{2,4,6,7\}$ \\2 MIs}  
        \label{4}  
    \end{minipage}

\end{figure}

\begin{figure}[H] 
    \centering    
    
    \begin{minipage}{0.18\linewidth}
        \centering
        \includegraphics[width=2.1cm,height=2.1cm]{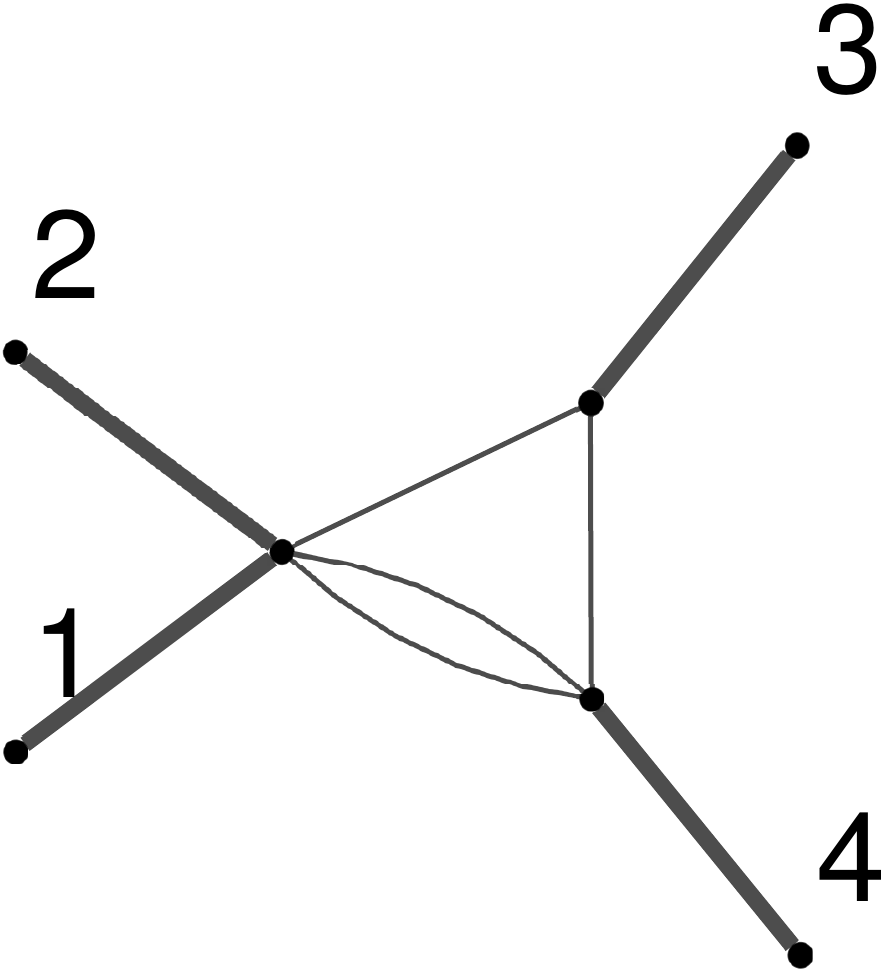}  
        \caption*{$\{1,4,5,7\}$ \\2 MIs}  
        \label{5}  
    \end{minipage}
    \begin{minipage}{0.18\linewidth}
        \centering
        \includegraphics[width=2.1cm,height=2.1cm]{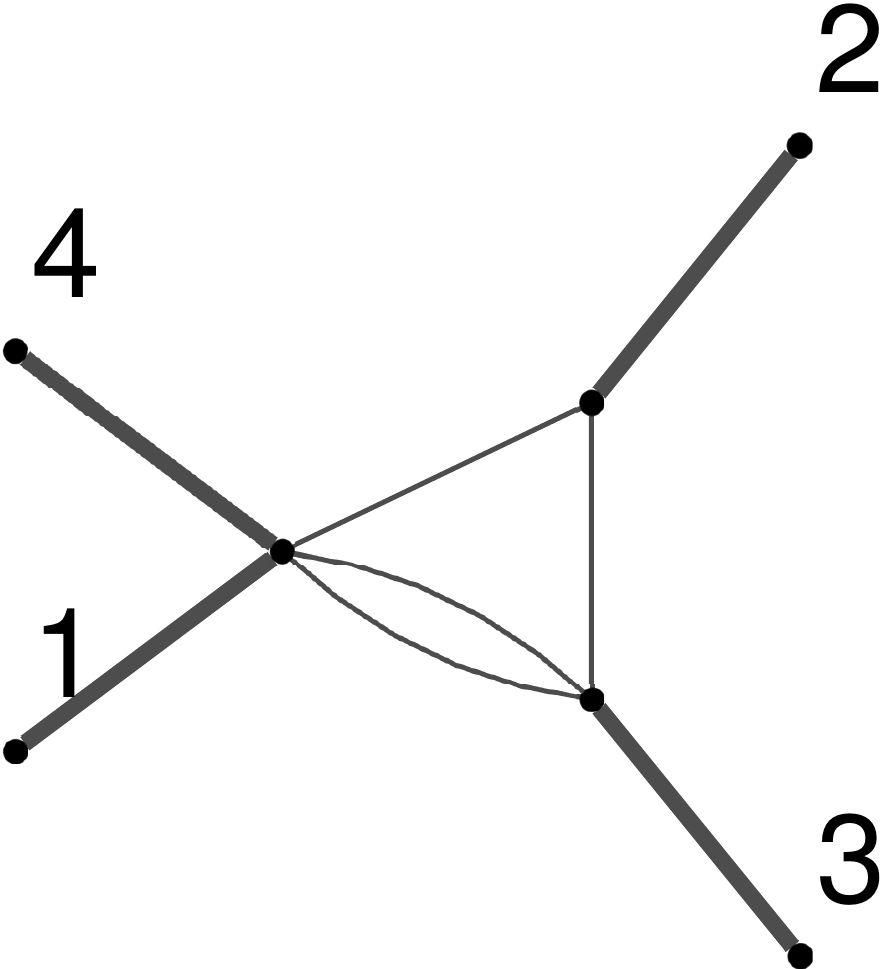}  
        \caption*{$\{2,3,5,7\}$ \\2 MIs}  
        \label{4}  
    \end{minipage}
    \begin{minipage}{0.18\linewidth}
        \centering
        \includegraphics[width=2.1cm,height=2.1cm]{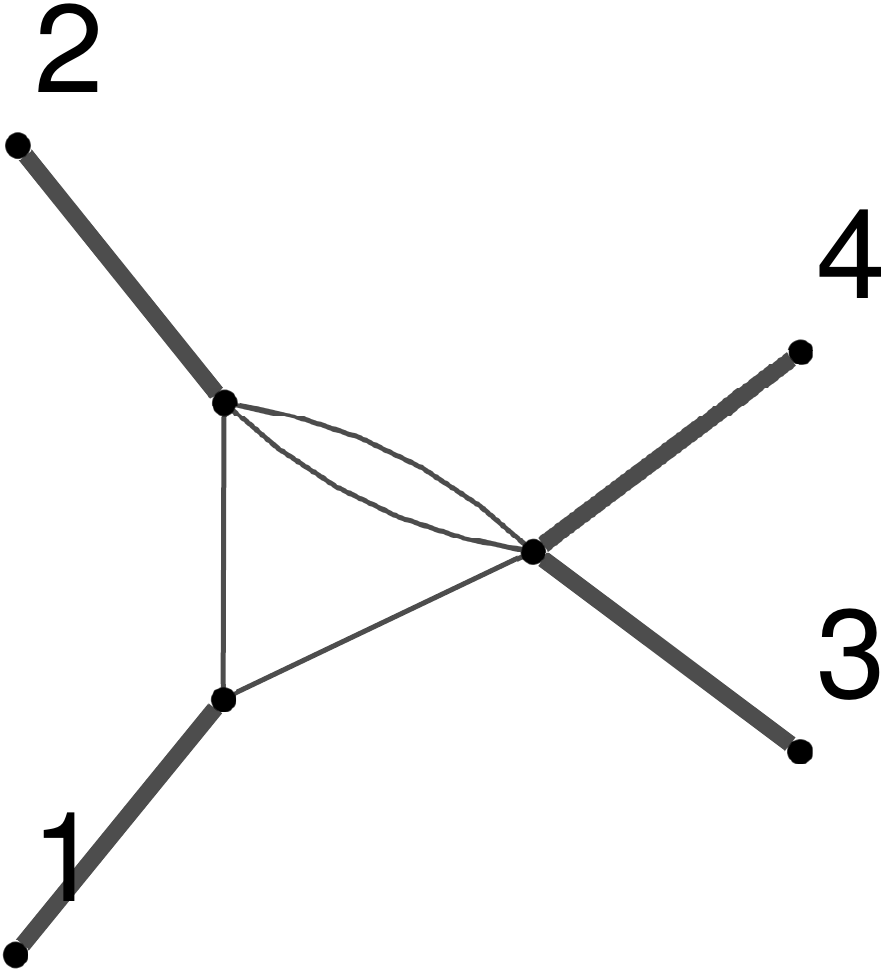}  
        \caption*{$\{1,2,4,7\}$ \\2 MIs}  
        \label{6}  
    \end{minipage}
    \begin{minipage}{0.18\linewidth}
        \centering
        \includegraphics[width=2.1cm,height=2.1cm]{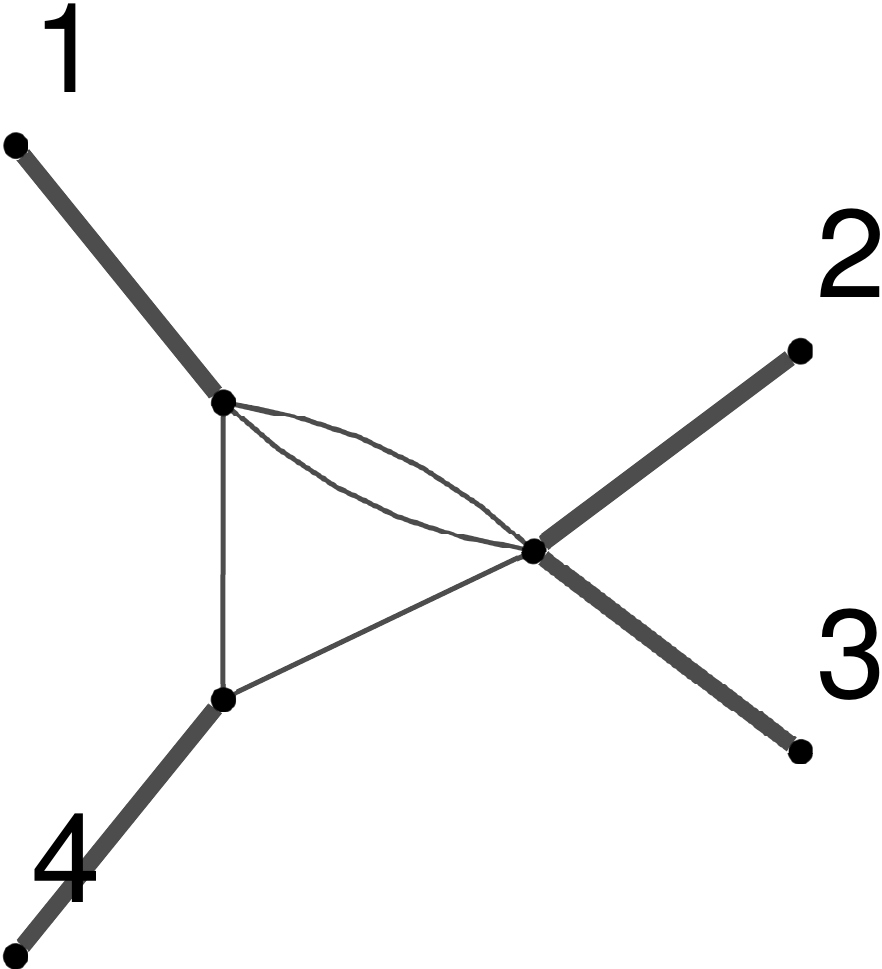}  
        \caption*{$\{2,5,6,7\}$ \\2 MIs}  
        \label{6}  
    \end{minipage}
    \begin{minipage}{0.18\linewidth}
        \centering
        \includegraphics[width=2.1cm,height=2.1cm]{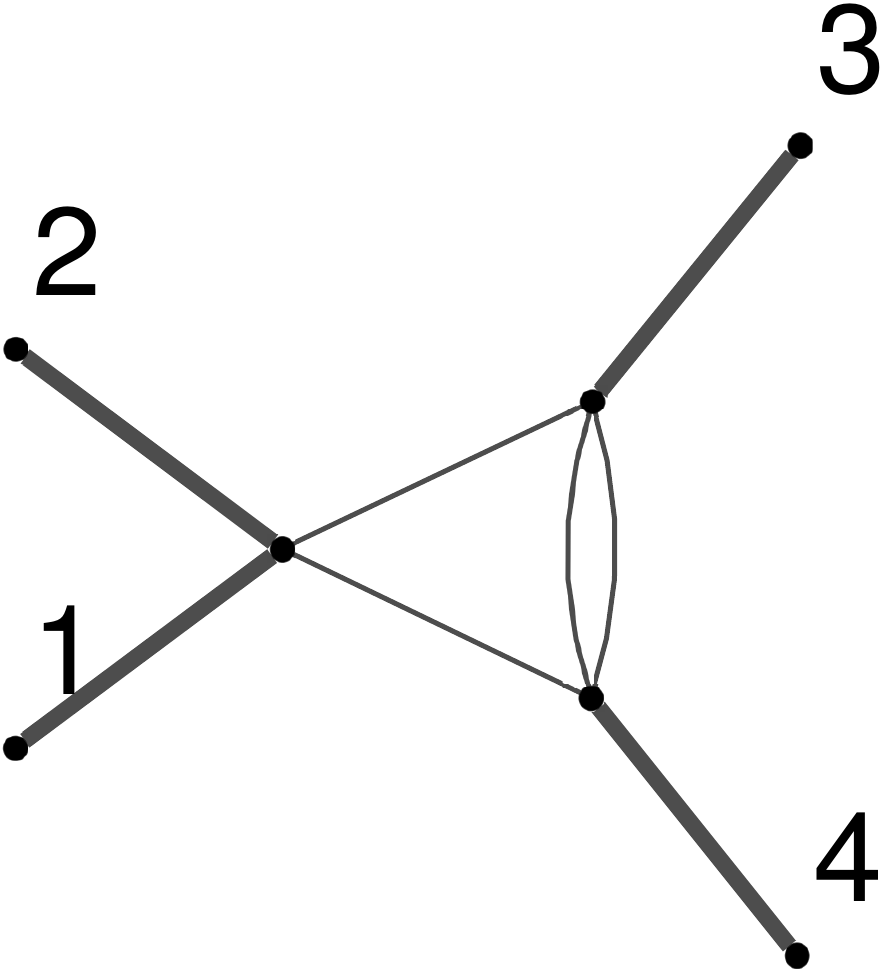}  
        \caption*{$\{1,3,5,7\}$ \\2 MIs}  
        \label{6}  
    \end{minipage}
    
\end{figure}

\begin{figure}[H] 
    \centering
    \begin{minipage}{0.18\linewidth}
        \centering
        \includegraphics[width=2.3cm,height=1.6cm]{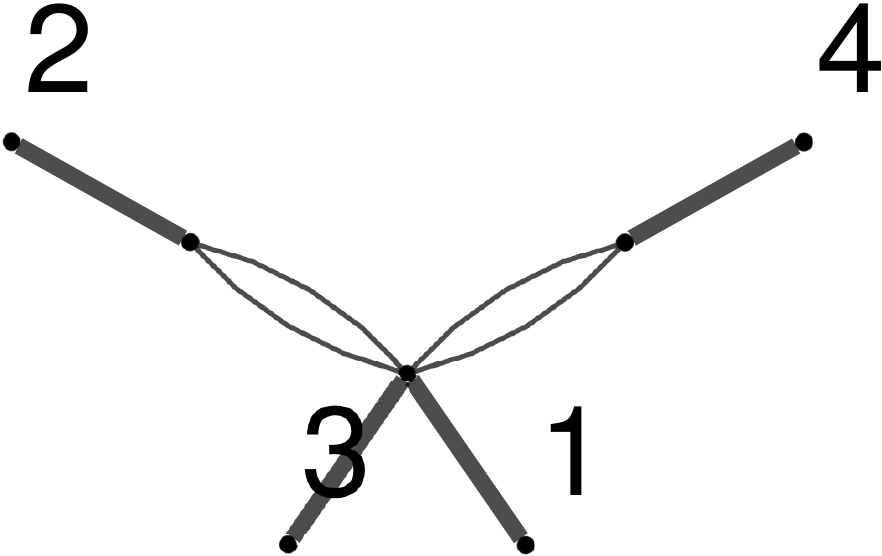}  
        \caption*{$\{2,3,5,6\}$ \\1 MI}  
        \label{5}  
    \end{minipage}
    \begin{minipage}{0.18\linewidth}
        \centering
        \includegraphics[width=2.3cm,height=1.6cm]{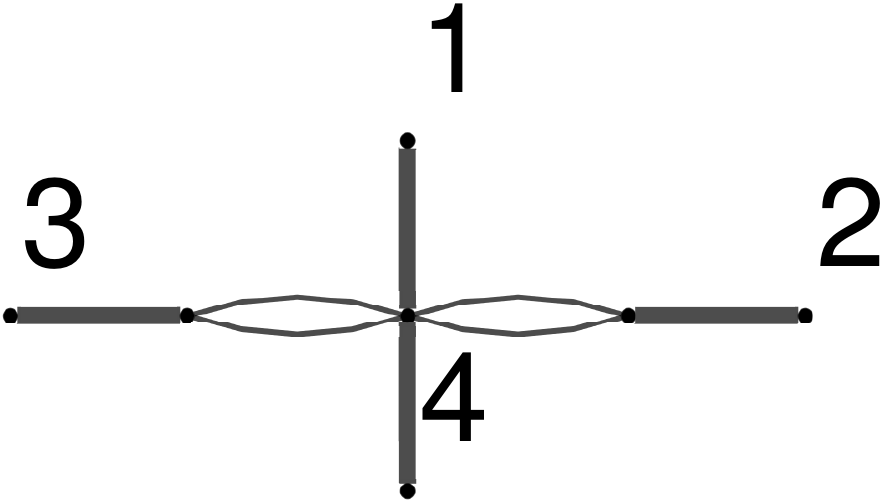}  
        \caption*{$\{2,3,4,5\}$ \\1 MI}  
        \label{4}  
    \end{minipage}
    \begin{minipage}{0.18\linewidth}
        \centering
        \includegraphics[width=2.3cm,height=1.6cm]{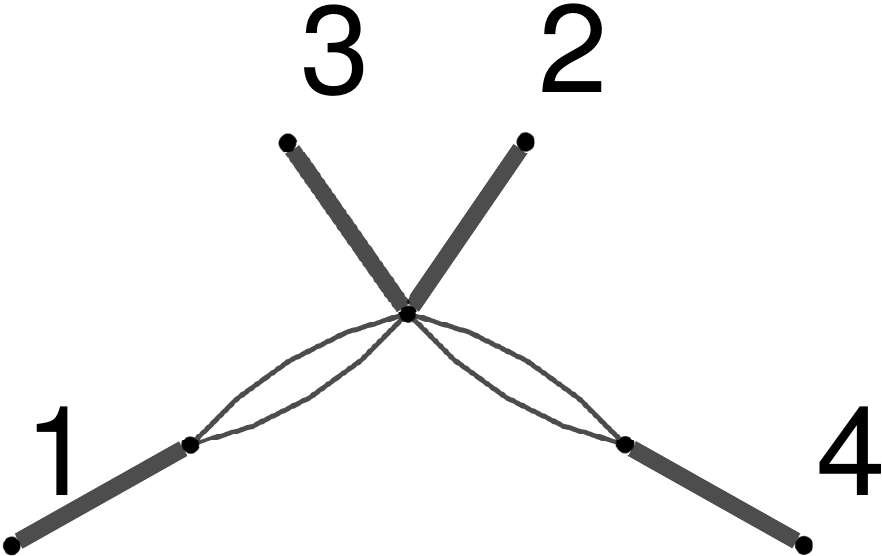}  
        \caption*{$\{1,2,5,6\}$ \\1 MI}  
        \label{5}  
    \end{minipage}
    \begin{minipage}{0.18\linewidth}
        \centering
        \includegraphics[width=2.3cm,height=1.6cm]{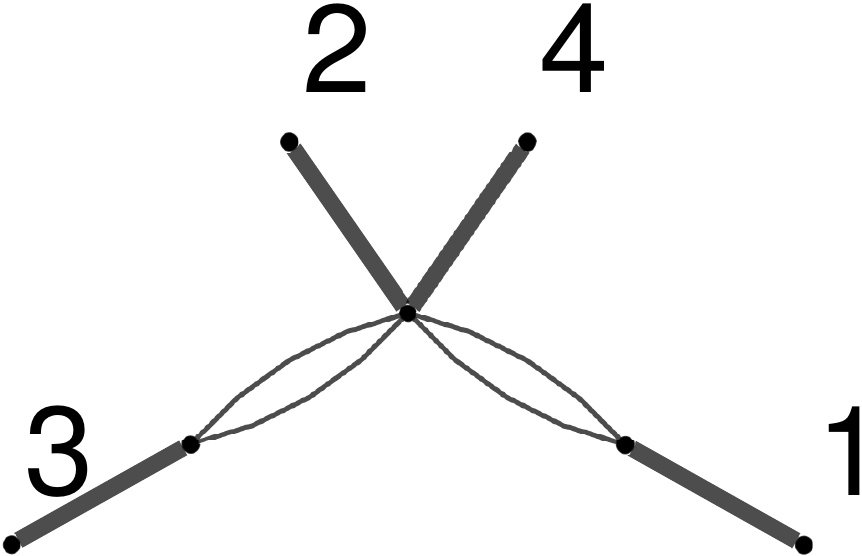}  
        \caption*{$\{1,2,4,5\}$ \\1 MI}  
        \label{5}  
    \end{minipage}
    \begin{minipage}{0.18\linewidth}
        \centering
        \includegraphics[width=2.3cm,height=1.6cm]{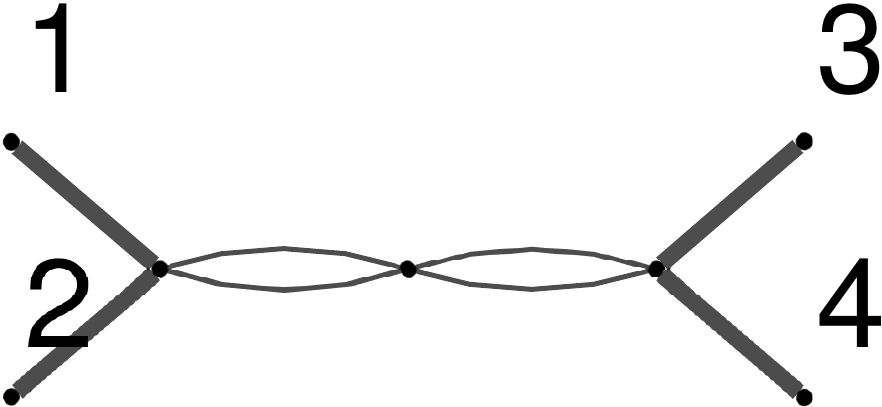}  
        \caption*{$\{1,3,4,6\}$ \\1 MI}  
        \label{5}  
    \end{minipage}
\end{figure}

\begin{figure}[H] 
    \centering   
   
    \begin{minipage}{0.18\linewidth}
        \centering
        \includegraphics[width=2.3cm,height=1.6cm]{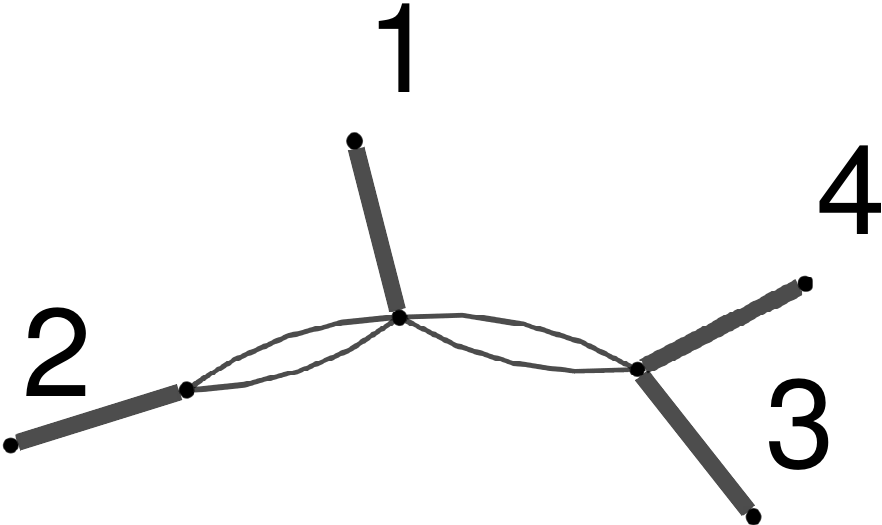}  
        \caption*{$\{2,3,4,6\}$ \\1 MI}  
        \label{6}  
    \end{minipage}
    \begin{minipage}{0.18\linewidth}
        \centering
        \includegraphics[width=2.3cm,height=1.6cm]{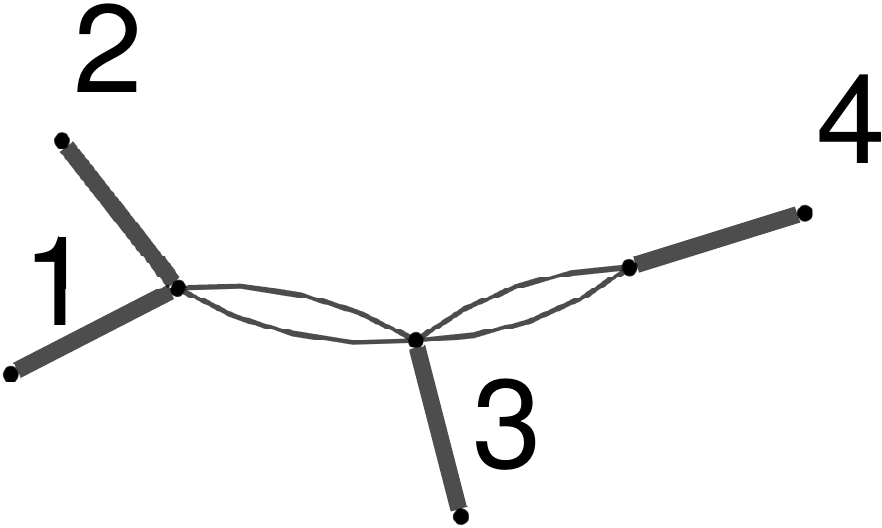}  
        \caption*{$\{1,3,5,6\}$ \\1 MI}  
        \label{4}  
    \end{minipage}
    \begin{minipage}{0.18\linewidth}
        \centering
        \includegraphics[width=2.3cm,height=1.6cm]{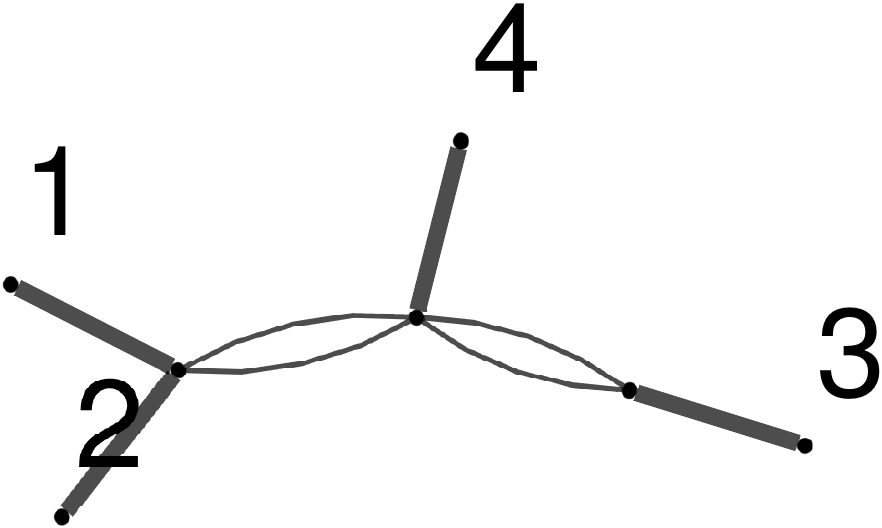}  
        \caption*{$\{1,3,4,5\}$ \\1 MI}  
        \label{6}  
    \end{minipage}
    \begin{minipage}{0.18\linewidth}
        \centering
        \includegraphics[width=2.3cm,height=1.6cm]{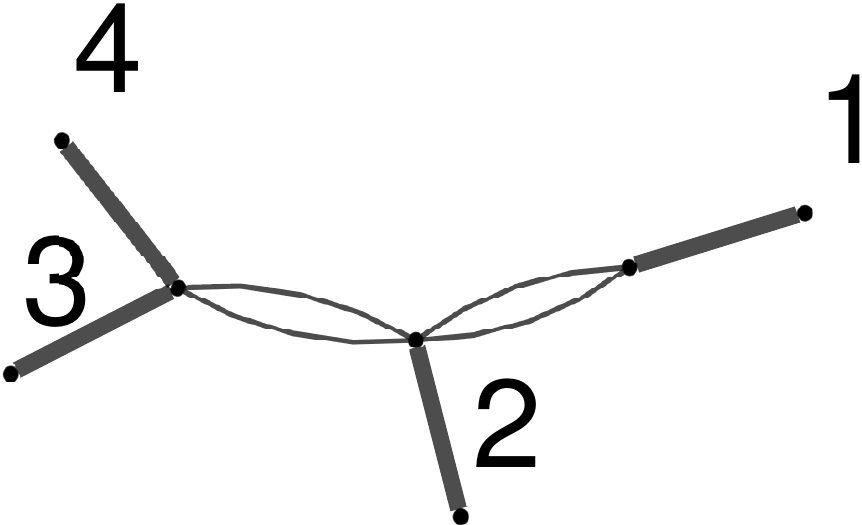}  
        \caption*{$\{1,2,4,6\}$ \\1 MI}  
        \label{4}  
    \end{minipage}
\end{figure}

\begin{figure}[H] 
    \centering    
    
    \begin{minipage}{0.18\linewidth}
        \centering
        \includegraphics[width=2.3cm,height=1.6cm]{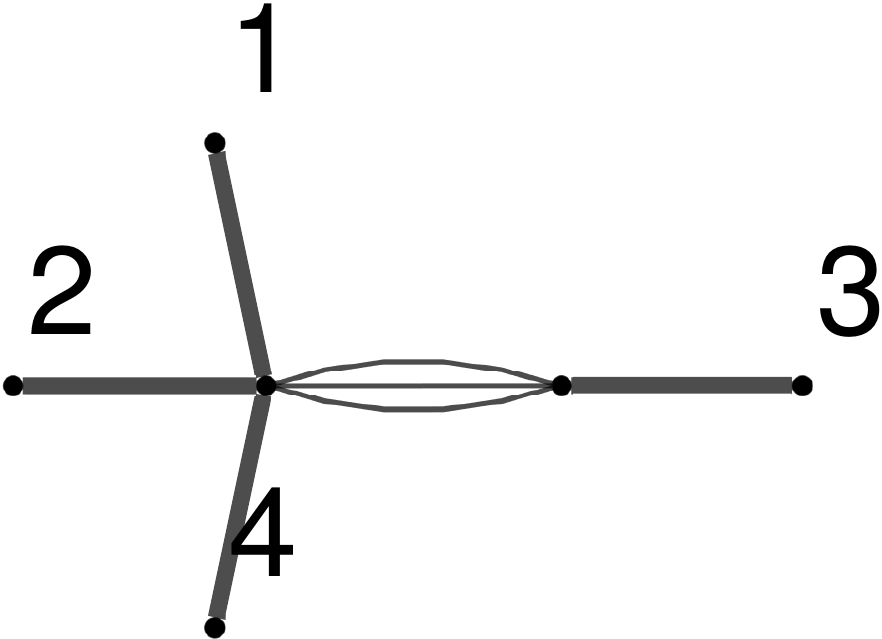}  
        \caption*{$\{3,5,7\}$ \\1 MI}  
        \label{6}  
    \end{minipage}
    \begin{minipage}{0.18\linewidth}
        \centering
        \includegraphics[width=2.3cm,height=1.6cm]{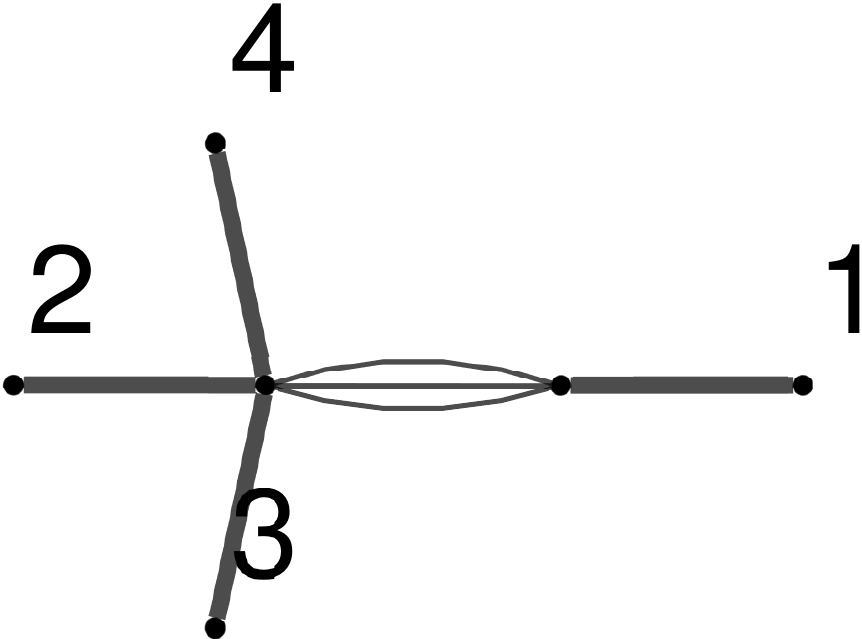}  
        \caption*{$\{2,5,7\}$ \\1 MI}  
        \label{4}  
    \end{minipage}
    \begin{minipage}{0.18\linewidth}
        \centering
        \includegraphics[width=2.3cm,height=1.6cm]{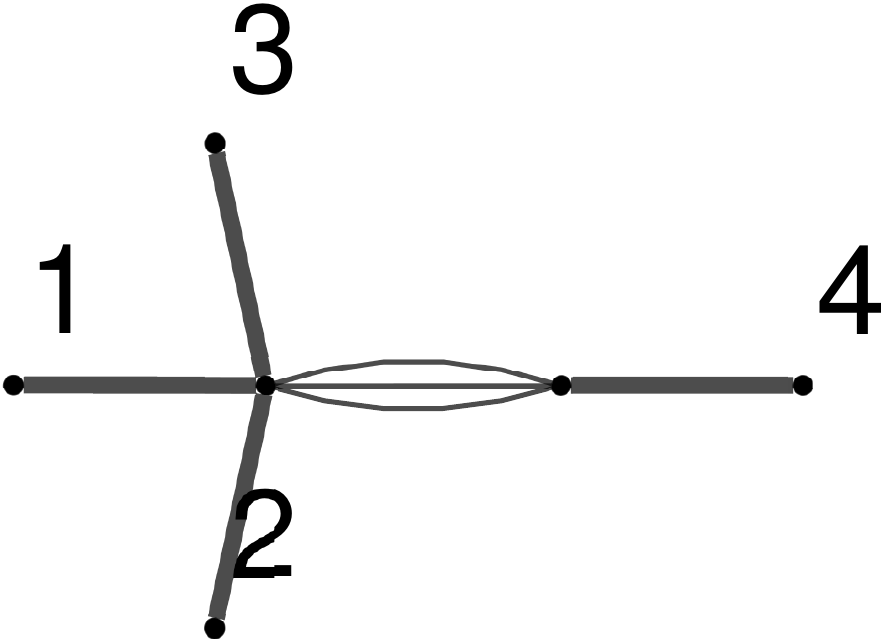}  
        \caption*{$\{1,5,7\}$ \\1 MI}  
        \label{4}  
    \end{minipage}
    \begin{minipage}{0.18\linewidth}
        \centering
        \includegraphics[width=2.3cm,height=1.6cm]{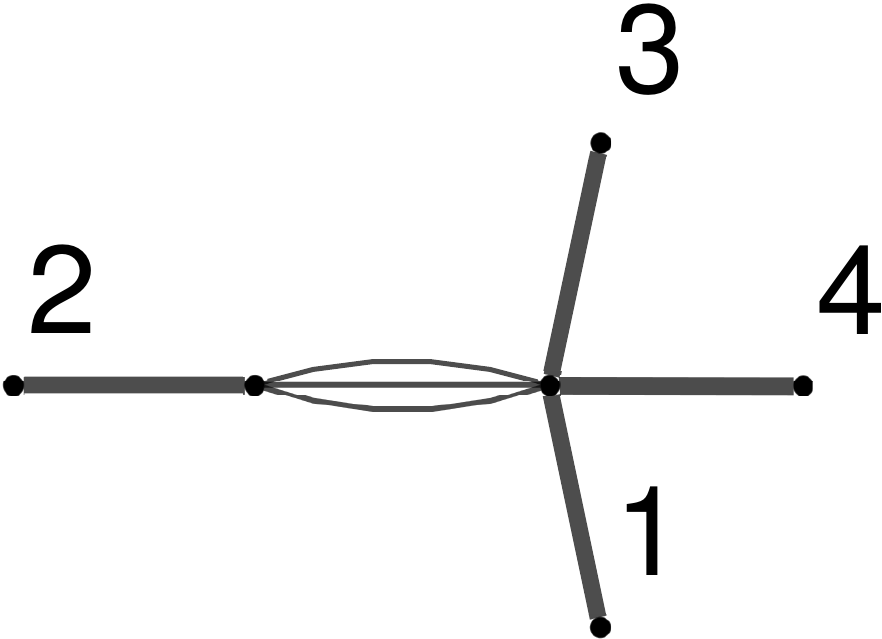}  
        \caption*{$\{2,4,7\}$ \\1 MI}  
        \label{6}  
    \end{minipage}
    
\end{figure}

\section{Polynomials for the definition of the symbol letters}
\label{appendix:f_polynomial}
The definition of polynomials $f_{35}\sim f_{68}$ that appear in the odd letters with 2 square roots, shown in \eqref{letterW35} $\sim$ \eqref{letterW68}, are as follows.

\begin{align}
f_{35}=& -s^2-2 s t+s m_1^2+s m_2^2+s m_3^2+m_1^2 m_3^2-m_2^2 m_3^2+s m_4^2-m_1^2 m_4^2+m_2^2 m_4^2,\\
f_{36}=& -s t+t m_1^2-s m_2^2-t m_2^2-m_1^2 m_2^2+m_2^4+s m_3^2-m_1^2 m_3^2-m_2^2 m_3^2+2 m_2^2 m_4^2,\\
f_{37}=& -s^2-s t+2 s m_1^2+t m_1^2-m_1^4+s m_2^2-t m_2^2+m_1^2 m_2^2+s m_3^2-m_1^2 m_3^2\nonumber\\&-m_2^2 m_3^2+2 m_2^2 m_4^2,\\
f_{38}=& -s t-s m_1^2-t m_1^2+m_1^4+t m_2^2-m_1^2 m_2^2+2 m_1^2 m_3^2+s m_4^2-m_1^2 m_4^2-m_2^2 m_4^2,\\
f_{39}=& -s^2-s t+s m_1^2-t m_1^2+2 s m_2^2+t m_2^2+m_1^2 m_2^2-m_2^4+2 m_1^2 m_3^2+s m_4^2\nonumber\\&-m_1^2 m_4^2-m_2^2 m_4^2,\\
f_{40}=& -s t+s m_1^2+t m_3^2-m_1^2 m_3^2-s m_4^2-t m_4^2-m_1^2 m_4^2+2 m_2^2 m_4^2-m_3^2 m_4^2+m_4^4,\\
f_{41}=& -s^2-s t+s m_1^2+2 s m_3^2+t m_3^2-m_1^2 m_3^2-m_3^4+s m_4^2-t m_4^2-m_1^2 m_4^2\nonumber\\&+2 m_2^2 m_4^2+m_3^2 m_4^2,\\
f_{42}=& -s t+s m_2^2-s m_3^2-t m_3^2+2 m_1^2 m_3^2-m_2^2 m_3^2+m_3^4+t m_4^2-m_2^2 m_4^2-m_3^2 m_4^2,\\
f_{43}=& -s^2-s t+s m_2^2+s m_3^2-t m_3^2+2 m_1^2 m_3^2-m_2^2 m_3^2+2 s m_4^2+t m_4^2-m_2^2 m_4^2\nonumber\\&+m_3^2 m_4^2-m_4^4,\\
f_{44}=& -s^2 t+s t m_1^2-s^2 m_2^2-s t m_2^2-s m_1^2 m_2^2+s m_2^4+s m_1^2 m_3^2-m_1^4 m_3^2+s m_2^2 m_3^2\nonumber\\&+2 m_1^2 m_2^2 m_3^2-m_2^4 m_3^2+2 s m_2^2 m_4^2,\\
f_{45}=& -s^2 t-s^2 m_1^2-s t m_1^2+s m_1^4+s t m_2^2-s m_1^2 m_2^2+2 s m_1^2 m_3^2+s m_1^2 m_4^2-m_1^4 m_4^2\nonumber\\&+s m_2^2 m_4^2+2 m_1^2 m_2^2 m_4^2-m_2^4 m_4^2,\\
f_{46}=& -s^2 t+s t m_1^2+s t m_2^2-2 s m_1^2 m_2^2+s m_1^2 m_3^2-m_1^4 m_3^2+m_1^2 m_2^2 m_3^2+s m_2^2 m_4^2\nonumber\\&+m_1^2 m_2^2 m_4^2-m_2^4 m_4^2,\\
f_{47}=& -s^2 t+s t m_1^2+s t m_2^2-2 s m_1^2 m_2^2+s^2 m_3^2-s m_1^2 m_3^2-2 s m_2^2 m_3^2-m_1^2 m_2^2 m_3^2\nonumber\\&+m_2^4 m_3^2+s m_2^2 m_4^2+m_1^2 m_2^2 m_4^2-m_2^4 m_4^2,\\
f_{48}=& -s^2 t+s t m_1^2+s t m_2^2-2 s m_1^2 m_2^2+s m_1^2 m_3^2-m_1^4 m_3^2+m_1^2 m_2^2 m_3^2+s^2 m_4^2\nonumber\\&-2 s m_1^2 m_4^2+m_1^4 m_4^2-s m_2^2 m_4^2-m_1^2 m_2^2 m_4^2,\\
f_{49}=& -s^2 t+s t m_3^2+s m_1^2 m_3^2-m_1^2 m_3^4-s^2 m_4^2-s t m_4^2+s m_1^2 m_4^2+2 s m_2^2 m_4^2\nonumber\\&-s m_3^2 m_4^2+2 m_1^2 m_3^2 m_4^2+s m_4^4-m_1^2 m_4^4,\\
f_{50}=& -s^2 t-s^2 m_3^2-s t m_3^2+2 s m_1^2 m_3^2+s m_2^2 m_3^2+s m_3^4-m_2^2 m_3^4+s t m_4^2+s m_2^2 m_4^2\nonumber\\&-s m_3^2 m_4^2+2 m_2^2 m_3^2 m_4^2-m_2^2 m_4^4,\\
f_{51}=& -s^2 t+s t m_3^2+s m_1^2 m_3^2-m_1^2 m_3^4+s t m_4^2+s m_2^2 m_4^2-2 s m_3^2 m_4^2+m_1^2 m_3^2 m_4^2\nonumber\\&+m_2^2 m_3^2 m_4^2-m_2^2 m_4^4,\\
f_{52}=& -s^2 t+s^2 m_1^2+s t m_3^2-s m_1^2 m_3^2+s t m_4^2-2 s m_1^2 m_4^2+s m_2^2 m_4^2-2 s m_3^2 m_4^2\nonumber\\&-m_1^2 m_3^2 m_4^2+m_2^2 m_3^2 m_4^2+m_1^2 m_4^4-m_2^2 m_4^4,\\
f_{53}=& -s^2 t+s^2 m_2^2+s t m_3^2+s m_1^2 m_3^2-2 s m_2^2 m_3^2-m_1^2 m_3^4+m_2^2 m_3^4+s t m_4^2\nonumber\\&-s m_2^2 m_4^2-2 s m_3^2 m_4^2+m_1^2 m_3^2 m_4^2-m_2^2 m_3^2 m_4^2,\\
f_{54}=& -2 s t-t^2+t m_1^2+t m_2^2-m_1^2 m_2^2+t m_3^2+m_1^2 m_3^2+t m_4^2+m_2^2 m_4^2-m_3^2 m_4^2,\\
f_{55}=& -s t-t^2+t m_1^2-s m_2^2+t m_2^2-m_1^2 m_2^2+s m_3^2+2 t m_3^2-m_1^2 m_3^2+m_2^2 m_3^2\nonumber\\&-m_3^4+2 m_2^2 m_4^2,\\
f_{56}=& -s t-t^2+s m_1^2+2 t m_1^2-m_1^4+t m_3^2-m_1^2 m_3^2-s m_4^2+t m_4^2+m_1^2 m_4^2\nonumber\\&+2 m_2^2 m_4^2-m_3^2 m_4^2,\\
f_{57}=& -s t-t^2+s m_2^2+2 t m_2^2-m_2^4-s m_3^2+t m_3^2+2 m_1^2 m_3^2+m_2^2 m_3^2+t m_4^2\nonumber\\&-m_2^2 m_4^2-m_3^2 m_4^2,\\
f_{58}=& -s t-t^2-s m_1^2+t m_1^2+t m_2^2-m_1^2 m_2^2+2 m_1^2 m_3^2+s m_4^2+2 t m_4^2+m_1^2 m_4^2\nonumber\\&-m_2^2 m_4^2-m_4^4,\\
f_{59}=& -s t^2+s t m_2^2+s t m_3^2+t m_1^2 m_3^2-2 t m_2^2 m_3^2+m_1^2 m_2^2 m_3^2-m_1^2 m_3^4+t m_2^2 m_4^2\nonumber\\&-m_2^4 m_4^2+m_2^2 m_3^2 m_4^2,\\
f_{60}=& -s t^2+2 s t m_2^2-s m_2^4+s t m_3^2+t m_1^2 m_3^2+s m_2^2 m_3^2-t m_2^2 m_3^2-m_1^2 m_2^2 m_3^2\nonumber\\&+m_2^4 m_3^2-m_1^2 m_3^4-m_2^2 m_3^4+2 m_2^2 m_3^2 m_4^2,\\
f_{61}=& -s t^2+s t m_2^2+2 s t m_3^2+s m_2^2 m_3^2-t m_2^2 m_3^2+2 m_1^2 m_2^2 m_3^2-m_2^4 m_3^2-s m_3^4\nonumber\\&+m_2^2 m_3^4+t m_2^2 m_4^2-m_2^4 m_4^2-m_2^2 m_3^2 m_4^2,\\
f_{62}=& -s t^2+s t m_1^2+t m_1^2 m_3^2-m_1^4 m_3^2+s t m_4^2-2 t m_1^2 m_4^2+t m_2^2 m_4^2+m_1^2 m_2^2 m_4^2\nonumber\\&+m_1^2 m_3^2 m_4^2-m_2^2 m_4^4,\\
f_{63}=& -s t^2+2 s t m_1^2-s m_1^4+s t m_4^2+s m_1^2 m_4^2-t m_1^2 m_4^2+m_1^4 m_4^2+t m_2^2 m_4^2\nonumber\\&-m_1^2 m_2^2 m_4^2+2 m_1^2 m_3^2 m_4^2-m_1^2 m_4^4-m_2^2 m_4^4,\\
f_{64}=& -s t^2+s t m_1^2+t m_1^2 m_3^2-m_1^4 m_3^2+2 s t m_4^2+s m_1^2 m_4^2-t m_1^2 m_4^2-m_1^4 m_4^2\nonumber\\&+2 m_1^2 m_2^2 m_4^2-m_1^2 m_3^2 m_4^2-s m_4^4+m_1^2 m_4^4,\\
f_{65}=& -s^2 t^2+s^2 t m_1^2+s t m_1^2 m_3^2-s m_1^4 m_3^2-s t m_1^2 m_4^2+2 s t m_2^2 m_4^2-s m_1^2 m_2^2 m_4^2\nonumber\\&+2 s m_1^2 m_3^2 m_4^2-m_1^4 m_3^2 m_4^2+m_1^2 m_2^2 m_3^2 m_4^2+m_1^2 m_2^2 m_4^4-m_2^4 m_4^4,\\
f_{66}=& -s^2 t^2+s^2 t m_2^2+2 s t m_1^2 m_3^2-s t m_2^2 m_3^2-s m_1^2 m_2^2 m_3^2-m_1^4 m_3^4+m_1^2 m_2^2 m_3^4\nonumber\\&+s t m_2^2 m_4^2-s m_2^4 m_4^2+2 s m_2^2 m_3^2 m_4^2+m_1^2 m_2^2 m_3^2 m_4^2-m_2^4 m_3^2 m_4^2,\\
f_{67}=& -s^2 t^2+s^2 t m_3^2+s t m_1^2 m_3^2-s t m_2^2 m_3^2+2 s m_1^2 m_2^2 m_3^2-s m_1^2 m_3^4-m_1^2 m_2^2 m_3^4\nonumber\\&+2 s t m_2^2 m_4^2-s m_2^2 m_3^2 m_4^2+m_1^2 m_2^2 m_3^2 m_4^2+m_2^4 m_3^2 m_4^2-m_2^4 m_4^4,\\
f_{68}=& -s^2 t^2+2 s t m_1^2 m_3^2-m_1^4 m_3^4+s^2 t m_4^2-s t m_1^2 m_4^2+s t m_2^2 m_4^2+2 s m_1^2 m_2^2 m_4^2\nonumber\\&-s m_1^2 m_3^2 m_4^2+m_1^4 m_3^2 m_4^2+m_1^2 m_2^2 m_3^2 m_4^2-s m_2^2 m_4^4-m_1^2 m_2^2 m_4^4
\end{align}

\bibliographystyle{JHEP}
\bibliography{main}

\end{document}